\documentclass[twocolumn,twocolappendix]{aastex631}
\usepackage{amsmath}
\usepackage{amsfonts}
\usepackage{graphicx}
\usepackage{rotating}
\usepackage{natbib}
\usepackage{txfonts}
\usepackage{color}
\usepackage{wrapfig}
\usepackage{amssymb}
\usepackage{color}
\usepackage{tikz}

\usepackage{placeins}  % Add this in the preamble
\usepackage{float}  % Ensures better float control
\usepackage{soul} 
\usepackage[normalem]{ulem}
\usepackage{lipsum}
\shorttitle{} 
\shortauthors{} 

\begin{document}

\title{ALMA Observations of Cold Methanol Gas in the Large Magellanic Cloud (LMC): N79 South GMC}

\correspondingauthor{Suman Kumar Mondal} 
\email{sumanphys39@gmil.com} 

\author[0000-0002-7657-1243]{Suman Kumar Mondal}
\affiliation{S. N. Bose National Centre for Basic Sciences, Salt Lake, JD Block, Sector 3, Bidhannagar, Kolkata, 700106, West Bengal, India}
\author[0000-0002-0095-3624]{Takashi Shimonishi} 
\affiliation{Institute of Science and Technology, Niigata University, Ikarashi-nihoncho 8050, Nishi-ku, Niigata 950-2181, Japan}
\author{Soumen Mondal}
\affiliation{S. N. Bose National Centre for Basic Sciences, Salt Lake, JD Block, Sector 3, Bidhannagar, Kolkata, 700106, West Bengal, India}
\author[0000-0003-1602-6849]{Prasanta Gorai}
\affiliation{Rosseland Centre for Solar Physics, University of Oslo, PO Box
1029 Blindern, 0315 Oslo, Norway}
\affiliation{Institute of Theoretical Astrophysics, University of Oslo, PO Box
1029 Blindern, 0315 Oslo, Norway}
\author[0000-0002-6907-0926]{Kei E. I. Tanaka}
\affiliation{Department of Earth and Planetary Sciences, Institute of Science Tokyo, Meguro, Tokyo, 152-8551, Japan}
\author{Kenji Furuya}
\affiliation{RIKEN Pioneering Research Institute, 2-1 Hirosawa, Wako-shi, Saitama 351-0198, Japan}
\author[0000-0003-4615-602X]{Ankan Das}
\affiliation{Institute of Astronomy Space and Earth Science, P-177, CIT Road, Scheme 7m, Ultadanga Station, Kolkata 700054, West Bengal, India}
%%%%%%%%%%%%%%%%%%%%%%%%%%%%%%%%%%%%%%%%%%%%%%%%%%%%% 
\begin{abstract}
We report ALMA continuum and molecular line observations at 0.1 pc resolution toward the super star cluster (SSC) candidate H72.97-69.39 in the N79 region of the LMC. The continuum emission has a sharp peak around the SSC candidate but is also widely distributed. We identify two continuum sources at the northern (N79S-1) and northwestern (N79S-2) positions of the  SSC continuum peak, associated with CH$_3$OH emission. In addition to CH$_3$OH, we also detect H$_2$CO, H$_2$CS, CS, SO, CO, CN, and CCH at the positions of N79S-1 and N79S-2. The rotation diagram analysis of CH$_3$OH and SO lines yields an average gas temperature of 13 $\pm$ 0.4 K for N79S-1 and 15 $\pm$ 0.9 K for N79S-2. Most emission lines exhibit line widths of less than 2.8 km s$^{-1}$, consistent with emissions from cold, dense molecular cloud cores. The abundance of cold CH$_3$OH gas is estimated to be (2.1 $\pm$ 1.1)$\times$ 10$^{-9}$ at N79S-1 and (4.5 $\pm$2.5)$\times$ 10$^{-10}$ at N79S-2. Despite the lower metallicity in the LMC, the CH$_3$OH abundance at N79S-1 is comparable to that of similar cold sources in our Galaxy. However, the formation of organic molecules is inhibited throughout the N79 regions, as can be seen in the non-detection of CH$_3$OH in most of the regions. The two positions N79S-1 and N79S-2 would be exceptional positions, where CH$_3$OH production is efficient. The possible origins of cold CH$_3$OH gas in these dense cores are discussed, along with a possible explanation for the non-detection of CH$_3$OH in the SSC candidate.  
\end{abstract} 
%%%%%%%%%%%%%%%%%%%%%%%%%%%%%%%%%%%%%%%%%%%%%%%%%%%%%
%%%%%%%%%%%%%%%%%%%%%%%%%%%%%%%%%%%%%%%%%%%%%%%%%%%%%%%%%
\section{Introduction \label{sec_introduction}} 
%Low metallicity 
% COMs and Hot cores
In astrochemistry, molecules that contain six or more atoms, including carbon, are classified as complex organic molecules (COMs, \citealt{herb09}). These molecules are expected to play a crucial role in prebiotic chemistry and are believed to be fundamental ingredients for the emergence of life \citep{garr13,nova14,jime20,das24}. COMs were detected at different stages of star formation, from dense molecular clouds to protoplanetary disks in the Galaxy\footnote{\url{(https://cdms.astro.uni-koeln.de/classic/molecules)}}\citep{mcgu18,gora21,mond21,mond23,brat23,boot24}. 
Hot molecular cores (HMCs) represent an early stage of high-mass star formation, whereas their low-mass analogs, called hot corinos (HCs), show rich molecular lines in the infrared (IR) and radio wavelengths \citep{kurt00,van04,herb09}. To date, several HMCs and HCs have been discovered in star-forming regions (e.g., Orion-KL; \citealt[][and references therein]{feng15}, Sgr B2(N); \citealt{bell09,bell13,bell16,mull16,li24}) and in low-mass protostellar environments (e.g., IRAS 16293–2422; \citealt{mani20}, NGC 1333-IRAS 4A/B; \citealt{lope7}, B1-C and S68N; \citealt{van20}).
%Hot molecular cores (HMCs) are one of the early stages of star formation and show rich molecular lines in the infrared and radio wavelengths \citep{kurt00,van04,herb09}.
These COMs formed on grain mantles during the cold prestellar phase and through gas-phase reactions after the parent species were desorbed from the grains by stellar radiation or shock \citep{van98,garr08,case12,aika13}. Methanol (CH$_3$OH) is one of the simplest COM and is frequently observed in star-forming regions of the Galaxy \citep{ball70,sutt85,taka00,mare05,wals16}. It is mainly formed on grain surfaces by successive hydrogenation of CO and released into the gas via sublimation \citep{wata02,hida04,wirs11}. CH$_3$OH  plays a crucial role in forming other COMs in star-forming regions and may serve as the precursor of large organic species \citep{garr06,nomu04,ober09a,chua16,taqu16}. \\
%LMC
Low metallicity significantly affects the physical and chemical processes in the molecular cloud \citep{rich16,sewi19,guad22,shim24}. Thus, understanding the chemical complexity in low metallicity environments is crucial to unveil the chemical and physical processes relevant to star formation at earlier epochs of cosmic evolution \citep{bale07,rafe12,narl22}. The Large Magellanic Cloud (LMC) is an excellent target for studying the formation and survival of COMs in low metallicity regions. The LMC is located at a distance of 49.97 $\pm$ 1.1 kpc \citep{piet13}, with a metallicity approximately 30-50\% of the solar value \citep {west90,russ92,roll02,chou16,chou21,lah24,omku25}. %The interstellar ultraviolet (UV) radiation field in the LMC is about 10 to 100 times stronger than typical Galactic values, \textbf{ integrated photon flux of approximately $\sim 10^{8}$ photons cm$^{-2}$ s$^{-1}$ in the far-UV (FUV) band \citep{brow03}. 
The interstellar ultraviolet (UV) radiation field in the LMC is about 10 to 100 times stronger than typical Galactic values \citep{brow03}, with a typical Galactic integrated photon flux of approximately $\sim10^{8}$ photons cm$^{-2}$ s$^{-1}$ in the far-UV band \citep{drai78, parr03, bial20}.
Additionally, gamma-ray observations indicate that the cosmic-ray density in the LMC is $\sim$ 25\% of that in the solar neighborhood \citep{abdo10}. The interstellar radiation field is less attenuated 
%\citep[A$_v$ $\sim$ 0.4,][]{roma14} 
due to the low dust-to-gas ratio \citep[0.0027,][]{bern08}, which warms the dust grain.
The above environmental divergences would result in a distinct chemical and physical history of star-forming regions in the LMC. \\
%This diverse environment presents obstacles for the synthesis of molecules in  star-forming regions of LMC.  
%Previous studies  
\cite{nish16} reported a deficiency of CH$_3$OH gas in molecular clouds of the LMC based on spectral line surveys of seven molecular clouds. The lower detection rate of CH$_3$OH masers (Class II maser at 6.7 GHz and 12.2 GHz) was reported in star-forming regions of LMC than in the Galaxy \citep[e.g.,][]{gree08,elli10}. Based on IR observations of ices, the solid CH$_3$OH around embedded high-mass Young Stellar Objects (YSOs) in the LMC is less abundant than similar Galactic objects \citep{shim16a}. They suggest that the warm ice chemistry is responsible for the low abundance of CH$_3$OH in the LMC. %They argue that mobile CO on warm dust surfaces can efficiently react with OH to produce CO$_2$, which leads to the inefficient production of CH$_3$OH.
The hypothesis suggests that high dust temperatures inhibit the hydrogenation of carbon monoxide (CO) on the grain surface, which leads to the inefficient production of CH$_3$OH ice. At elevated dust temperatures (around 20 K), hydrogen atoms tend to evaporate quickly from the surface, resulting in a reduced number of available hydrogen atoms on the surface. Conversely, the increased mobility of CO at these temperatures may enhance the formation of carbon dioxide (CO$_2$) through the reaction CO + OH $\rightarrow$ CO$_2$ + H. Astrochemical models developed for the LMC environment quantitatively characterize these temperature effects and provide theoretical support for this hypothesis \citep{acha15, acha18, paul18,shim20}.
Previous observations reported increased CO$_2$/H$_2$O ice ratios toward LMC YSOs \citep{shim08,shim10,oliv09,seal11}.\\
Based on recent ALMA (Atacama Large Millimeter/submillimeter Array) observations of HMCs in the LMC, the abundance of CH$_3$OH gas shows a large variation (a factor of $\sim$ 25) among the LMC HMCs \citep{shim16b,shim20,sewi18,sewi22}.
The abundances of CH$_3$OH gas in these LMC HMCs were estimated with respect to the hydrogen column density ($N_{H_2}$), which was measured from dust continuum emission.
There are organic-poor HMCs such as ST11 and ST16, while N113 A1, N113 B3, N105-2A and N105-2B are organic-rich. The abundance of CH$_3$OH  is underabundant by one to three orders of magnitude in organic-poor HMCs compared to Galactic HMCs \citep[typical abundances relative to H$_2$ in  Galactic HMCs of 10$^{-7}$ –10$^{-8}$,][]{hern19}.
%\textbf{(typical abundances relative to H$_2$ of 10$^{-7}$ –10$^{-8}$, although sometimes the abundance reaches 10$^{-6}$).} 
In contrast, it is roughly scaled with metallicity in organic-rich HMCs compared to Galactic HMCs. \\ 
%This paper
 In this study, we report the detection of cold ($\leq$ 15 K) CH$_3$OH gas towards two dense cores, which are located $\sim$ 1.2 pc offset from the super star cluster (SSC) candidate (H72.97-69.39) in the N79 region of the LMC. This paper is organized as follows. The details of the target source, observations, and data reduction are described in Section \ref{sec_target} and Section \ref{sec_observations}. The obtained spectra and images are presented in Section \ref{sec_results}. The analysis of the continuum and spectral line data is given in Section \ref{sec_analysis}. Section \ref{sec_discussion} describes the distribution of molecular emission, a discussion of molecular abundance, and the properties of sources. Finally, the conclusion is given in Section \ref{sec_conclution}. 

%%%%%%%%%%%%%%%%%%%%%%%%%%%%%%%%%%%%%%%%%%%%%%%%%%%%%%%%%%%%%%%%%%%%
\section{Overview of Target Source} \label{sec_target} 
%We observe the massive \textbf{($>$ 8 M$_\odot$)} YSOs, HSOBMHERICC J72.971176-69.391112 (hereafter H72.97-69.39), located in the N79 South GMC (giant molecular cloud) of the LMC (\citealt{seal14}, based on Herschel observations).
We observe the massive ($>$ 8 M$_\odot$) YSOs, HSOBMHERICC J72.971176–69.391112 (hereafter H72.97–69.39), located in the N79 South GMC (giant molecular cloud) of the LMC, as identified by Herschel observations \citep{seal14}.
It is the most luminous ($2.2 \times 10^6$ L$\odot$) IR compact source, and this source has the potential to evolve into an SSC candidate ( M $>$ 10$^5$ M$_\odot$) \citep{ochs17}. Prior observations of H72.97-69.39 suggest that the source is young (timescale of the outflows to be $\sim$ 6.5$\times$10$^4$ yr) and just beginning to ionize the surrounding gas \citep{seal09,naya19,naya21}. This source shows an extremely rich near-IR spectrum with many emission lines from multiple species \citep{reit19}. \cite{naya19} revealed with ALMA observations, this source is located at the center of two colliding filaments, and it has a higher outflow rate (0.02 M$_\odot$ yr$^{-1}$ based on the estimation of $^{13}$CO observation) compared to massive YSOs in the Milky Way. In this work, we study the relatively quiescent molecular cloud that is not directly associated with the HII region. 
%%%%%%%%%%%%%%%%%%%%%%%%%%%%%%%%%%%%%%%%%%%%%%%%%%%%%%%% 

 \section{ALMA Observations} \label{sec_observations}
Observations were carried out with ALMA as part of the Cycle 5 (2017.1.01323.S) and Cycle 6 (2018.1.01366.S) programs (PI: T. Shimonishi). The target was observed using 46 antennas with the shortest and longest baseline lengths of 15.1 - 1213.4 m for Band 6 and 50 antennas with shortest and longest baseline lengths of 15.1 - 1397.8 m for Band 7. The telescopes were pointed to R.A. = 04$^\mathrm{h}$51$^\mathrm{m}$53$\fs$290 and Decl. = -69$^\circ$23$\arcmin$28$\farcs$600 (J2000). The total on-source integration time was 27.2 minutes for Band 6 data and 36.3 minutes for Band 7. The 12-m Array Configuration C43-4  was used in the observations. Four spectral windows covering the sky frequencies of 241.23–243.1, 243.59 - 245.46, 256.72–258.60, and 258.58–260.46 GHz for Band 6 and 336.95–338.82, 338.75-340.62, 348.83–350.71, and 350.63–352.51 GHz for Band 7 were set up to observe the target source. The spectral resolution was 0.98 MHz, which corresponds to 1.1 km s$^{-1}$ for Band 6 (250.84186 GHz) and 0.86 km s$^{-1}$ for Band 7 (344.72671 GHz). The primary beam full width at half maximum (FWHM) was about 25 $\arcsec$ at 250.84186 GHz and 18 $\arcsec$ at 344.72671 GHz. The maximum recoverable angular scale was about 5.0 $\arcsec$ at 250.84186 GHz and 3.7 $\arcsec$ at 344.72671 GHz. The mean precipitable water vapor (PWV) is 1.6 - 1.8 mm for Band 6 and 0.7 - 0.9 mm for Band 7.\\
The raw visibility data were calibrated using \textit{Common Astronomy Software Applications} (CASA) package (CASA 5.1.1 was used for Band 6, and CASA 5.4.0 was used for Band 7). The bandpass and flux calibrator was J0635-7516 for Band 6 and  J0519-4546 for Band 7, while the phase calibrator was J0437-7148 for Band 6 and J0529-7245 for Band 7.
 %For imaging, we use CASA 5.4.0 for all data. 
 The measurement sets are cleaned by the CASA task ``tclean'' with the Briggs weighting and the robustness parameter of 0.5, using a common circular restoring beam size of 0$\farcs$4 for Band 6 and 7 data in order to accommodate the spectral analyses in separated frequency regions. This beam size corresponds to 0.097 pc at the distance of the LMC (49.97 kpc). The continuum image is created from line-free channels of the four spectral windows, and the clean process is then carried out. We use the ``impbcor'' task in CASA for the primary beam correction. The self-calibration is not applied. We use the ``uvcontsub'' task in CASA to subtract the continuum emission and construct the line image of each spectrum window. The rms noise in the primary beam (PB)-corrected image ranges from 135 to 159 mK for Band 6 and 39 to 46 mK for Band 7. The rms noise is measured at CH$_3$OH peak positions, and it is non-uniform across the field. 
%%%%%%%%%%%%%%%%%%%%%%%%%%%%%%%%%%%%%%%%%%%%%%%%%%%%%%%%%%%%%%%%%%%%%%%%%%%
\begin{deluxetable}{ l c c c c }
\tablecaption{Positions of sources and CH$_3$OH peak   \label{tab_coor}}
\tablecolumns{4}
\tablewidth{0pt}
\tabletypesize{\footnotesize} 
\tablehead{
%\colhead{}  &      \multicolumn{2}{c}{Center}& \colhead{}  \\
%\cline{2-3}   \\
\colhead{Position} & \colhead{R.A. (J2000)} & \colhead{Decl.(J2000)}& \colhead{Remarks} }
\startdata
H72.97-69.39&04$^\mathrm{h}$51$^\mathrm{m}$53$\fs$08&-69$^\circ$23$\arcmin$28$\farcs$00&Herschel coordinates$^a$ \\
Continuum peak&04$^\mathrm{h}$51$^\mathrm{m}$53$\fs$319&-69$^\circ$23$\arcmin$28$\farcs$764&toward SSC\\
N79S-1&04$^\mathrm{h}$51$^\mathrm{m}$53$\fs$280&-69$^\circ$23$\arcmin$24$\farcs$700&peak position \\
N79S-2&04$^\mathrm{h}$51$^\mathrm{m}$52$\fs$913&-69$^\circ$23$\arcmin$24$\farcs$142&peak position\\
CH$_3$OH peak&04$^\mathrm{h}$51$^\mathrm{m}$53$\fs$270&-69$^\circ$23$\arcmin$24$\farcs$382&at N79S-1\\
CH$_3$OH peak&04$^\mathrm{h}$51$^\mathrm{m}$52$\fs$905&-69$^\circ$23$\arcmin$24$\farcs$140&at N79S-2 \\
 \enddata
 \tablecomments{
$^a$ \citep[Herschel coordinates of H72.97-69.39,][]{seal14}}
\end{deluxetable}
%%%%%%%%%%%%%%%%%%%%%%%%%%%%%%%%%%%%%%%%%%%%%%%%%%%%%%%%%%%%%%%%%%%%%%%%%%%%%%%%%%
\section{Results} \label{sec_results}
\subsection{Continuum Emission} \label{sec:cont-emis}
\begin{figure*}
\begin{minipage}{0.30cm}
\vspace*{-0.3cm}
\small
\rotatebox{90}{\textcolor{black}{Declination (J2000)}}
\end{minipage}
\begin{minipage}{\textwidth}
 \begin{tikzpicture}
\node[anchor=south west, inner sep=0] (image1) at (0,0) 
 %\hspace{-5.0cm}
{\includegraphics[width=\textwidth]{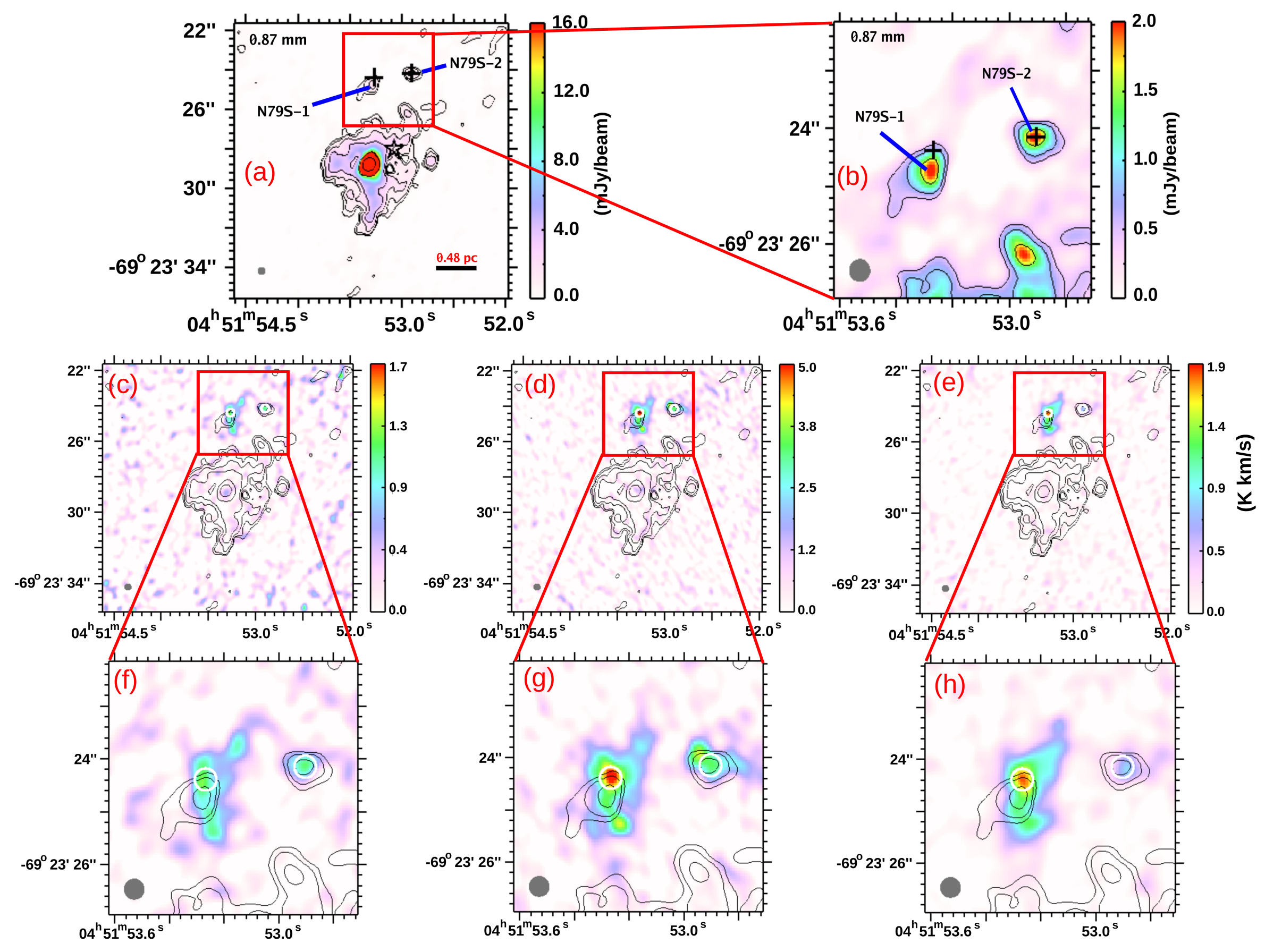}};
% Draw red lines to zoom-in areas
\begin{scope}[x={(image1.south east)},y={(image1.north west)}]
\node[black, font=\scriptsize\bfseries] at (0.823, 0.635) {CH$_3$OH (64 K$<$E$_u$$>$91 K)};
      \node[black, font=\scriptsize\bfseries] at (0.51, 0.635) {CH$_3$OH (34 K$<$E$_u$$>$61 K)};
      \node[black, font=\scriptsize\bfseries] at (0.19, 0.635) {CH$_3$OH (E$_u$= 16 K)};
\end{scope}
\end{tikzpicture}
\vspace*{0.1cm}\hspace*{8.0cm}\small \textcolor{black}{Right Ascension (J2000)}
\end{minipage}
\caption{Flux distributions of the 0.87 mm continuum (top-left panel, a) and integrated intensity distributions of CH$_3$OH (middle row panels, c, d, and e). Black contours represent the continuum emission, with contour levels at 3$\sigma$, 5$\sigma$, 10$\sigma$, 20$\sigma$, 100$\sigma$, and 300$\sigma$ of the rms noise ($\sigma$ = 0.158 $,\mathrm{mJy,beam^{-1}}$). The black cross in panels (a, b) indicates the positions of the CH$_3$OH peak emission. The black star marks the position of the SSC candidate (Herschel coordinates). The blue line in panels (a, b) indicates the positions of N79S-1 and N79S-2. The synthesized beam size (0$\farcs$4) is shown as a gray-filled circle in each panel. Two white open circles mark the positions from which spectra are extracted in this work. The top-right panel (b) presents a zoomed-in view of N79S-1 and N79S-2, while the bottom panels (f, g, h) show zoomed-in views of the integrated intensity distributions of CH$_3$OH in N79S-1 and N79S-2.}
\label{Fig:cont-imag}
\end{figure*}
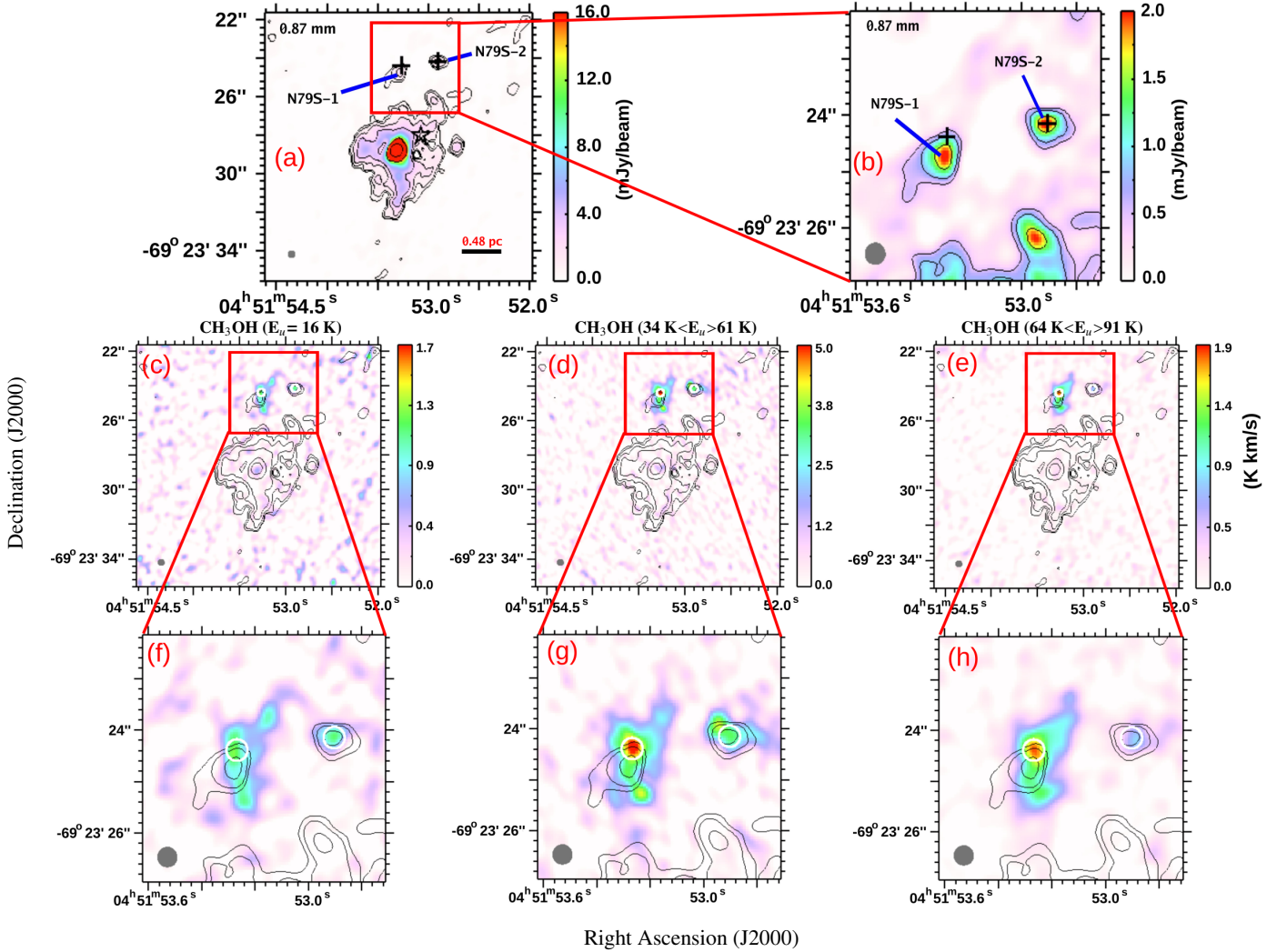

Figure \ref{Fig:cont-imag} shows the 0.87 mm continuum emission toward the target region. The rms noise of the continuum image is 0.158 $\,\mathrm{mJy\,beam^{-1}}$. The continuum emission is widely distributed ($\sim$ 1.5 pc) within the observed field, as shown by the 3$\sigma$ contour in Figure \ref{Fig:cont-imag}. The peak intensity of the continuum emission is 123.78 $\,\mathrm{mJy\,beam^{-1}}$. 
The position (Hersel coordinates, R.A. = 04$^\mathrm{h}$51$^\mathrm{m}$53$\fs$08 and Decl. = -69$^\circ$23$\arcmin$28$\farcs$00 (J2000)) of SSC candidate H72.97-69.39 is shown as a blue star (see Figure \ref{Fig:cont-imag}). The SSC candidate's position is slightly offset by 1.34$\arcsec$ (0.32 pc) from the continuum peak (hereafter SSC continuum peak). However, this positional offset is negligible compared to the Herschel resolution. A similar positional offset (1.8$\arcsec$ (0.43 pc)) was also reported by \cite{naya19}.
 \cite{Inde04} detected radio source, BO452-6927, at 3 and 6 cm with ATCA at the location of  SSC continuum peak %continuum peak toward the SSC candidate H72.97-69.39, at the continuum peak toward the SSC candidate. 
Also, hydrogen recombination lines were detected at the  SSC continuum peak position \citep{naya19}. Therefore, at that location, the dust continuum emission is contaminated by free-free emission. Many molecular line emissions, along with two hydrogen recombination
lines H36$\beta$ (260.032 GHz) and H41$\gamma$ (257.635 GHz) are detected toward the SSC candidate H72.97-69.39 and/or  SSC continuum peak. In this work, we only discuss the emission lines of CN, CCH, NO, H$^{13}$CO$^+$, and recombination lines (see Section \ref{sec:non_methanol} and Section \ref{sec:origin_methanol}). The details of molecular emission around the SSC candidate H72.97- 69.39 will be discussed in a follow-up work. \\
 We have observed CH$_3$OH emission toward the north and north-west of the  SSC continuum peak position (see Figure \ref{Fig:cont-imag}). Their coordinates are listed in Table \ref{tab_coor}. These positions are associated with continuum emissions, which are likely separate sources. In this work, the north and north-west positions are referred to as N79S-1 and N79S-2, respectively. The continuum signal-to-noise ratio is about 13 for both sources. N79S-1 is separated by about 4.3 $\arcsec$ (1.04 pc) from the  SSC continuum peak, while N79S-2 is about 5.0 $\arcsec$ (1.21 pc). These two cores are separated by about 2.1$\arcsec$ (0.50 pc). We estimate the source size of the dust continuum using a two-dimensional Gaussian fitting within a region that encloses 50\% of the peak intensity (see Table \ref{table:cont_prop}).

%%%%%%%%%%%%%%############################################
\begin{deluxetable*}{lccccc}
\tabletypesize{\footnotesize}
\label{tab:contprop}
\tablewidth{0pt}
\small
\tablecaption{Summary of the continuum image.\label{table:cont_prop}}
\tablehead{\colhead{Continuum}&\colhead{Wavelegth}&\colhead{Flux density$^a$}&\colhead{Size (FWHM$^b$)}&\colhead{Deconvolved size$^c$}\\
\colhead{}&\colhead{(mm)}&\colhead{($\,\mathrm{mJy\,beam^{-1}}$)}&\colhead{($''$/pc)} &\colhead{($''$/pc)}}
\startdata
N79S-1%&1.20&0.71$\pm$0.20&0.90 (0.22)&0.81 (0.20)&{$0.200$}\\
&0.87&0.84$\pm$0.16&0.55 ( 0.13)&0.38 (0.09)\\
%&&&&&\\
N79S-2%&1.20&0.96$\pm$0.20&0.73 (0.18)&0.61 (0.15)&{$0.200$}\\
&0.87&1.64$\pm$0.16&0.53 (0.13)&0.34 (0.08)\\
\enddata
\tablecomments{
$^a$ See section \ref{sec_H2}. 
$^b$ Mean of major and minor FWHM sizes.
$^c$ Mean of major and minor deconvolved sizes.}
\end{deluxetable*}
%%%%%%%%%%%%%%%%%%%%%%%%%%%%%%%%%%%%%%%%%%%%%%%%%%%%%%%%%%%%%%%%%%%%%%%%%%%%%%%%%%%

\subsection{Spectra Toward the CH$_3$OH Peak Positions}  \label{sec_spectra}
 The spectra are extracted from the 0$\farcs$42 (0.1 pc) diameter region centered at CH$_3$OH peak positions toward N79S-1 and N79S-2. Line identification is carried out with CASSIS (this software has been developed by IRAP-UPS/CNRS, \url{ http://cassis.irap.omp.eu}) together with the spectroscopic database `Cologne Database for Molecular Spectroscopy' \citep[CDMS,][]{mull01,mull05}\footnote{\url{(https://www.astro.uni-koeln.de/cdms)}} and `Jet Propulsion Laboratory' \citep[JPL,][]{pick98}\footnote{\url{(http://spec.jpl.nasa.gov/)}} database. The typical radial velocity of the observed molecular lines is $\sim$ 232.5 km s$^{-1}$ for N79S-1 whereas, it is $\sim$ 234 km s$^{-1}$ for N79S-2. We detect any line based on the following criteria: a signal-to-noise ratio (S/N) $\geq$ 3; the velocity coincides with the systemic velocity, and the line is not severely blended with another molecular line. Blended lines are identified by checking the position of the line with other lines, depending on the value of, Einstein coefficient (A$_{ij}$) and upper-state energy (E$_u$). In addition to CH$_3$OH, we detect the molecular emission of H$_2$CO, H$_2$CS, SO, CS, CN, and CCH toward the target positions.\\
The observed transitions of various molecules together with their quantum numbers, E$_u$, and line parameters such as peak brightness temperature (T$_{b}$), FWHM, LSR velocity, and integrated intensity are noted in Tables \ref{tab_line_N79S_1} and \ref{tab_line_N79S_2}. The line parameters are measured by fitting a single Gaussian profile to the observed spectral profile of each unblended transition. The Gaussian fitting is performed for each transition in the velocity range between 229 and 236 km s$^{-1}$  for N79S-1 and between 230 and 238 km s$^{-1}$  for N79S-2.
 The results of the Gaussian fitting are shown in Figures \ref{Fig:1specS1}, \ref{Fig:2specS1}, and  \ref{Fig:3specS2}. We observe multiple hyperfine components for CN and CCH molecules. The transitions were not clearly resolved because of the low spectral resolution of the present spectra. The integrated intensity of the hyperfine lines was determined using Gaussian fitting, which was then divided by their $S\mu^{2}$ values (S is the line strength, and $\mu$ is the electric dipole moment). The transition with the highest intensity among these hyperfine components was selected to calculate the column density.

%%%%%%%%%%%%%%%%%%%%%%%%%%%%%%%%%%%%%%%%%%%%%%%%%%%%%%%%%%%%%%%%%%%%%%%%%%%%%%
\subsection{Line Images \label{sec:line_image}}
The moment zero maps of each molecular line are created by integrating each spectrum in the velocity range where the emission line is seen, typically between 229 and 238 km s$^{-1}$. Figure \ref{fig:mom1} shows the spatial distribution of observed molecular lines (a zoomed-in view of the molecular distributions in N79S-1 and N79S-2 is shown in Figure \ref{fig:mom2}). We estimate the deconvolved source size of each molecular line using a two-dimensional Gaussian fitting within a region that encloses 50\% of the peak intensity (See Table \ref{table:line-colu}). Details of the distribution of molecular line emission are discussed in Section \ref{dist}.

%%%%%%%%%%%%%%%%%%%%%%%%%%%%%%%%%%%%%%%%%%%%%%
 \section{Analysis \label{sec_analysis}}
%\subsection{Rotation Diagram Analysis} \label{sec_rd}
\subsection{Gas-phase Molecular Column Density} \label{sec_rd}
\begin{figure*}
\begin{minipage}{0.5\textwidth}
\includegraphics[width=\textwidth]{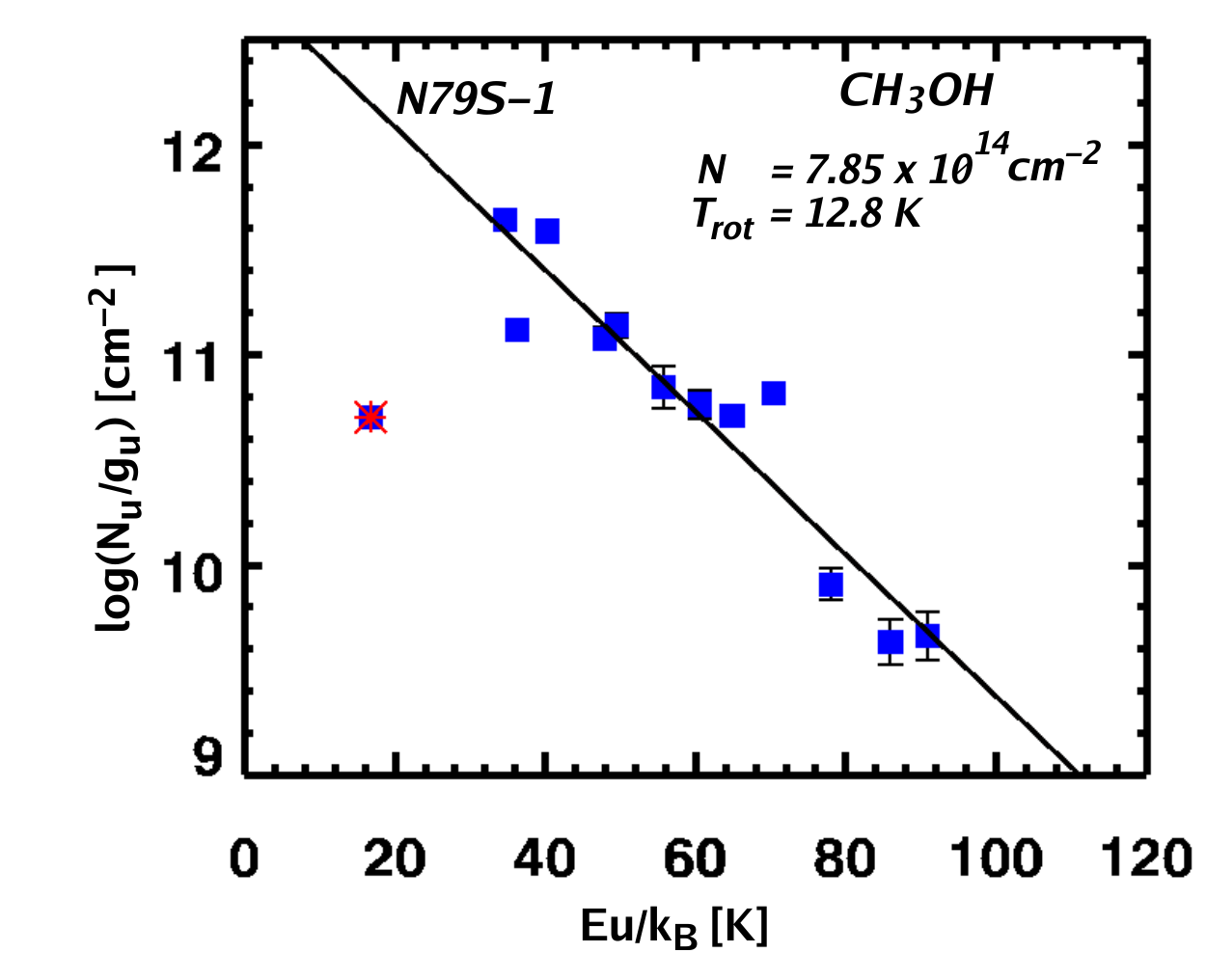}
\end{minipage}
\hskip -1.0 cm
\begin{minipage}{0.5\textwidth}
\includegraphics[width=\textwidth]{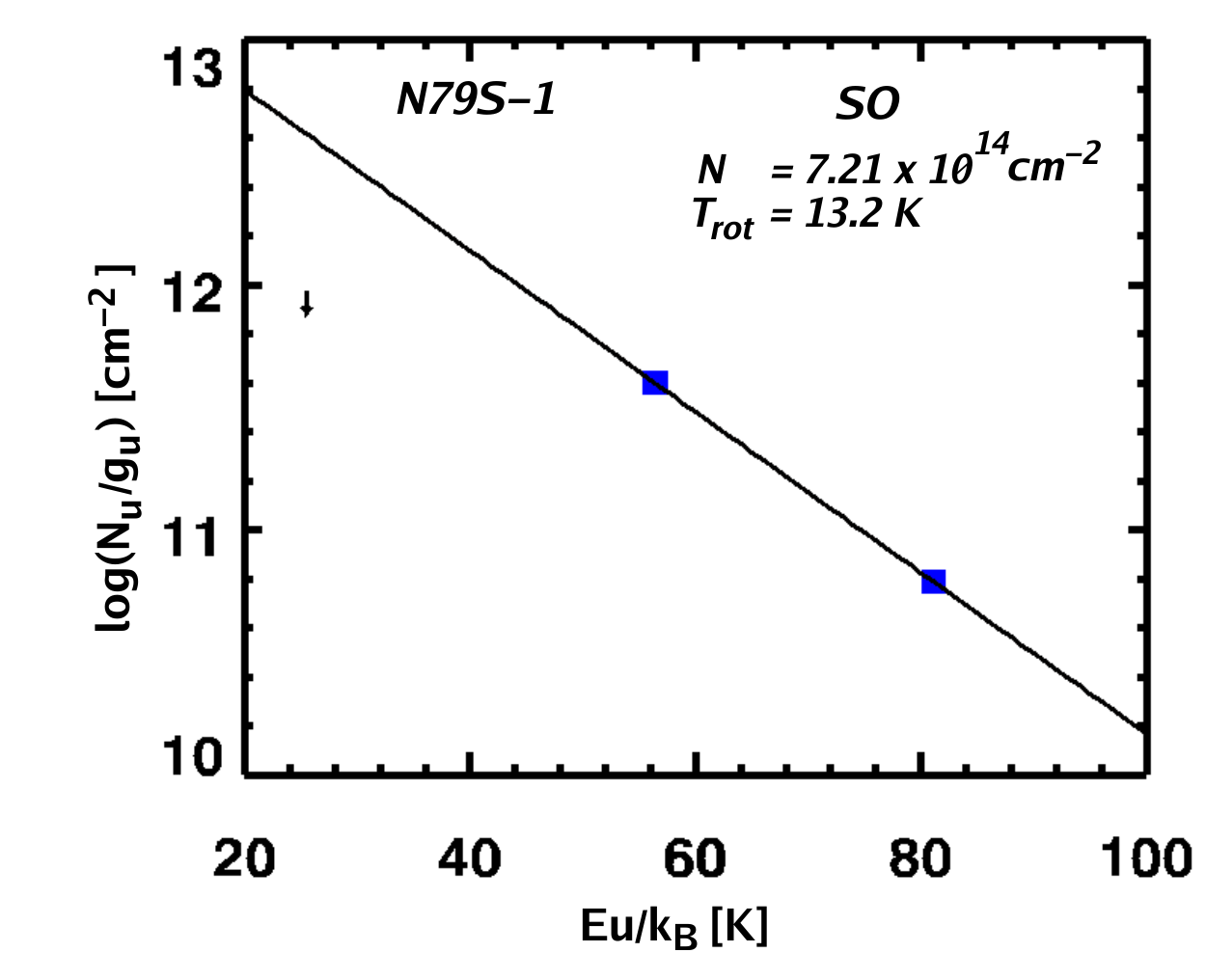}
\end{minipage}
\begin{minipage}{0.5\textwidth}
\includegraphics[width=\textwidth]{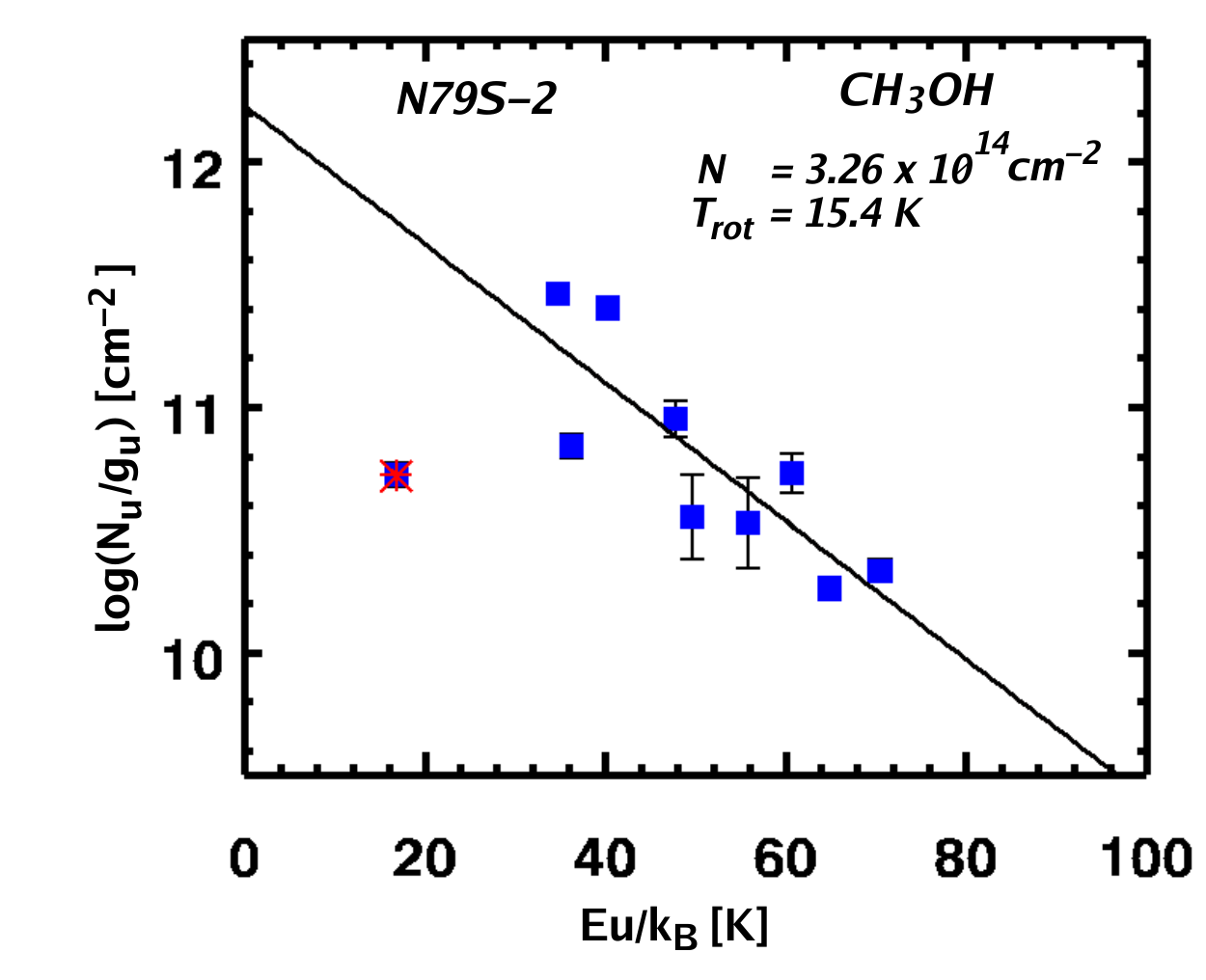}
\end{minipage}
\begin{minipage}{0.5\textwidth}
\includegraphics[width=\textwidth]{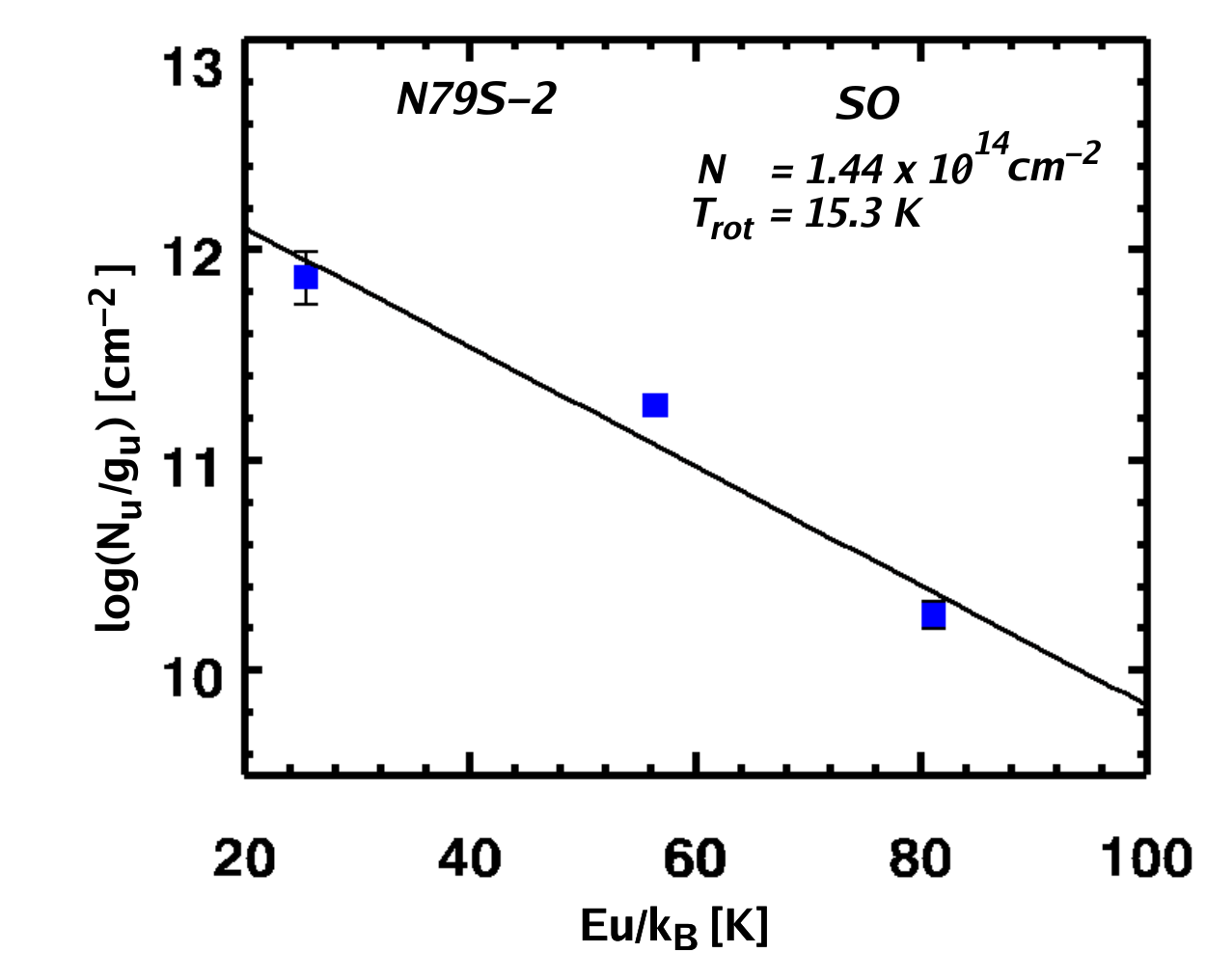}
\end{minipage}
\caption{Rotation diagram of CH$_3$OH and SO at N79S-1 (top) and N79S-2 (bottom) are shown. The downward arrow represents the upper limit point. The solid line represents the fitted straight line. One CH$_3$OH transition {1$_1$-0$_0$}A$^+$ (indicated by the red cross) is excluded from the fit because of its high optical depth. 
%it significantly deviates from other data points.
The derived rotational temperature and column density are given in each panel.}
\label{fig:Rot}
\end{figure*}

We observe multiple transitions of CH$_3$OH and SO with different E$_u$. Thus, we perform the rotation diagram analysis to obtain the rotational temperatures by assuming the observed transitions are optically thin and in local thermodynamic equilibrium (LTE). For the optically thin transitions, upper state column density (N$_u^{thin}$) can be expressed as \citep{gold99},
\begin{equation}
\frac{N_u^{thin}}{g_u}=\frac{3k_B\int{T_{b}dV}}{8\pi^{3}\nu S\mu^{2}},
\label{eqn:clmn}
\end{equation}
where g$_u$ is the degeneracy of the upper state, k$_B$ is the Boltzmann constant, $\rm{\int T_{b}dV}$ is the integrated intensity,
$\nu$ is the rest frequency, and $\mu$ is the electric dipole moment. Under the LTE conditions, the total
column density ($N_{total}$) can be written as,
\begin{equation}
\frac{N_u^{thin}}{g_u}=\frac{N_{total}}{Q(T_{rot})}\exp(-E_u/k_BT_{rot}),
\label{Eq2}
\end{equation}
where $T_{rot}$ is the rotational temperature, E$_u$ is the upper state energy, $\rm{Q(T_{rot})}$ is the partition function at rotational
temperature. This can be rearranged as,
\begin{equation}
log\Bigg(\frac{N_u^{thin}}{g_u}\Bigg)=-\Bigg(\frac{loge}{T_{rot}}\Bigg)\Bigg(\frac{E_u}{k}\Bigg)+log\Bigg(\frac{N_{total}}{Q(T_{rot})}\Bigg).
\end{equation}
There is a linear relationship between the upper state energy and column density at the upper level \citep{gold99,herb09}. The column density and rotational temperature are obtained by fitting a straight line. The rotation diagrams for CH$_3$OH and SO are presented in Figure \ref{fig:Rot}. The derived temperatures, column densities, and their uncertainties (2$\sigma$ level) are summarized in Table \ref{table:line-colu}.\\
We have also conducted non-LTE calculations using RADEX for CH$_3$OH in two sources, N79S-1 and N79S-2, to test potential differences from an LTE approach \citep{vand07}. As input parameters, we use a background temperature of 2.73 K and H$_2$ gas densities of 1.8 $\times$ 10$^{6}$ cm$^{-3}$ for N79S-1 and 3.6 $\times$ 10$^{6}$ cm$^{-3}$ for N79S-2 (see Subsection \ref{sec_H2}). The kinetic temperatures are assumed to be the same as those derived from the rotational diagram analysis (see Table \ref{table:line-colu}), and the average line width is set to 2.0 km s$^{-1}$ for N79S-1 and 2.3 km s$^{-1}$ for N79S-2 (see Tables \ref{tab_line_N79S_1} and \ref{tab_line_N79S_2}). 
For N79S-1, the non-LTE line intensities match well with the observed intensities when using a CH$_3$OH column density of 7.5 $\times$ 10$^{14}$ cm$^{-2}$, which is consistent with the LTE estimation. However, our analysis shows that four lines {5$_{-1}$}-{4$_{-1}$} E, {5$_0$}-4{$_0$} A$^+$, {4$_0$-3$_{-1}$}E and {1$_1$-0$_0$} A$^+$ are optically thick. Among these, three have optical depths of approximately 1, while the {1$_1$-0$_0$} A$^+$ line has a significantly higher optical depth of 5.3. This line was already excluded from the rotational diagram analysis in our study.
For N79S-2, the good match to the observed line intensities is obtained with a column density of 2.3 $\times$ 10$^{14}$ cm$^{-2}$, which is also consistent with the LTE estimation. However, our analysis shows that one line {1$_1$-0$_0$} A$^+$, is optically thick with an optical depth of approximately 1.2; this line was already excluded from our rotational diagram analysis.

%%%%%%%%%%%%%%%%%%%%%%%%%%%%%%%%%%%%%%%%%%%%%%%%%%%%%%%%%%%%%%%%%%%%%%%%%%%%%%%%%%%
%%%%%%%%%%%%%%%%%%%%%%%%%%%%%%%%%%%%%%%%%%%%%%%%%%%%%%%%%%%%%%%%%%%%%%%%%%%%%%%%%%%%%%%%%%%%%%%%%%%%
%%%%%%%%%%%%%%%%%%%%%%%%%%%%%%%%%%%%%%%%%%%%%%%%%%%%%%%%%%%%%%%%%%%%%%%%%%%%%%%%%%%%%
%\subsection{Column Densities of Other Molecules \label{oth_col}}
Column densities of other molecular species (CS, H$_2$CS, H$_2$CO, CN, and CCH) are derived by solving Eqn. \ref{Eq2} for $N_{total}$, assuming LTE and optically thin condition. The gas temperature at N79S-1 and N79S-2 is assumed to be 13 K and 15 K, respectively, based on the rotation diagram analysis of CH$_3$OH and SO. The peak emission of all observed molecules coincides with the CH$_3$OH peak, and the radial velocity range of these molecules is similar to CH$_3$OH, supporting the validity of this assumption. However, we vary the T$_{rot}$ from 10 K to 20 K to account for the uncertainty of the column density.    
%However, the observed CH$_3$OH emission could be sub-thermally excited because the assumed gas temperature in N79S1 and N79S-2 is lower than that of the desorption temperature of CH$_3$OH ($\sim$ 100 K). Therefore,  we derive the column densities, assuming that the excitation temperature ranges from 10 K to 20 K. 
We consider the column density value for excitation temperature 10 K as an upper limit and the column density value for excitation temperature 20 K as a lower limit. The average value of the upper limit and lower limit is used as the final column density. The derived column densities are summarized in Table \ref{table:line-colu}.  
We observe multiple transitions of CN and CCH with the same upper-state energy. For these two molecules, we estimate the column density for each transition, and the average value is used as the final column density. 
%%%%%%%%%%%%%%%%%%%%%%%%%%%%%%%%%%%%%%%%%%%%%%%%%%%%%%%%%%%%
\subsection{Column Density of H$_2$ \label{sec_H2}}
The H$_2$ column density ($N_{H_2}$) is estimated from the optically thin dust continuum emission using  the following equation: 
\begin{equation}
N_{\mathrm{H_2}} = \frac{F_{\nu} / \Omega}{2 \kappa_{\nu} B_{\nu}(T_{d}) Z \mu m_{\mathrm{H}}}. \label{Eq7}
\end{equation}
where F$_{\nu} / \Omega$ is the flux density per beam solid angle, $\kappa_{\nu}$ is the mass absorption coefficient, B$_{\nu}(T_{d})$ is the Plank function, $T_{d}$ is dust temperature, Z is the dust-to-gas mass ratio, $\mu$ is a mean atomic mass per hydrogen, and $m_{H}$ is a hydrogen mass. We use  0.87 mm (344.72671 GHz) continuum emission data and the mass absorption coefficient of  1.89 cm$^2$ g$^{-1}$ for 0.87 mm, assuming the Mathis-Rumpl-Nordsieck or MRN distribution with thin ice mantles at a
gas density of 10$^5$ cm$^{-3}$ \citep{osse94}. Here we use $\mu$ = 1.41 \citep{Cox00}. In this work, we use the dust-to-gas mass ratio of 0.0027, which is derived from scaling the Galactic value of 0.008 by the metallicity of the LMC ($\sim$1/3 Z$\odot$) \citep{bern08}.\\
We measure the average flux density within a circular region of 0$\farcs$42 (0.1 pc) diameter centered at CH$_3$OH peak positions toward N79S-1 and N79S-2 (see Table \ref{tab:contprop}). The dust temperature at N79S-1 and N79S-2 is assumed to be 13 K and 15 K, based on the rotational temperatures of CH$_3$OH under the assumption of LTE. In cold molecular clouds, gas and dust are thermally coupled at densities above $\sim$ 10$^5$ cm$^{-3}$ due to efficient collisional energy exchange \citep{gold01,glov12,lin22}. Therefore, in the present source, we can consider gas and dust as thermally coupled. The estimated $N_{H_2}$ values are noted in Table \ref{tab_hydrogen}. The dust temperature is a key assumption for deriving N$_{H_2}$. Thermal desorption of CH$_3$OH is less likely at our assumed dust temperature. The kinetic temperature of a gas is often higher than the excitation temperature under sub-thermal conditions. Therefore, we estimate the \(N_{H_2}\) at two different temperatures: 10 K and 20 K (see Table \ref{tab_hydrogen}). The calculated value of \(N_{H_2}\) changes by a factor of approximately 3 when the dust temperature is adjusted from 10 K to 20 K. We consider the \(N_{H_2}\) value at 10 K to be an upper limit, while the value at 20 K is a lower limit. The final $N_{H_2}$ is determined by taking the average of these upper and lower limit values. We also estimate a lower limit of the gas density around N79S-1 and N79S-2. Assuming a source diameter of 0.1 pc and a spherical distribution of gas, the gas density and visual extinction are derived to be $n_{\mathrm{H_2}} = 1.85 \times 10^6$ cm$^{-3}$ and $A_V \sim 128$ mag for N79S-1, and $n_{\mathrm{H_2}} = 3.59 \times 10^6$ cm$^{-3}$ and $A_V \sim 196$ mag for N79S-2.
 
%%%%%%%%%%%%%%%%%%%%%%%%%%%%%%%%%%%%%%%%%%%%%%%%%%%%%%%%%%%%%%%%%%%%%%%%%%%%%%%%%%
\begin{table*}
\centering
\footnotesize
\caption{Estimated column densities\label{table:line-colu}}
\begin{tabular}{cccccccccc}
\hline
Species & \multicolumn{4}{c}{N79S-1} & & \multicolumn{4}{c}{N79S-2} \\
\cline{2-5} \cline{7-10}
& T$_{\text{rot}}^a$ (K) & Column density$^b$ (cm$^{-2}$) & Size$^c$ ($\arcsec$) & Size$_{\text{decon}}^d$ ($\arcsec$) 
& & T$_{rot}$ (K) & Column density$^b$ (cm$^{-2}$) & Size ($\arcsec$) & Size$_{\text{decon}}$ ($\arcsec$) \\
\hline
CH$_3$OH$^*$&12.8$\pm$0.3&7.8(14)$\pm$1.0(14)&0.94&0.85&
&15.4$\pm$0.8&3.3(14)$\pm$5.5(13)&0.74&0.62\\ 
SO&13.2$\pm$0.4&7.2(14)$\pm$1.1(14)&0.80&0.69&
&15.3$\pm$0.9&1.4(14)$\pm$3.2(13)&0.67&0.54\\
\hline
H$_2$CO&\nodata&2.6(14)$\pm$1.9(14)&1.18&1.04&
&\nodata&2.4(14)$\pm$1.8(14)&0.88&0.79\\
H$_2$CS&\nodata&2.4(14)$\pm$1.7(14)&0.58&0.42&
&\nodata&2.1(14)$\pm$1.5(14)&0.70&0.57 \\
CS&\nodata&2.2(14)$\pm$ 1.1(14)&1.12&1.03&
&\nodata&1.5(14)$\pm$7.7(13)&0.87&0.77\\
CCH&\nodata&1.6(15)$\pm$9.7(14) &1.07&0.99&
&\nodata&5.9(14)$\pm$3.6(14)&0.85&0.75\\
CN&\nodata&1.4(14)$\pm$6.2(13)&1.02&0.94&
&\nodata&6.0(13)$\pm$2.7(13)&0.71&0.58\\

 \hline
\end{tabular}
\tablecomments{$^a$ T$_{rot}$ varied between 10 and 20 K except for CH$_3$OH and SO (see Section \ref{sec_rd}) $^b$Numbers in parentheses indicate the power of ten, $^c$ Mean of major and minor FWHM sizes (see Section \ref{sec:line_image}) $^d$ Mean of major and minor deconvolved sizes, $^*$ The calculated non-LTE column density of CH$_3$OH using RADEX is 7.5 $\times$ 10$^{14}$ cm$^{-2}$ for N79S-1 and 2.3 $\times$ 10$^{14}$ cm$^{-2}$ for N79S-2.}
\end{table*}
%%%%%%%%%%%%%%%%%%%%%%%%%%%%%%%%%%%%%%%%%%%%%%%%%%%%%%%%%%%%%%%%%%%%%

\begin{deluxetable}{lcccccccc}
\tablecaption{Hydrogen column density ($N_{H_2}$) \label{tab_hydrogen}}
\tablecolumns{10}
\tablewidth{0pt}
\tabletypesize{\scriptsize} % Adjust font size
\tablehead{
    \colhead{} & 
    \multicolumn{3}{c}{N79S-1 (0.87 mm)} & 
    \colhead{} & 
    \multicolumn{3}{c}{N79S-2 (0.87 mm)} \\
    \cline{2-4} \cline{6-8}
    \colhead{$T_d$ (K)} & 
    \colhead{10 K} & 
    \colhead{13 K} & 
    \colhead{20 K} & 
    \colhead{}&
    \colhead{10 K} & 
    \colhead{15 K}&
    \colhead{20 K}
}
\startdata
$N_{H_2}$ ($10^{23}$ cm$^{-2}$)&5.81&3.53 & 1.76 & &11.37&5.41 & 3.45 \\
%$A_{v}$ (mag)&211& 128 & 64 & &413& 196 & 125 \\
\enddata
\tablecomments{Hydrogen column density is estimated from CH$_3$OH at peak positions (see Section \ref{sec_H2}). In this study, we use $N_{H_2}$ = (3.8$\pm$2.0)$\times$ $10^{23}$ cm$^{-2}$ for N79S-1 and $N_{H_2}$ = (7.4$\pm$4.0)$\times$ $10^{23}$ cm$^{-2}$ N79S-2  as a typical value (see Section \ref{sec_H2}).} 
\end{deluxetable}

%%%%%%%%%%%%%%%%%%%%%%%%%%%%%%%%%%

\section{Discussions \label{sec_discussion}}
\begin{deluxetable*}{ l c c c c c c c c c }
\tablecaption{Relative abundances  \label{tab_X}}
\tablecolumns{7}
\tablewidth{0pt}
\renewcommand{\tablenotemark}[1]{{\scriptsize\textsuperscript{#1}}}
%\tabletypesize{\tiny} %manuscript
\tabletypesize{\tiny} %preprint
\tablehead{ \colhead{ }   &  \multicolumn{8}{c}{$N$(X)/$N$(H$_2$)}   \\
            \cline{2-10}  
\colhead{Molecule}     & \colhead{N79S-1{\tablenotemark{1}}} &\colhead{N79S-2\tablenotemark{1}}& \colhead{ST11\tablenotemark{2}} & \colhead{ST16\tablenotemark{3}}  & \colhead{N113A1\tablenotemark{4}}& \colhead{N105-2A\tablenotemark{5}}& \colhead{N105-1C\tablenotemark{5}} & \colhead{N105-2E\tablenotemark{5}} & \colhead{N105-3C\tablenotemark{5}}\\ 
\noalign{\vskip -1.8em}
\colhead{Temperature (K)$^a$}  
& \colhead{$\sim 13$} &\colhead{$\sim 15$}& \colhead{$>100$} & \colhead{$>100$}  & \colhead{$\sim 130$}& \colhead{$>100$}& \colhead{$\sim 17$} & \colhead{$\sim 13$} & \colhead{$\sim 11$}\\
\noalign{\vskip -1.8em}
\colhead{Beam size ($\arcsec$)} 
& \colhead{$\sim 0.42$}
& \colhead{$\sim 0.42$}
& \colhead{$\sim 0.50$} 
&\colhead{$\sim 0.40$}  
& \colhead{$\sim 0.70$} 
& \colhead{$\sim 0.51$} 
& \colhead{$\sim 0.50$} 
& \colhead{$\sim 0.51$} 
& \colhead{$\sim 0.50$}}

\startdata 
 CH$_3$OH &(2.1$\pm$ 1.1)$\times$ 10$^{-9}$ &(4.5$\pm$2.5)$\times$ 10$^{-10}$ & $<$ 8$\times$ 10$^{-10}$ & (4.8$\pm$0.9)$\times$ 10$^{-9}$ & (2.0$\pm$0.3)$\times$10$^{-08}$ & (1.9 $\pm$0.2) $\times$ 10$^{-08}$ & (3.3 $\pm$ 0.9) $\times$ 10$^{-10}$ & (3.1 $\pm$ 0.8) $\times$ 10$^{-10}$ &(7.9 $\pm$ 4.0) $\times$ 10$^{-10}$ \\
 SO &(1.9$\pm$ 1.0)$\times$ 10$^{-9}$ &(1.9$\pm$1.1) $\times$ 10$^{-10}$ & (2.4$\pm$ 0.8) $\times$ 10$^{-8}$ & (1.3 $\pm$ 0.2)$\times$ 10$^{-8}$ & (9.3$\pm$1.8)$\times$10$^{-9b}$ & (7.5 $\pm$0.9) $\times$ 10$^{-9}$ & (2.9 $\pm$ 1.2 )$\times$ 10$^{-10}$ & (5.9 $\pm$ 2.0) $\times$ 10$^{-10}$ & (2.3 $\pm$ 1.1) $\times$ 10$^{-9}$ \\
 H$_2$CO &(6.8$\pm$6.1)$\times$ 10$^{-10}$ &(3.2$\pm$3.0) $\times$ 10$^{-10}$ & (2.2$\pm$0.2)$\times$ 10$^{-10}$ & (3.8$\pm$0.6)$\times$ 10$^{-10}$ & \nodata & \nodata & \nodata & ... & ... \\ 
 H$_2$CS &(6.3$\pm$5.5) $\times$ 10$^{-10}$ &(2.8$\pm$2.5) $\times$ 10$^{-10}$ & (6.2$\pm$2.0)$\times$ 10$^{-10}$ & (3.4$\pm$1.1)$\times$ 10$^{-11}$ & \nodata & (1.6 $\pm$0.2)$\times$ 10$^{-9}$ & (7.2 $\pm$ 3.2) $\times$ 10$^{-11}$ & \nodata & \nodata \\ 
 CS &(5.8$\pm$4.2) $\times$ 10$^{-10}$ & (2.0$\pm$1.5) $\times$ 10$^{-10}$ & $<$3$\times$ 10$^{-10}$ & (2.3$\pm$0.4)$\times$ 10$^{-10}$ & (3.9$\pm$0.6)$\times$10$^{-9c}$ & (2.3 $\pm$0.3) $\times$ 10$^{-9}$ & (1.4 $\pm$ 0.5) $\times$ 10$^{-10}$ & (2.1$\pm$ 0.7) $\times$ 10$^{-10}$&(1.4 $\pm$ 0.6) $\times$ 10$^{-10}$\\ 
 CCH &(4.2$\pm$3.4) $\times$ 10$^{-9}$ & (8.0$\pm$6.5) $\times$ 10$^{-10}$ & \nodata & (2.3$\pm$0.4) $\times$ 10$^{-10}$ & \nodata & \nodata & \nodata & \nodata & \nodata \\
 CN &(3.7$\pm$2.5) $\times$ 10$^{-10}$ & (8.1$\pm$5.7)$\times$ 10$^{-11}$ & \nodata & (8.8$\pm$1.8) $\times$ 10$^{-11}$ & \nodata & \nodata & \nodata & \nodata & \nodata\\
\enddata
\tablecomments{The H$_2$ column density of N79S-1 and N79S-2 is $(3.8\pm2.0)\times 10^{23}$ cm$^{-2}$ and $(7.4\pm4.0)\times 10^{23}$ cm$^{-2}$ respectively.
$^1$This work, $^2$\cite{shim16b},$^3$\cite{shim20}, $^4$\cite{sewi18},$^5$\cite{sewi22},$^a$Temperature of the molecular gas, $^b$Estimated from $^{33}$SO with $^{32}$S/$^{33}$S = 53 \citep{shim20}, $^c$Estimated from $^{13}$CS with $^{12}$C/$^{13}$C = 49 \citep{wang09}}
\end{deluxetable*}
%%%%%%%%%%%%%%%%%%%%%%%%%
\subsection{ Physical Properties of N79S-1 and N79S-2 Core \label{phy_pro}}
The molecular cloud cores N79S-1 and N79S-2 are not associated with IR and radio sources at 3 and 6 cm. The derived gas density and dust extinction are $n_{\mathrm{H_2}} = 1.85 \times 10^6$ cm$^{-3}$ and $A_V \sim 128$ mag for N79S-1, and $n_{\mathrm{H_2}} = 3.59 \times 10^6$ cm$^{-3}$ and $A_V \sim 196$ mag for N79S-2. The H$_2$ density corresponds to a total gas mass of 67 M$\odot$ for N79S-1 and 131 M$\odot$ for N79S-2. From the rotation diagram analysis of CH$_3$OH and SO, the temperature of molecular gas is estimated to be $\sim$ 13 K at N79S-1 and $\sim$ 15 K at N79S-2. The spectra of CO emission from N79S-1 and N79S-2 show a self-absorption profile (see Figure \ref{fig:CO_mom}). Deeply embedded sources often show such a self-absorption profile due to the presence of a significant amount of cold gas in their envelope. These properties suggest that both sources contain dense and cold molecular gas.
The observed widths of the emission lines ($\leq 2.8$ km s$^{-1}$; as detailed in Tables \ref{tab_line_N79S_1} and \ref{tab_line_N79S_2}) from both sources are less than the typical line widths (4-10 km s$^{-1}$) of molecular emissions found in Galactic hot core regions \citep[e.g.,][]{helm97,feng15,mull16,Alle17,bonf17,brou22}. In contrast, they are significantly wider than the line widths (typically 0.1-0.5 km s$^{-1}$) of molecular emissions observed in Galactic low-mass starless cores \citep{tafa06,soma15, beut15,lu18,scib20,zhou22}. Therefore, both sources may be massive starless cores or early-stage high-mass embedded YSOs without a hot core region \citep{saka08,saka10}.\\
The presence or absence of an embedded IR object or molecular outflow in dense cores plays a crucial role in distinguishing between prestellar and protostellar objects \citep{schn12,feng16b}. The dense cores L1448
IRS2E, Per-Bolo 45, and L1451-mm core have previously been classified as starless based on non-detections in the Spitzer near-IR and mid-IR images from the c2d survey \citep{evan03}. The presence of a protostar or first hydrostatic core in these sources was then confirmed based on the detection of molecular outflow \citep{chen10b,pine11}. Another molecular clump, G28.34 S, located around the southern edge of the filamentary IRDC G28.34+0.06, has been considered a high-mass starless core \citep{chen10a,tan13,kong16}. Recently, in G28.34 S, the detection of the strong emission of HNCO, the HCO+ asymmetric line profile implying significant infall, and SiO with broad line wings, indicates that G28.34 S is the host of potential protostellar objects \citep{feng16a,feng16b}. However, the molecular outflows are not observed in the present molecular cloud cores N79S-1 and N79S-2, and these sources are not associated with an IR object. Nevertheless, the detection of cold CH$_3$OH gas may indicate the presence of shocks by young protostellar outflows, as seen in both high-mass and low-mass star-forming regions. The velocity resolution in present observations (0.9-1.1 km s$^{-1}$) may not be sufficient to resolve multiple velocity components, possibly smoothing the observed linewidths. Future high spectral and spatial resolution observations are desirable to distinguish between pre-stellar cores and young protostellar objects.
The chemical compositions of both dense cores are similar. However, the molecular species in N79S-1 show higher abundances compared to N79S-2. This difference in abundance may result from the level of non-thermal desorption, or it may be attributed to a lower gas density in N79S-1 compared to N79S-2.
%%%%%%%%%%%%%%%%
\subsection{Spatial Distribution of Molecular Emission \label{dist}}
All the observed molecules toward N79S-1 and N79S-2 exhibit multiple peaks throughout the region (see Figures \ref{fig:mom1} and \ref{fig:mom2}). The most intense emission of all molecules is observed toward the  SSC continuum peak or SSC candidate, with the exception of CH$_3$OH and H$_2$CS. Emissions from CH$_3$OH and H$_2$CS are not detected at the  SSC continuum peak; they are only observed at locations N79S-1 and N79S-2. Additionally, a weaker peak for all molecules is observed at N79S-2. The spatial distribution of H$_2$CS differs from that of all other molecules. The brightest CH$_3$OH peak is observed toward N79S-1, while a fainter peak is at N79S-2. The emission of CH$_3$OH and SO is slightly extended and similar. The deconvolved source size of CH$_3$OH is about 0\farcs7 (0.17 pc) for N79S-1 and 0\farcs6 (0.14 pc) for N79S-2. For SO, the deconvolved source size of 0\farcs6 (0.14 pc) for N79S-1 and 0\farcs5 (0.12 pc) for N79S-2. The emission of H$_2$CO, CS, CCH, and CN is more extended compared to CH$_3$OH and SO at N79S-1. The distribution of these four molecules at  N79S-1 and  N79S-2 is similar, and line peaks coincide with the CH$_3$OH peaks. The deconvolved source size of these four molecules is about 1$\arcsec$ (0.24 pc) for N79S-1 and 0\farcs7 (0.17 pc) for N79S-2. Since the observed molecules are widely distributed in cold regions, they might have originated via non-thermal desorption or shock.\\
%%%%%%%%%%%%%%%%%%%%%%%%%%%%%%%%%%%%%%%%%%%%%%%%%%%%%%%%%%%%
%\subsection{Spatial Distribution of CO Emission}
 To investigate whether the observed molecules are related to shock chemistry, we compare their spatial distribution with CO outflows observed in ALMA Cycle 7 (ID: 2019.1.01770.S).
%We use ALMA Cycle 7 (2019.1.01770.S) Band 7 observations (PI: Kei Tanaka) for this source to observe CO emission. We investigate the direction of outflow linked with the SSC candidate and examine whether this outflow could potentially initiate shock chemistry in regions N79S-1 and N79S-2. 
Figure \ref{fig:CO_mom} (a) shows the integrated emission (integrated over 205-257 km s$^{-1}$) of CO(3-2) line in the left panel, and the lower panel of Figure \ref{fig:CO_mom} (c) shows the spectra of CO(3-2) line extracted from the  SSC continuum peak position. The emission of CO is widely distributed throughout the region. The peak emission of CO is observed around the SSC continuum peak, and it is relatively faint (the peak intensity differs by a factor of $\sim$ 3) at N79S-1 and N79S-2. Figure \ref{fig:CO_mom} (b) shows the spatial distribution of blue-shifted emission integrated over 205-220 km s$^{-1}$ and red-shifted emission integrated over 242-257 km s$^{-1}$. Protostellar molecular outflows are observed in the SSC candidate, and outflows have a velocity span of 15 km s$^{-1}$. The molecular outflow is oriented in an east-west direction. The molecular outflows in this source were reported by \cite{naya19} with ALMA observation of the $^{13}$CO(2-1) line. However, they observed the outflow axis along a northeast to southwest direction, and outflows have a velocity span of 5 km s$^{-1}$. This difference may be due to the difference in optical depth of the two lines. 
%%%%%%%%%%%%%%%%%%%%%%%%%%%%%%%%%%%%%%%%%%%%%%%%%%%%%%%%%%%%
\subsection{Relative Molecular Abundance}
Relative abundances of molecules around N79S-1 and N79S-2 are summarized in Table \ref{tab_X}. The relative abundance of molecules is derived by dividing the molecular column density by $N_{H_2}$. The derived abundances of the observed molecules may be significantly affected by several factors, including spatial filtering effects, assumptions about excitation temperature, optical depth, and beam filling factors. In this work, we consider only the temperature uncertainty when estimating H$_2$ column densities from dust continuum observations. 
The chemical compositions of these two sources appear similar, but the molecular species in N79S-1 show higher abundances compared to N79S-2 by a factor of two to ten, depending on the species. We compare the relative molecular abundances of these two sources with LMC HMCs (see Table \ref{tab_X}).
%There are organic-poor HMCs, such as ST11 and ST16, and there are organic-rich HMCs, including N113 A1, N113 B3, and N105 2A.
 Organic-rich HMCs are associated with larger COMs, and their molecular abundances are roughly proportional to metallicity. In contrast, the low abundance of organic molecules in organic-poor HMCs cannot be accounted for by a reduced C and O.
 %We consider only N113 A1 because N113 B3 has a similar molecular abundance to N113 A1 \citep{sewi18}. 
 The temperature of molecular gas (estimated from CH$_3$OH and SO) is 13 K around N79S-1 and 15 K around N79S-2, whereas the temperature of LMC HMCs is $\geq$ 100 K. 
Thus, we also compare the observed relative abundances with sources associated only with cold ($\leq$ 20 K) molecular gas. \cite{sewi22} reported three 1.2 mm continuum sources in the star-forming region N105: N105-1C, N105-2E, and N105-3C, which are associated only with cold gas. 
The molecular abundances in N79S-1 are higher than those of 1.2 mm continuum sources (except SO in N105-3C), while the molecular abundances in N79S-2 are similar to those of 1.2 mm continuum sources (except SO and CH$_3$OH in N105-3C)  (see Table \ref{tab_X}).
\subsubsection{Relative abundance of CH$_3$OH \label{sub:metha}}
The low abundance of CH$_3$OH in the LMC has been reported in previous studies \citep{nish16, shim16a,shim16b}. \cite{shim16a} proposed that higher dust temperatures \citep[2-3 times,][]{bern08,ott10,gala13} inhibit the formation of CH$_3$OH in dense clouds of LMC. Astrochemical simulations for the LMC HMCs suggest that the dust temperature at the early ice-forming stage significantly influences the abundance of CH$_3$OH \citep{acha18,paul18, shim20}. However, the abundance of CH$_3$OH shows large variation among the LMC HMCs \citep{shim20}. This variation suggests that warm ice chemistry does not always dominate organic chemistry in the LMC, and the CH$_3$OH abundance is not simply regulated by the elemental abundance of carbon and oxygen. \cite{shim20,shim21} suggested that organic-rich LMC HMCs undergo different chemical and physical conditions during the ice formation stage that significantly affect the grain-surface chemistry. They also suggested the observed variation may be due to the difference in the hot core’s evolutionary stage, as high-temperature gas-phase chemistry might reduce CH$_3$OH abundance at later stages. \\
The relative abundance of CH$_3$OH gas is (2.1$\pm$1.1)$\times 10^{-9}$ at N79S-1 and (4.5 $\pm$2.5)$\times 10^{-10}$ at N79S-2. CH$_3$OH abundance in N79S-1 is about two times smaller than that of an organic-poor LMC hot core, ST16, and one order of magnitude smaller than that of organic-rich LMC HMCs. This difference is likely due to gas temperature. In HMCs, CH$_3$OH ice mantles are mostly sublimated because of the high dust temperature ($>$100 K), while in N79S-1, the low temperature of CH$_3$OH gas around N79S-1 suggests that CH$_3$OH gas originates via non-thermal desorption, and the abundance of CH$_3$OH gas depends on the degree of non-thermal desorption. Thus, it will be fine if we compare the observed CH$_3$OH gas abundances with those of such cold and embedded sources. The observed CH$_3$OH abundance in N79S-1 is $\sim$ 7 times higher than in N105-1C and N105-2E, and  $\sim$ 3 times higher than in  N105-3C.
\cite{font22} detected CH$_3$OH in 15 dense molecular cloud cores towards the outer Galaxy. They estimated that the excitation temperatures are in the range $\sim$7-16 K and the relative abundances of CH$_3$OH range from $\sim$ (0.6-7.4) $\times 10^{-9}$. However, they could not constrain the evolutionary stage of these targets. The abundances of CH$_3$OH gas in Galactic infrared dark clouds are estimated to be $\leq$ 1.0$\times10^{-9}$ \citep{gern14}. Similar abundances have been observed for Galactic embedded YSOs without HMCs/HCs \citep[e.g.,][]{van00, mini02, wata12}. The CH$_3$OH gas abundance at N79S-1 is comparable to that of Galactic counterparts without being corrected for metallicity, and that at N79S-2 is comparable to the typical Galactic value after being corrected for the metallicity. \cite{shim18} reported that the abundance of cold CH$_3$OH gas for a cold molecular cloud core in the Small Magellanic Cloud (SMC) is comparable or marginally higher than those of similar cold sources in our Galaxy.\\
The present observational results suggest that the CH$_3$OH can be produced adequately in a low metallicity environment if the region is cold, dense, and well-shielded. Low-metallicity hot-core simulations in \cite{shim20} show that the CH$_3$OH abundance in a hot-core stage approaches to the metallicity-corrected Galactic  CH$_3$OH abundances if the initial ice-forming stage is well shielded. They also suggest that organic-rich HMCs had experienced such an initial condition before the hot core stage.
%Thus, the question remains whether the observed chemical composition reflects the common properties of cold and dense molecular clouds in metal-poor environments.
%There are only two organic-poor hot cores reported in the literature, and they may be in a different hot core’s evolutionary stage than organic-rich hot cores.

%However, in these cold and dense cores in metal-poor environments, no COMs larger than CH$_3$OH are found, whereas COMs larger than CH$_3$OH were detected in Galactic infrared dark clouds. So, we need to study a large sample of HCs and cold and embedded sources or early-stage  YSOs without a hot core region in LMC.
\subsubsection{Relative abundance of H$_2$CO}
H$_2$CO is a ubiquitous molecule in the interstellar medium \citep{mang93,mang13,gins16,Tang17a,Tang17b,pegu20,bell25}.
H$_2$CO abundance in N79S-1 is $\sim$ two times higher than that in N79S-2. This molecule was observed only in organic-poor HMCs (ST11 and ST16). The observed abundance of H$_2$CO in N79S-1 is 2 to 3 times higher than that in organic-poor HMCs. H$_2$CO has possible formation routes both in the gas phase and solid phase. The H$_2$CO ice formed through successive hydrogenation of CO ice \citep[e.g.,][and references there in]{hama13}. However, high dust temperatures in the LMC suppress the hydrogenation of CO on grain surfaces, according to the warm ice chemistry hypothesis \citep{shim16a}. The H$_2$CO line is often optically thick in star-forming regions \citep{mang93, mcca11, gins16, mahm24}. We evaluated the optical depth of the H$_2$CO line using non-LTE calculations with RADEX. We use two different sets of input parameters. Our estimates show optical depths ranging from 0.4 to 1.6 for N79S-1 and from 0.5 to 1.7 for N79S-2. Therefore, the H$_2$CO line is moderately optically thick in both sources. In this study, the derived column density of H$_2$CO may be considered a lower limit for both sources. \\
%\st{The thermal desorption of H$_2$CO occurs at 50 K; therefore,  non-thermal desorption plays a significant role if it is produced on a grain surface. However, the abundance of H$_2$CO is similar to that in organic-poor hot cores, and the spatial distribution of H$_2$CO is similar to the CS, CN, and CCH, and widely distributed compared to CH$_3$OH emission, supporting gas phase formation.}
In the present sources, non-thermal desorption or shock may play a significant role in the formation of H$_2$CO. The spatial distribution of H$_2$CO is extended as compared to CH$_3$OH. The difference in desorption temperature between H$_2$CO and CH$_3$OH may lead to H$_2$CO being abundant in a larger area, as its desorption temperature \citep[$\sim$50 K;][]{tani20} is lower than that of CH$_3$OH \citep[$\sim$100 K;][]{aika97}. However, H$_2$CO could also efficiently form in the gas phase. Additionally, the spatial distribution of H$_2$CO resembles the CS emission, suggesting that H$_2$CO might trace the shock gas.

\subsubsection{Relative abundances of SO, H$_2$CS and CS}
The abundance of SO in N79S-1 is significantly lower than in the organic-rich ($\sim$ four times) and organic-poor ($\sim$ one order of magnitude) LMC HMCs. SO and SO$_2$ are mainly produced in HMCs via high-temperature gas phase reaction, from H$_2$S, which is sublimated from ice mantles \citep{char97,nomu04,vidal18}.  ALMA observations of the ST11 and ST16 HMCs suggest that SO$_2$ is a useful molecular tracer for studying hot core chemistry in low metallicity \citep{shim16b,shim20}. \cite{shim23} compared the column density of SO, SO$_2$, and SiO for the LMC, SMC, and Galactic HMCs. They obtained a strong correlation with SO$_2$ and SO, while the shock tracer SiO did not correlate with SO$_2$ and SO.
SO and SO$_2$ can be produced in cold gas via neutral processes if H$_2$S is desorbed from ices \citep{sewi22}. SO and SO$_2$ are good tracers of shocked gas, and the enhancement of SO and SO$_2$ abundances is thought to be the result of shock-dominated chemistry triggered by protostellar outflows \citep{saka14,oya19,van21,tych21}.
\cite{tang24} investigated the correlations between the abundance and nonthermal velocity dispersion of SO and SO$_2$ in 248 dense cores from 11 massive star-forming clumps. They found that nonthermal motions may enhance the abundance of SO and SO$_2$ in low-mass cores. Notably, in high-mass cores, both the abundance and nonthermal velocity dispersion of SO$_2$ were significantly increased, suggesting that feedback from high-mass star formation significantly elevates the abundance of SO$_2$.\\
In this work,  the observed abundance of SO in N79S-1 is $\sim$ one order of magnitude higher than in N79S-2, $\sim$ 6 times higher than in N105-1C and $\sim$ 3 times higher than in N105-2E. %So, their formation in N79S-1 may be related to shock chemistry. but they can also trace large-scale gas The observed SO lines have high critical densities (~10⁶ cm⁻³), indicating that SO traces very dense gas.  Hence, SO may trace shocks rather than large-scale gas.
This suggests that the formation of SO in N79S-1 may be related to shock chemistry, although SO can also trace large-scale gas. The observed SO lines have high critical densities ($\sim$ 10$^6$ cm$^{-3}$), indicating that SO primarily traces very dense gas. Therefore, in the present observations, SO is more likely tracing shocks rather than rather than large-scale gas. However, the non-detection of the well-known shock tracer SiO indicates the presence of weak shock. This could be the reason for the non-detection of SO$_2$ in N79S-1 and N79S-2. Because weak shocks yield lower sputtering of ice, resulting in less H$_2$S and H$_2$O being released into the gas phase.\\
The abundance of other sulfur-bearing molecules in N79S-1, such as CS, is approximately two times higher than that found in organic-poor LMC HMCs and significantly lower than that in organic-rich LMC HMCs. However, it is approximately three to four times higher than that of three 1.2 mm continuum sources in the N105 star-forming region. In contrast, H$_2$CS exhibits different behavior; its abundance in N79S-1 is comparable to that in ST11, one order of magnitude lower in ST16, approximately three times higher in N105-2A, and approximately eight times lower in N105-1C. \\

\subsubsection{Relative abundances of CCH and CN}
The abundance of CCH and CN in N79S-1 is approximately five times and four times higher, respectively, than in N79S-2. In contrast, the abundance of CCH in ST16 is an order of magnitude lower compared to N79S-1, while the abundance of CN is reduced by a factor of four.
CCH and CN are efficiently formed in the presence of UV radiation, because the photodissociation of CO in moderate UV fields efficiently produces atomic carbon \citep[e.g.,][]{fuen93, jans95, rodr98, pety17}. In the present source, we observe strong CCH emission, and the spatial distribution of CCH is similar to that of CN emission. So, CCH and CN emissions may trace the outflow cavity walls that are irradiated by the UV light from the central protostar and/or shock-heated regions \citep{wals10,zhan18,miro21}. However, the high visual extinction at the CCH/CN emission region attenuated the interstellar radiation field. It is also possible that UV radiation produced locally in shocks causes the enhancement of CCH and CN emission \citep{tych21}. 
In addition, both molecules have similar distributions as the H$_2$CO and CS emission. So, in the present source, the emission of these four molecules traces similar physical properties.
%%%%%%%%%%%%%%%%%%%%%%%%%%%%%%%%%%%%%%%%%%%%%%%%%%%%%%%%%%%%%%%%%%%%%%%%%%%%%%%%%%%%%%%%%

\subsection{Non-detection of CH$_3$OH Toward SSC Continuum Peak \label{sec:non_methanol}}
No transitions of CH$_3$OH are detected around the SSC continuum peak. The CH$_3$OH transitions have been detected only at N79S-1 and N79S-2. CH$_3$OH is mainly produced during the cold stage via hydrogenation of CO on grain surfaces, as there are no effective gas phase pathways \citep[e.g.,][]{wata07, garr06,herb09}. They are released into the gas phase via thermal or non-thermal desorption. 
%The elevated dust temperature in the LMC suppresses the hydrogenation of CO on grain surfaces due to the high volatility of atomic hydrogen \citep{shim20,shim16a}. However,
Despite the LMC conditions that suppress CO hydrogenation on grains, the CH$_3$OH abundance in N79S-1 is comparable to those of similar Galactic objects. Therefore, as mentioned in Subsection \ref{sub:metha}, the different chemical and physical histories during the initial ice-forming stage may account for the non-detection of CH$_3$OH in SSC candidate. 
\cite{shim16a} suggests that the formation of CH$_3$OH ice depends on the column density ratio of  CO$_2$/H$_2$O ice because CO on warm dust surfaces could efficiently reacts with OH to form CO$_2$, which inhibits the production of CH$_3$OH on the grain surface. The CO$_2$/H$_2$O ice abundance ratio is unknown toward the SSC candidate H72.97-69.39, N79S-1, and N79S-2 and remains to be investigated. However, these sources, N79S-1, N79S-2, and SSC Candidate H72.97-69.39, are part of the same molecular cloud, with a separation of approximately 1.2 pc. Therefore, the chemical and physical conditions during the initial ice-forming stage for these sources are most likely similar.\\
 Another possibility is that the evolutionary stage of the embedded YSO may be responsible for the non-detection of CH$_3$OH gas in this source. High-temperature gas-phase chemistry can also reduce the CH$_3$OH abundance at a later stage \citep[e.g.,][]{nomu04,garr06,vasy13}. \cite{gern14} observed 59 sources with different evolutionary sequences to study and characterize the physical and chemical evolution. They reported that the intensity of more complex and heavy molecules reaches its maximum in the HMC phase and declines for the UCHII stage because these molecules are probably destroyed by the UV radiation from the forming stars. Using the gas grain chemical model, they also reported that the abundance of CH$_3$OH and CH$_3$OCHO reaches a maximum in the HMC phase, and then the abundance decreases in the UCHII stage. \\
 The detection of [C II] and high-J CO lines toward the SSC candidate with SOFIA/GREAT observation reveals the presence of hot gas \citep{naya21}. \cite{naya19} reported that the central source is embedded in a high-density compact region. The detection of hydrogen recombination lines and fine-structure lines in the IR spectrum suggests that a HII region is formed around the source \citep{seal09,naya19}. 
In this work, we also detect the emission lines of CN, CCH, NO, and H$^{13}$CO$^+$ toward the SSC candidate (see Figure \ref{fig:mom1}). Among them, NO and H$^{13}$CO$^+$ are not detected toward N79S-1 and N79S-2. The emission of CN ( N = 3-2, peak intensity of 3.2 K) and CCH ( N = 4-3, peak intensity of 4.1 K) is very bright toward the SSC candidate (see Figure \ref{Fig:specac1}). This may indicate the presence of strong UV radiation radiating from newly born stars. Normally, HCO$^+$is formed mainly in gas phase reactions of CO with H$_{3}^{+}$ in dense molecular cloud where H$_{3}^{+}$ is formed by ionization of H$_2$ via cosmic rays \citep[e.g.,][]{turn95,case98}. However, the abundance of HCO$^{+}$ is strongly influenced by UV radiation or X-rays from their host star \citep[e.g.,][]{rawl04,clee17}. The emission of H$^{13}$CO$^+$  (J = 2-3, peak intensity of 4.9 K) is very strong in the SSC candidate, indicating the influence of UV and/or X-ray radiation from their host star. \cite{shim16b} reported a higher abundance of NO in a LMC hot core (ST11) than in Galactic counterparts. In this work, the peak intensity of NO transitions is similar to that observed in ST11. Normally, NO is formed by a neutral–neutral reaction of N and OH in the gas-phase \citep{herb73,pine90}. The chemical simulation suggests that the NO abundance is enhanced by a factor of 1.6  through that reaction as the temperature increases from 70 K to 250 K \citep[see Table 4 in][]{pine90}. As suggested in the FUV models of \cite{stau07}, gas temperatures above approximately 300 K were required to explain the observed NO abundances. \\
The above results suggest that strong radiation might be present, which could significantly affect the chemical composition of the SSC candidate. The possible energy source is the most luminous IR source or shock-induced radiation field. Ice photolysis experiments reported that CH$_3$OH readily dissociates on desorption by UV irradiation \cite{bert16,mart16}. At high temperatures, the CH$_3$OH molecule is actively destroyed or chemically evolved into other species, as well as surface chemistry becomes insignificant at high temperatures. Further observations of intense radiation tracers (such as CO$^+$) will help to study the ionization degree, and the effect of radiation should be examined with the aid of astrochemical models.

 %%%%%%%%%%%%%%%%%%%%%%%%%%%%%%%%%%%%%%%%%%%%%%%%%%%%%%%%%%%%%%%%%%%%%%%%%%%%%%%%%
 \subsection{Possible Origin of CH$_3$OH in N79S-1 and N79S-2 Core \label{sec:origin_methanol}}
The major pathway for the formation of solid CH$_3$OH is thought to be produced by the hydrogenation of CO on grain surfaces \citep{wata02}. % As noted earlier, CH$_3$OH ice forms primarily through the hydrogenation of CO on grain surfaces. CH$_3$OH ice forms primarily through the hydrogenation of CO on grain surfaces, as noted earlier.CH$_3$OH ice forms primarily through the hydrogenation of CO on grain surfaces, as mentioned earlier. 
 The thermal desorption of CH$_3$OH ice is higher than $\sim$ 80 K \citep[e.g.,][]{tiel05}. The gas temperature is $\sim$13 K around  N79S-1 and $\sim$15 K around N79S-2, and the non-thermal desorption plays a significant role for the production of CH$_3$OH in both sources. There are several mechanisms for the non-thermal desorption of CH$_3$OH ice.\\ 
%{\color{red} Following lines are commented out by Kenji-san. Based on his comment we may write the following:}
One possible desorption mechanism is photodesorption by UV photons. However, recent experiments by \cite{bert16} suggested nominal CH$_3$OH photodesorption yields, which raises its importance in these circumstances. The possible sources of UV photons in N79S-1 and N79S-2 include cosmic-ray-induced UV photons and UV radiation from a nearby forming star. Based on Gamma-ray observations, the average cosmic-ray density in the LMC is approximately 25\% of that in the solar neighborhood \citep{abdo10}. However, the cosmic-ray flux is not uniform throughout the LMC. The molecule H$^{13}$CO$^+$, most sensitive to the cosmic-ray ionization rate, is not detected in both sources.
%The SSC candidate H72.97-69.3 is the most luminous YSO in the LMC, but it is located far from N79S-1 ($\sim$ 1 pc) and N79S-2 ($\sim$ 1.2 pc). However, the high visual extinction at  N79S-1 and N79S-2 attenuated the interstellar radiation field. So, the production of cold CH$_3$OH emission by the photodesorption mechanism is ruled out in these sources.
In the N79 region, the SSC candidate H72.97-69.3 is the most luminous IR compact source of the LMC. However, it is far from N79S-1 ($\sim$ 1 pc) and N79S-2 ($\sim$ 1.2 pc). Additionally, CCH and CN, which are good tracers of UV-irradiated regions, show emission peaks that coincide with the peaks of CH$_3$OH. The abundance of CN and CCH is significantly higher in N79S-1 compared to a LMC hot core (ST16). This indicates that the presence of CH$_3$OH at N79S-1 might be a result of photodesorption. However, the high visual extinction at the CH$_3$OH emission peak in N79S-1 and N79S-2 attenuated the interstellar radiation field.\\
%\sout{From dust continuum data, we also estimate that the visual extinction from the outer edge of the continuum emitting region to the CH$_3$OH emission region is at least greater than 7 mag for N79S-1 and greater than 60 mag for N79S-2 (see Section \ref{sec_av}.}
There is a possibility that UV radiation from the SSC candidate H72.97-69.3 is responsible for the photodesorption of CH$_3$OH. 
%The ATCA radio source (BO452-6927) has been detected at the location of the SSC candidate, indicating that the 0.87 mm continuum emission is contaminated with the free–free emission \citep{Inde04}. 
The observed hydrogen recombination lines H36$\beta$ (260.032 GHz) and H41$\gamma$ (257.635 GHz) at the location of SSC center H72.97- 69.39 show compact emission (see Figure \ref{fig:mom1}). \cite{naya19} also reported that the recombination line H30$\alpha$ (231.995 GHz) at the location of the SSC candidate is compact in size. Thus, we expect the outer edge of the continuum emission at the location of the SSC candidate H72.97-69.3 to be the thermal emission from dust. We estimate that the visual extinction (Av) between the outer edge of the SSC center and the CH$_3$OH emission region is greater than 50 mag. We assumed a molecular gas mass of at least 10$^5$ M$\odot$ within a radius of 2 pc in the N79 region \citep[see Figure 5 in][]{ochs17}. The dust mass is estimated considering the dust-to-gas ratio of 0.0027 \citep{roma14,ande21}. Using this dust mass, we estimate a dust mass density ($\rho_d$) of 5.4 $\times$ 10$^{-22}$ g cm$^{-3}$ within the same 2 pc radius. We use the relation $\rho_d$ = Z$\mu_H$N$_{H_2}$2m$_H$ /L, where N$_{H_2}$ is the column density of molecular hydrogen, m$_H$ is the hydrogen mass and $\mu_H$ is the mean atomic mass per hydrogen (1.41), Z is dust to gas mass ratio (0.0027) and L is the path length (1.04 pc for N79S-1 and 1.21 pc for N79S-1), to estimate the column density of molecular hydrogen. We use the N$_{H_2}$/Av ratio of  2.8 $\times$ 10$^{21}$ to convert the column density of molecular hydrogen to visual extinction.
\cite{ande21} studied the stellar content within 2 parsecs of the cluster H72.97-69.39 using near-IR imaging and derived that the average extinction is Av $>$ 10 of H72.97-69.39. This suggests that the N79S-1 and N79S-2 cores are well shielded (Av $>$ 50) from the external radiation field.\\
%However, the value of visual extinction may vary depending on the assumption of dust temperature along the line of sight. Therefore, to understand the contribution of photodesorption, it is necessary to estimate the visual extinction along the line of sight more comprehensively or examine it with the help of detailed chemical models.\\
The gas-phase CH$_3$OH is also formed through the sputtering of ice mantles by shocks. Protostellar outflows are detected in the SSC candidate. However, the direction of the outflows differs from the position of the CH$_3$OH emission, and the velocity span of the outflows is small. Additionally, the position of cold CH$_3$OH emissions is too far from the SSC candidate for small velocity magnetohydrodynamic (MHD) waves to propagate and cause mantle disruption \citep{mark00}. Further, a common shock tracer, SiO, is not detected in positions of CH$_3$OH emission, and the line widths of CH$_3$OH transitions are narrow and comparable to those of the other observed molecules. However, observations toward several protostellar outflows and shocked regions have revealed the presence of cold CH$_3$OH gas and enhanced CH$_3$OH abundances, attributed to the release of CH$_3$OH into the gas phase via shock-induced sputtering of icy grain mantles. Such conditions have been reported in sources such as L1157 \citep{bach97,code10,bene13,feng22}, %NGC 1333–IRAS 4A (Jørgensen et al. 2004; Santangelo et al. 2015), Serpens SMM1 (Kristensen et al. 2010),
L1448-mm \citep{jim08}, and IRS7B \citep{saba04}, among others. Therefore, the possible energy sources for triggering shock chemistry are still unknown from the present observations. Future observations of shock and outflow tracers will be necessary to better constrain the role of shock-induced processes.
%Therefore, the desorption mechanism sputtering ice mantles by shocks is less likely to produce cold CH$_3$OH, or possible energy sources for triggering the shock chemistry are unknown from the present observation. Future observations of shock and outflow tracers will need to consider the role of shock chemistry. 
One possibility is that desorption is caused by a localized kinematic \citep[e.g.,][ and references therein]{{wirs14,sewi22}}.
Relative grain–grain streaming could occur through collisions between small gas clumps or filaments that are interacting and merging \citep{taka03,buck06}. In this process, collisions between low-velocity individual grains cause transient heating of dust grains, and grain temperatures rise, which then cools by sublimation of surface molecules. The continuous structure of gas clumps between the two cores (N79S-1 and N79S-2) is seen in the velocity maps of CS and H$_2$CO emission lines (see Figure \ref{fig:chan_map}). So, the continuous velocity structure from the core N79S-1 to the core N79S-2 would support the physical interaction between the gas clumps. \\
The sublimation of solid CH$_3$OH may also occur through reactive desorption. This involves ice mantle molecules using the part of the energy released in an exothermic grain surface reaction to overcome the physisorption binding energy \citep{garr07,mini16,chua18, das15,furu15}.
%\textbf{ The relative contribution of the chemical desorption process can be evaluated with the aid of astrochemical models \citep{garr13,taqu14,furu15}.}
However, the efficiency of the chemical desorption process is poorly understood because of a lack of experimental study \citep{furu22}. If chemical desorption of CH$_3$OH from dust surfaces is taking place in the present sources, we would expect to observe widespread cold CH$_3$OH emission rather than a compact distribution. The observed compact emission of CH$_3$OH suggests that grain collisions may play a significant role in the desorption of CH$_3$OH from dust grain mantles.
%However, experimental and  computational chemistry are required to know the efficiency of the chemical desorption process \citep{furu22}. the main difficulty for reactive desorption scenario is the lack of experiments.
%However, in several Galactic dark clouds, the origin of methanol and other complex organic molecules in dense clumps, which are located far from a protostar or any outflows, remains unknown. \citep[e.g.,][]{bacm12,vast14,wirs14,taqu17,soma18,agun21}.
%%%%%%%%%%%%%%%%%%%%%%%%%%%%%%%%%%%%%%%%%%%%%%%%%%%%%%%%%%%%

%%%%%%%%%%%%%%%%%%%%%%%%%%%%%%%%%%%
\section{Conclusions \label{sec_conclution}}
In this work, we present the results of 0.1 pc scale observations with ALMA toward the SSC candidate H72.97-69.39 in the N79 South GMC of the LMC. We identified two dense molecular cloud cores, named N79S-1 and N79S-2, which are separated by $\sim$ 1.0-1.2 pc from the SSC candidate. We discuss the physical and chemical properties of these two sources and obtain the following conclusions.\\
\begin{enumerate}
   
 \item We have detected emission lines of CH$_3$OH, H$_2$CO, H$_2$CS, CS, SO, CO, CN, and CCH toward the positions of N79S-1 and N79S-2. The dust continuum and all molecules except CH$_3$OH and H$_2$CS have the brightest emission around the SSC candidate and a weak peak around N79S-1 and N79S-2. No CH$_3$OH transitions are detected toward the SSC candidate. The protostellar outflows are observed in SSC candidate H72.97-69.39. 
 \item We derive the physical properties of two molecular cloud cores and found that these are very dense ($n_{\mathrm{H_2}}$ $\sim$ 10$^6$ cm$^{-3}$), well shielded (A$_v$ $>$ 100 mag) and are not associated with an IR source. Based on the rotation diagram analysis of CH$_3$OH and SO lines, we derive a gas temperature of 13 K for N79S-1 and 15 K for N79S-2. The observed line widths of molecular lines are $\leq$ 2.8 km s$^{-1}$. The emission spectra of CO show a self-absorption profile indicating the presence of cold gas. These properties are consistent with the properties of the cold and dense molecular cloud cores in our Galaxy.
 \item The present observation shows that the chemical compositions of both dense cores are similar. However, the molecular species in N79S-1 show 2 to 10 times higher abundances than N79S-2. The relative abundance of cold CH$_3$OH gas is estimated to be (2.1$\pm$1.1)$\times$10$^{-9}$ at  N79S-1 and (4.5$\pm$2.5)$\times$10$^{-10}$ at N79S-2. Despite the lower metallicity in the LMC, the relative abundance of CH$_3$OH at N79S-1 is comparable to those of similar cold sources in our Galaxy. However, the formation of organic molecules is inhibited throughout the N79 region, as can be seen in the non-detection of CH$_3$OH in most of the regions. The two positions N79S-1 and N79S-2 would be exceptional positions, where CH$_3$OH production is efficient. The possible origins of cold CH$_3$OH gas in these dense cores are discussed, as well as the non-detection of CH$_3$OH toward the SSC candidate.
\end{enumerate}
 %%%%%%%%%%%%%%%%%%%%%%%%%%%%%%%%%%%%%%%%%%%%%%%%%%%%%

\section{Acknowledgment}
This paper makes use of the following ALMA data: ADS/JAO.ALMA$\#$017.1.01323.S, $\#$2018.1.01366.S and $\#$2019.1.01770.S . ALMA is a partnership of ESO (representing its member states), NSF (USA) and NINS (Japan), together with NRC (Canada), MOST and ASIAA (Taiwan), and KASI (Republic of Korea), in cooperation with the Republic of Chile. The Joint ALMA Observatory is operated by ESO, AUI/NRAO and NAOJ. This work is supported by JSPS KAKENHI grant no. JP21H01145. S.K.M. acknowledges the SNBNCBS Institute, Kolkata, for providing research infrastructure and the Department of Science and Technology, Government of India, for providing financial assistance to carry out research. K.F. acknowledges support from the JSPS KAKENHI grant no. JP21H01145.

 %{\bf P.G.} . {\bf A.D.} . 

\software{CASA \citep{mcmu07})}
%%%%%%%%%%%%%%%%%%%%
%%%%%%%%%%%%%%%%%%%%

\clearpage
\appendix
\restartappendixnumbering
\section{Measured Line Parameters\label{line_para} }
Table \ref{tab_line_N79S_1}  and Table \ref{tab_line_N79S_2} summarize the line parameters obtained by Gaussian fitting to the observed transition (see Section \ref{sec_spectra} for details). 

\startlongtable
\begin{deluxetable*}{ l c c c c c c c c }
\tablecaption{Summary of the line parameters of observed molecules towards N79S-1\label{tab_line_N79S_1}}
\tablewidth{0pt}
\tabletypesize{\scriptsize} 
\tablehead{
\colhead{Molecule}   & \colhead{Transition}                               & \colhead{Frequency} & \colhead{$Eu/k$} &\colhead{$V_{LSR}$} & \colhead{$T_{b}$} & \colhead{$\Delta$$V$} & \colhead{$\int T_{b} dV$} &  \colhead{Remarks} \\
\colhead{ } & \colhead{ }    & \colhead{(GHz)} & \colhead{(K)}         & \colhead{(km/s)} & \colhead{(K)} &\colhead{(km/s)}   & \colhead{(K km/s)} & \colhead{}
}
\startdata
SO&{6$_6$}-{5$_5$}&258.25583&56.5&232.529$\pm$   0.019&   3.761$\pm$     0.061&   2.101$\pm$     0.044&   8.412$\pm$     0.313&\\ 
&{8$_{7}$-7$_6$} &340.71416&81.2&232.428$\pm$   0.008&   1.791$\pm$     0.024&   1.061$\pm$     0.020&   2.024$\pm$     0.065&\\
&&&&&&&&\\
CS&{5}-{4}&244.93556&35.3&232.440$\pm$   0.011&   9.929$\pm$     0.067&   2.898$\pm$     0.025&   30.627$\pm$    0.473&\\
&&&&&&&&\\
H$_2$CS&{7$_{1,6}$-6$_{1,5}$}&244.04850&60.0&232.448$\pm$   0.103&   0.884$\pm$     0.058&   2.844$\pm$     0.253&   2.676$\pm$     0.413&\\ 
&&&&&&&&\\
H$_2$CO&{5$_{1,5}$-4$_{1,4}$}&351.76864&62.5&232.502$\pm$   0.006&   3.930$\pm$     0.018&   2.260$\pm$     0.013&   9.453$\pm$     0.098& \\   
&&&&&&&&\\
CH$_3$OH&{5$_0$-4$_0$} E&241.70016&47.9&232.766$\pm$   0.058&   1.418$\pm$     0.076&   1.840$\pm$     0.136&   2.777$\pm$     0.353&\\   
&{5$_{-1}$}-{4$_{-1}$} E&241.76723&44.4&232.426$\pm$   0.023&   3.770$\pm$     0.071&   2.157$\pm$     0.054&   8.657$\pm$     0.381&\\   
&{5$_0$}-4{$_0$} A$^+$&241.79135&34.8&232.371$\pm$   0.020&   4.407$\pm$     0.070&   2.168$\pm$     0.048&   10.169$\pm$    0.386&\\   
&{5$_1$-4$_{1}$} E& 241.87902&55.9&232.421$\pm$   0.124&   0.684$\pm$     0.070&   2.202$\pm$     0.292&   1.603$\pm$     0.377&\\
&{5$_{-2}$}-{4$_{-2}$} E&241.90415&60.7&231.949$\pm$   0.081&   0.979$\pm$     0.072&   2.176$\pm$     0.192&   2.268$\pm$     0.367&1\\   
&{5$_1$}-{4$_1$} A$^-$&243.91579&49.7&232.534$\pm$   0.079&   1.203$\pm$     0.061&   2.429$\pm$     0.189&   3.110$\pm$     0.399&\\   
&{7$_0$-6$_0$} E&338.12449&78.1&232.256$\pm$   0.075&   0.190$\pm$     0.015&   1.770$\pm$     0.178&   0.359$\pm$     0.065&\\   
&{7$_{-1}$-6$_{-1}$} E &338.34459 &70.6&232.497$\pm$   0.009&   1.846$\pm$     0.019&   1.499$\pm$     0.020&   2.944$\pm$     0.071&\\   
&{7$_0$-6$_0$} A$^+$& 338.40870 &65.0&232.496$\pm$   0.006&   2.685$\pm$     0.026&   1.029$\pm$     0.014&   2.942$\pm$     0.067&\\   
&{7$_1$-6$_{1}$} E& 338.61493&86.1&232.431$\pm$   0.078&   0.144$\pm$     0.016&   1.285$\pm$     0.184&   0.197$\pm$     0.049&\\   
&{7$_{-2}$-6$_{-2}$} E&338.72290&90.9&233.516$\pm$   0.151&   0.171$\pm$     0.013&   2.107$\pm$     0.397&   0.385$\pm$     0.102&1\\   
&{4$_0$-3$_{-1}$} E& 350.68766&36.3&232.648$\pm$   0.007&   1.573$\pm$     0.019&   1.014$\pm$     0.017&   1.697$\pm$     0.048&\\   
&{1$_1$-0$_0$} A$^+$ &350.90510&16.8&232.305$\pm$   0.029&   0.425$\pm$     0.013&   1.844$\pm$     0.071&   0.834$\pm$     0.058&\\   
&&&&&&&&\\
CN&N= 3- 2, J=5/2-5/2, F=3/2-3/2&339.44678&32.6&233.203$\pm$   0.091&   0.126$\pm$     0.014&   1.387$\pm$     0.229&   0.187$\pm$     0.052&\\   
&N= 3- 2, J=5/2-5/2, F=7/2-7/2&339.51664&32.6&232.103$\pm$   0.116&   0.164$\pm$     0.014&   2.751$\pm$     0.309&   0.480$\pm$     0.094&\\   
&N= 3- 2, J=5/2-3/2, F=5/2-5/2&340.00813&32.6&233.017$\pm$   0.084&   0.196$\pm$     0.018&   1.674$\pm$     0.197&   0.350$\pm$     0.074&\\   
&N= 3- 2, J=5/2-3/2, F=3/2-3/2&340.01963&32.6&232.248$\pm$   0.060&   0.210$\pm$     0.013&   1.664$\pm$     0.164&   0.371$\pm$     0.060&\\   
&N= 3- 2, J=5/2-3/2, F=7/2-5/2&340.03155&32.6&232.441$\pm$   0.006&   1.966$\pm$     0.014&   1.678$\pm$     0.016&   3.512$\pm$     0.058&2\\   
&N= 3- 2, J=5/2-3/2, F=5/2-3/2&340.03541&32.6&232.639$\pm$   0.008&   2.115$\pm$     0.017&   1.600$\pm$     0.020&   3.603$\pm$     0.073&2,3\\   
&N= 3- 2, J=7/2-5/2, F=9/2-7/2&340.24777&32.7&232.159$\pm$   0.006&   3.756$\pm$     0.018&   2.446$\pm$     0.014&   9.780$\pm$     0.101&4\\   
&N= 3- 2, J=7/2-5/2, F=5/2-5/2& 340.26177&32.7&232.877$\pm$   0.061&   0.171$\pm$     0.014&   1.468$\pm$     0.155&   0.268$\pm$     0.050&\\   
&N= 3- 2, J=7/2-5/2, F=7/2-7/2&340.26495&32.7&232.727$\pm$   0.088&   0.231$\pm$     0.014&   2.074$\pm$     0.225&   0.509$\pm$     0.087&\\   
&&&&&&&&\\
CCH&N= 4- 3, J=9/2-7/2, F= 5- 4&349.33771&41.9&231.899$\pm$   0.003&   6.498$\pm$     0.014&   2.770$\pm$     0.007&   19.159$\pm$    0.092&2\\   
&N= 4- 3, J=7/2-5/2, F= 4- 3&349.39928&41.9&231.965$\pm$   0.004&   5.519$\pm$     0.014&   2.844$\pm$     0.009&   16.707$\pm$    0.097&2\\   
&N= 4- 3, J=7/2-7/2, F= 4- 4&349.60361&41.9&232.413$\pm$   0.126&   0.150$\pm$     0.016&   2.242$\pm$     0.296&   0.359$\pm$     0.086&\\   
\enddata
\tablecomments{
\scriptsize{(1) Two CH$_3$OH lines with similar spectroscopic constants are blended. (2) Partial blend with CN (N= 3- 2, J=5/2-3/2, F=5/2-3/2 and N= 3- 2, J=5/2-3/2, F=7/2-5/2). (3) Blended with two hyperfine components. (4) Blended with three hyperfine components.
}}
\end{deluxetable*}
%%%%%%%%%%%%%%%%%%%%%%%%%%%%%%%%%%%%%
\startlongtable
\begin{deluxetable*}{ l c c c c c c c c }
\tablecaption{Summary of the line parameters of observed molecules towards N79S-2\label{tab_line_N79S_2}}
\tablewidth{0pt}
\tabletypesize{\scriptsize} 
\tablehead{
\colhead{Molecule}   & \colhead{Transition}                               & \colhead{Frequency} & \colhead{$Eu/k$} &\colhead{$V_{LSR}$} & \colhead{$T_{b}$} & \colhead{$\Delta$$V$} & \colhead{$\int T_{b} dV$} &  \colhead{Remarks} \\
\colhead{ } & \colhead{ }    & \colhead{(GHz)} & \colhead{(K)}         & \colhead{(km/s)} & \colhead{(K)} &\colhead{(km/s)}   & \colhead{(K km/s)} & \colhead{}
}
\startdata
SO&{6$_6$}-{5$_5$}&258.25583&56.5&234.271$\pm$   0.039&   1.937$\pm$     0.067&   1.869$\pm$     0.092&   3.853$\pm$     0.324&\\ 
&{3$_{3}$-2$_3$} &339.34146&25.5&235.008$\pm$   0.148&   0.143$\pm$     0.017&   2.187$\pm$     0.356&   0.333$\pm$     0.095&\\   
&{8$_{7}$-7$_6$} &340.71416&81.2&234.192$\pm$   0.079&   0.247$\pm$     0.017&   2.299$\pm$     0.185&   0.606$\pm$     0.090&\\
&&&&&&&&\\
CS&{5}-{4}&244.93556&35.3&234.241$\pm$   0.011&   7.637$\pm$     0.057&   2.679$\pm$     0.026&   21.780$\pm$    0.376&\\
&&&&&&&&\\
H$_2$CS&{7$_{1,6}$-6$_{1,5}$}&244.04850&60.0&234.398$\pm$   0.075&   0.981$\pm$     0.057&   2.296$\pm$     0.178&   2.396$\pm$     0.324&\\ 
&&&&&&&&\\
H$_2$CO&{5$_{1,5}$-4$_{1,4}$}&351.76864&62.5&234.094$\pm$   0.007&   3.957$\pm$     0.021&   2.124$\pm$     0.016&   8.946$\pm$     0.114& \\   
&&&&&&&&\\
CH$_3$OH&{5$_0$-4$_0$},E&241.70016&47.9&234.259$\pm$   0.104&   0.758$\pm$     0.055&   2.587$\pm$     0.251&   2.086$\pm$     0.353&\\   
&{5$_{-1}$}-{4$_{-1}$},E&241.76723&44.4&234.143$\pm$   0.037&   2.220$\pm$     0.063&   2.401$\pm$     0.088&   5.672$\pm$     0.369&\\   
&{5$_0$}-4{$_0$},A$^+$&241.79135&34.8&234.378$\pm$   0.028&   2.679$\pm$     0.058&   2.386$\pm$     0.066&   6.804$\pm$     0.335&\\   
&{5$_1$-4$_{1}$}E& 241.87902&55.9&234.112$\pm$   0.190&   0.384$\pm$     0.070&   1.897$\pm$     0.447&   0.776$\pm$     0.324&\\
&{5$_{-2}$}-{4$_{-2}$},E&241.90415&60.7&233.809$\pm$   0.121&   0.736$\pm$     0.059&   2.687$\pm$     0.284&   2.106$\pm$     0.393&1\\   
&{5$_1$}-{4$_1$},A$^-$&243.91579&49.7&234.502$\pm$   0.161&   0.455$\pm$     0.074&   1.667$\pm$     0.379&   0.807$\pm$     0.315&\\   
%&{7$_0$-6$_0$}E&338.12449&78.1&232.256$\pm$   0.075&   0.190$\pm$     0.015&   1.770$\pm$     0.178&   0.359$\pm$     0.065&\\   
&{7$_{-1}$-6$_{-1}$}E &338.34459 &70.6&234.100$\pm$   0.060&   0.360$\pm$     0.017&   2.518$\pm$     0.141&   0.966$\pm$     0.101&\\  &{7$_0$-6$_0$}A$^+$& 338.40870 &65.0&234.170$\pm$   0.050&   0.430$\pm$     0.018&   2.306$\pm$     0.117&   1.056$\pm$     0.098&\\   
%&{7$_1$-6$_{1}$}E& 338.61493&86.1&232.431$\pm$   0.078&   0.144$\pm$     0.016&   1.285$\pm$     0.184&   0.197$\pm$     0.049&\\   
%&{7$_{-2}$-6$_{-2}$}E&338.72290&90.9&233.516$\pm$   0.151&   0.171$\pm$     0.013&   2.107$\pm$     0.397&   0.385$\pm$     0.102&1\\   
&{4$_0$-3$_{-1}$}E& 350.68766&36.3&234.098$\pm$   0.071&   0.327$\pm$     0.017&   2.606$\pm$     0.167&   0.908$\pm$     0.104&\\   
&{1$_1$-0$_0$}A$^+$ &350.90510&16.8&234.013$\pm$   0.064&   0.365$\pm$     0.018&   2.259$\pm$     0.150&   0.878$\pm$     0.100&\\   
&&&&&&&&\\
CN%&N= 3- 2, J=5/2-5/2, F=3/2-3/2&339.44678&32.6&233.203$\pm$   0.091&   0.126$\pm$     0.014&   1.387$\pm$     0.229&   0.187$\pm$     0.052&\\   
%&N= 3- 2, J=5/2-5/2, F=7/2-7/2&339.51664&32.6&234.385$\pm$   0.203&   0.107$\pm$     0.017&   1.990$\pm$     0.477&   0.227$\pm$     0.091&CANCEL 3SIGMA \\   
%&N= 3- 2, J=5/2-3/2, F=5/2-5/2&340.00813&32.6&233.017$\pm$   0.084&   0.196$\pm$     0.018&   1.674$\pm$     0.197&   0.350$\pm$     0.074&\\   
%&N= 3- 2, J=5/2-3/2, F=3/2-3/2&340.01963&32.6&232.248$\pm$   0.060&   0.210$\pm$     0.013&   1.664$\pm$     0.164&   0.371$\pm$     0.060&\\   
&N= 3- 2, J=5/2-3/2, F=7/2-5/2&340.03155&32.6&233.508$\pm$   0.107&   0.399$\pm$     0.013&   4.711$\pm$     0.298&   1.999$\pm$     0.189&2\\   
%&N= 3- 2, J=5/2-3/2, F=3/2-1/2&340.03541&32.6&236.982$\pm$   0.126&   0.391$\pm$     0.014&   5.122$\pm$     0.328&   2.133$\pm$     0.214&2,3\\   
&N= 3- 2, J=7/2-5/2, F=7/2-5/2&340.24777&32.7&234.022$\pm$   0.008&   2.419$\pm$     0.018&   2.145$\pm$     0.018&   5.525$\pm$     0.088&2\\   
%&N= 3- 2, J=7/2-5/2, F=5/2-5/2& 340.26177&32.7&232.877$\pm$   0.061&   0.171$\pm$     0.014&   1.468$\pm$     0.155&   0.268$\pm$     0.050&\\   
&N= 3- 2, J=7/2-5/2, F=7/2-7/2&340.26495&32.7&234.208$\pm$   0.174&   0.149$\pm$     0.018&   1.971$\pm$     0.410&   0.313$\pm$     0.103&\\   
&&&&&&&&\\
CCH&N= 4- 3, J=9/2-7/2, F= 5- 4&349.33771&41.9&233.634$\pm$   0.007&   3.227$\pm$     0.017&   2.557$\pm$     0.016&   8.783$\pm$     0.101&3\\   
&N= 4- 3, J=7/2-5/2, F= 4- 3&349.39928&41.9&233.691$\pm$   0.007&   2.861$\pm$     0.021&   1.663$\pm$     0.016&   5.066$\pm$     0.085&3\\   
%&N= 4- 3, J=7/2-7/2, F= 4- 4&349.60361&41.9&232.413$\pm$   0.126&   0.150$\pm$     0.016&   2.242$\pm$     0.296&   0.359$\pm$     0.086&\\   
\enddata
\tablecomments{
\scriptsize{(1) Two CH$_3$OH lines with similar spectroscopic constants are blended. (2) Blended with three hyperfine components. (3) Blended with two hyperfine components.
}}
\end{deluxetable*}
%%%%%%%%%%%%%%%%%%%%%%%%%%%%%%%%%%%%%%%%%%%%%%%%%%%%%%%%
%\clearpage
\restartappendixnumbering
\section{Observed spectra \label{fig_spectra}}
\begin{figure*}
\begin{minipage}{0.25\textwidth}
\includegraphics[width=\textwidth]{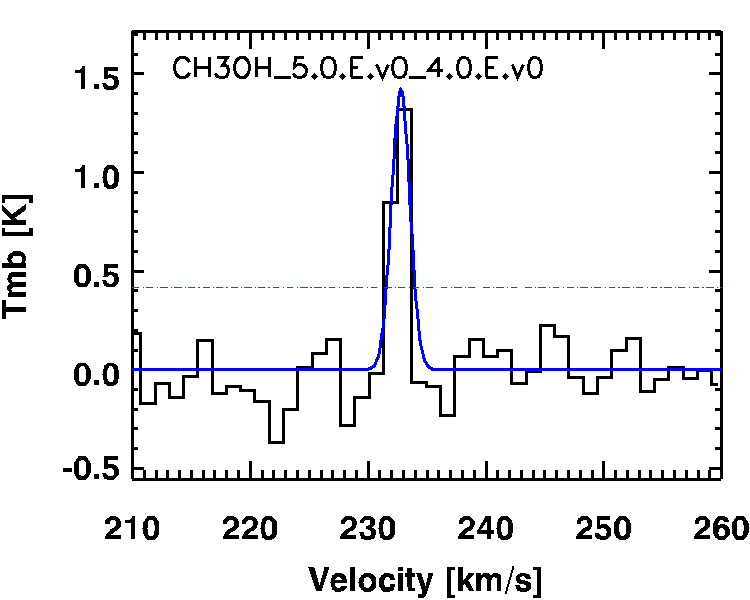}
\end{minipage}
\hskip -1.2 cm
\begin{minipage}{0.25\textwidth}
\includegraphics[width=\textwidth]{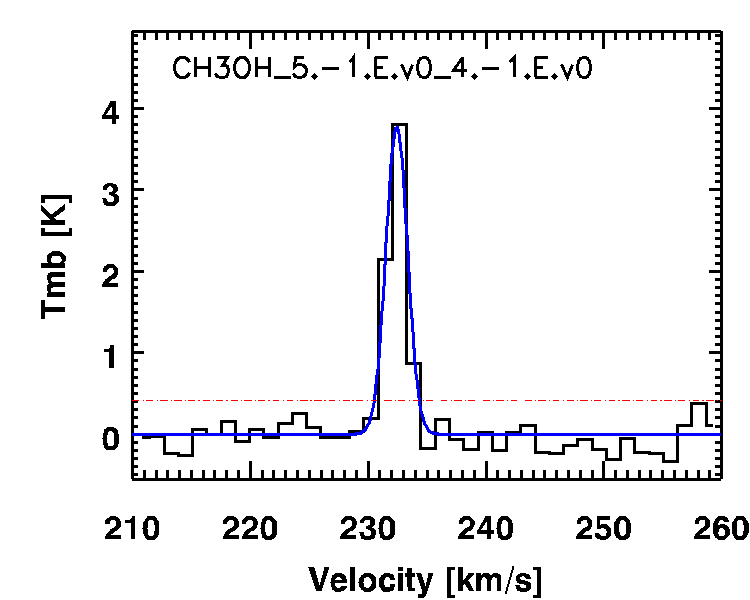}
\end{minipage}
\hskip -1.2 cm
\begin{minipage}{0.25\textwidth}
\includegraphics[width=\textwidth]{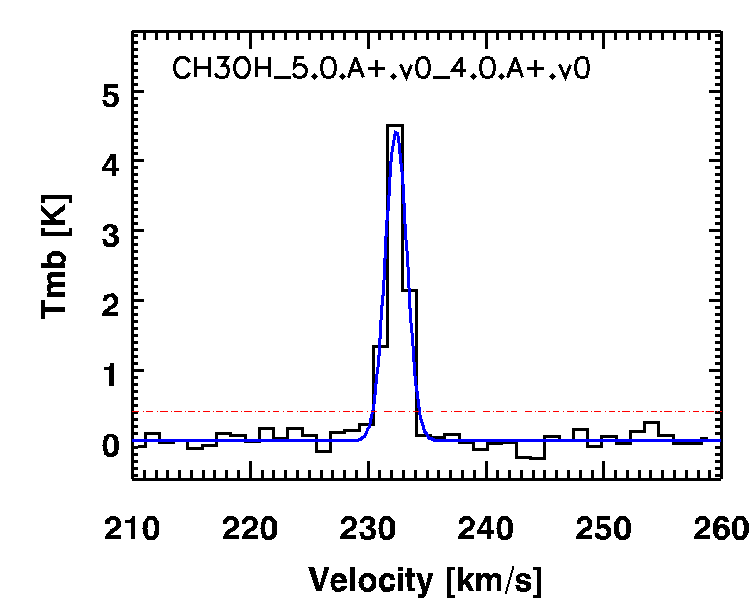}
\end{minipage}
\hskip -1.2 cm
\begin{minipage}{0.25\textwidth}
\includegraphics[width=\textwidth]{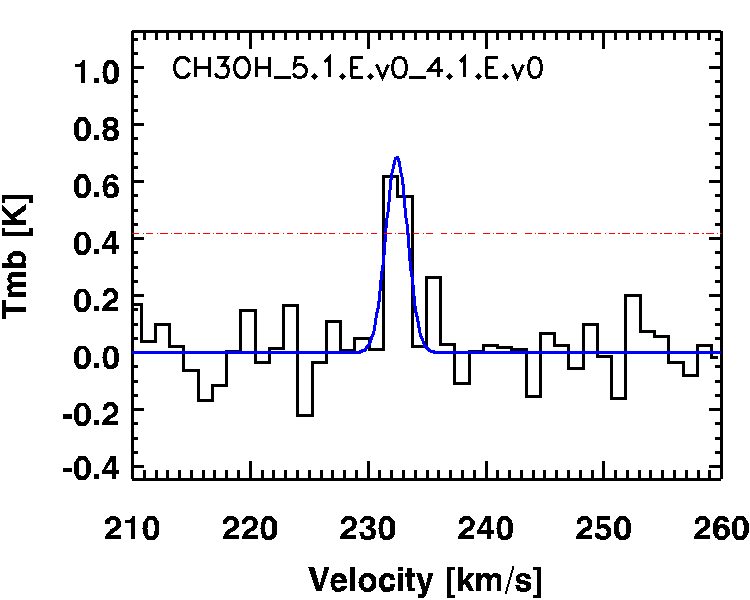}
\end{minipage}
\begin{minipage}{0.25\textwidth}
\includegraphics[width=\textwidth]{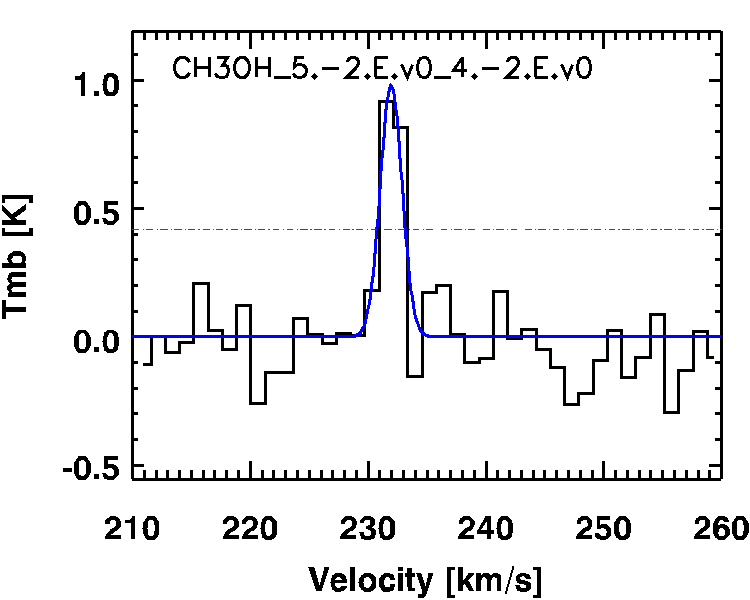}
\end{minipage}
\hskip -1.2 cm
\begin{minipage}{0.25\textwidth}
\includegraphics[width=\textwidth]{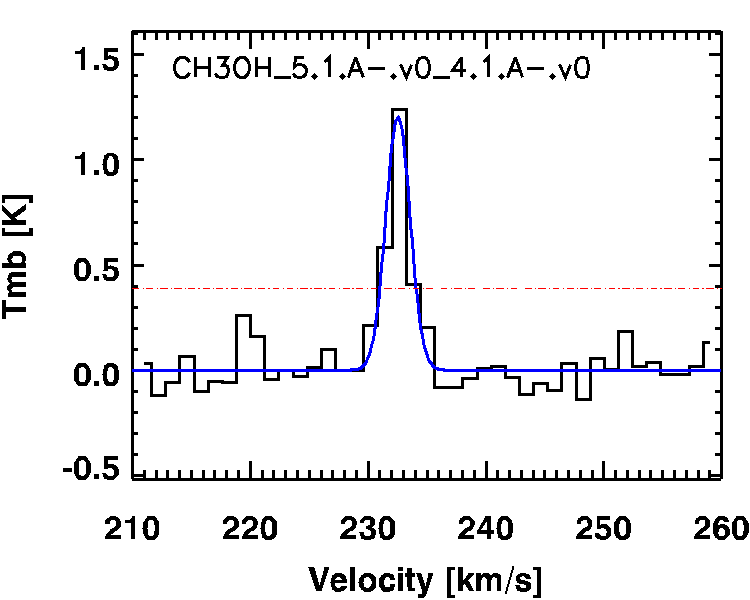}
\end{minipage}
\hskip -1.2 cm
\begin{minipage}{0.25\textwidth}
\includegraphics[width=\textwidth]{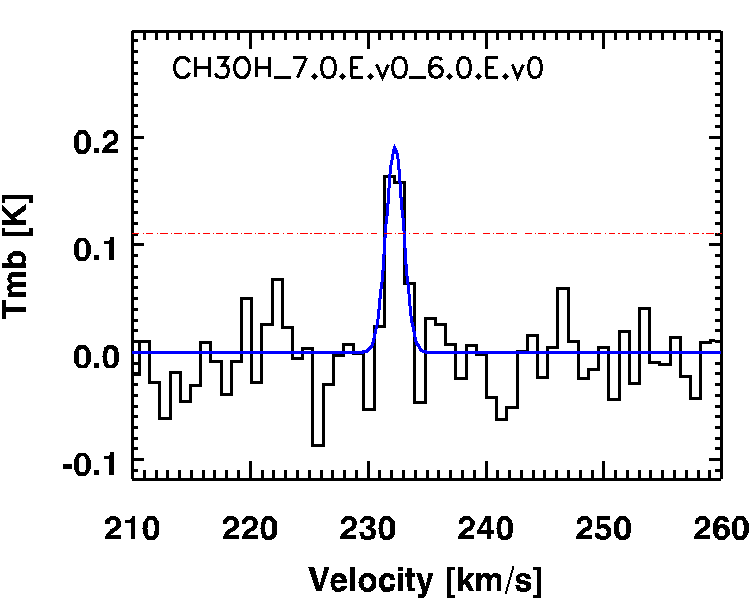}
\end{minipage}
\hskip -1.2 cm
\begin{minipage}{0.25\textwidth}
\includegraphics[width=\textwidth]{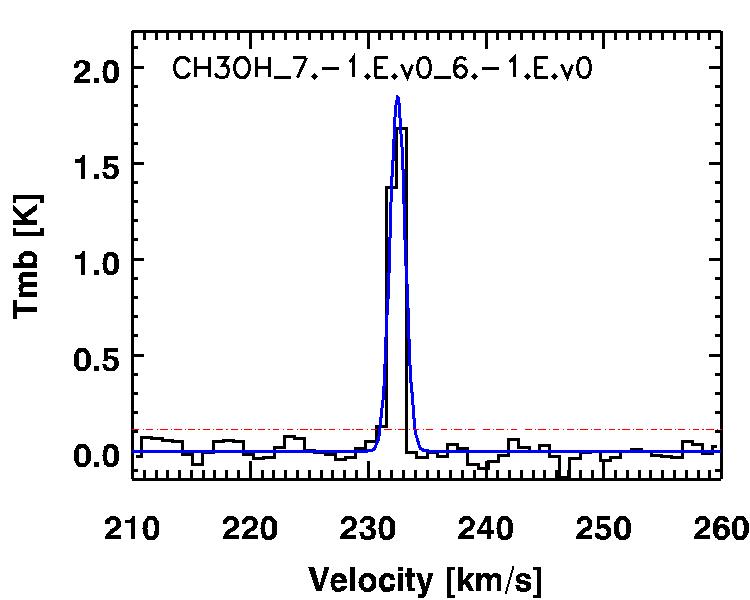}
\end{minipage}
\begin{minipage}{0.25\textwidth}
\includegraphics[width=\textwidth]{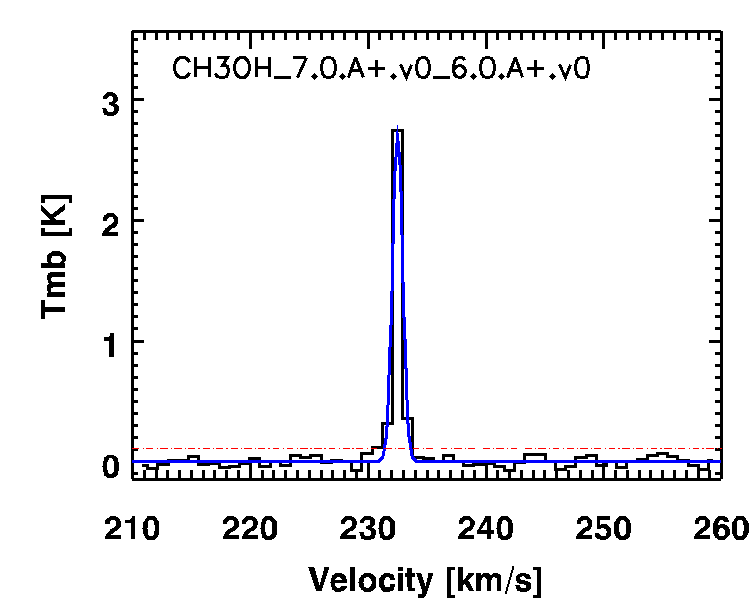}
\end{minipage}
\hskip -1.2 cm
\begin{minipage}{0.25\textwidth}
\includegraphics[width=\textwidth]{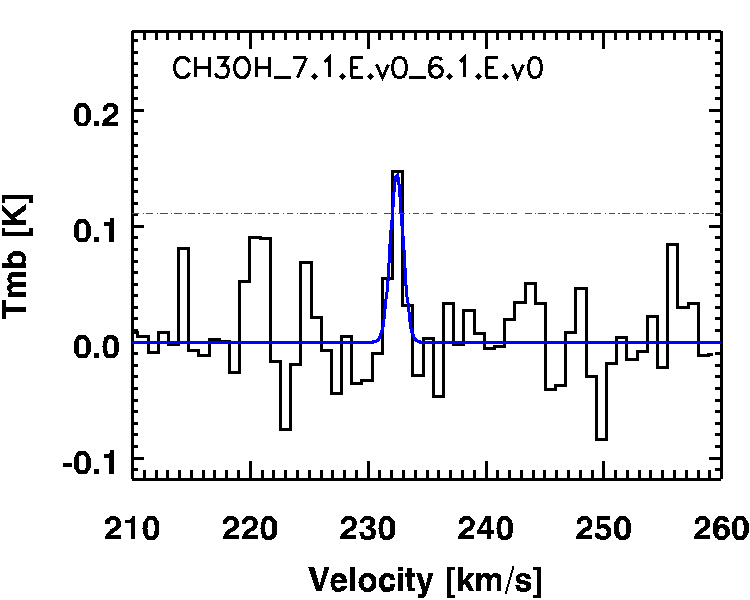}
\end{minipage}
\hskip -1.2 cm
\begin{minipage}{0.25\textwidth}
\includegraphics[width=\textwidth]{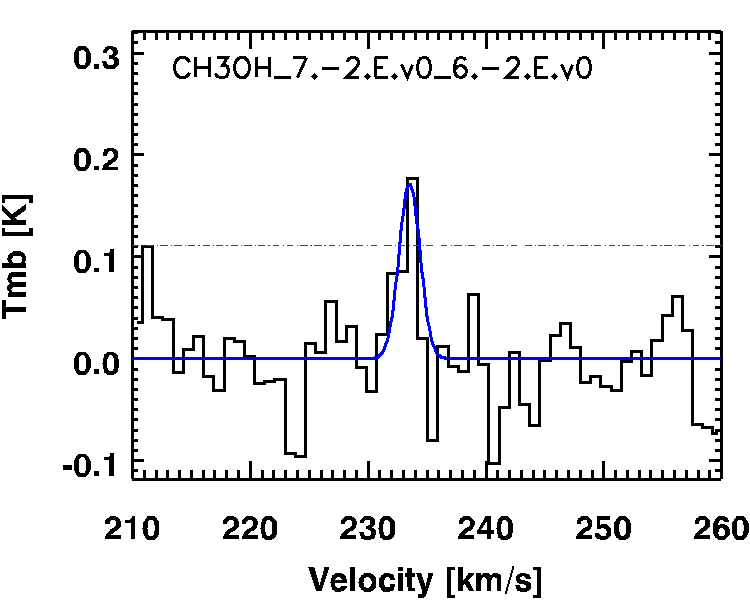}
\end{minipage}
\hskip -1.2 cm
\begin{minipage}{0.25\textwidth}
\includegraphics[width=\textwidth]{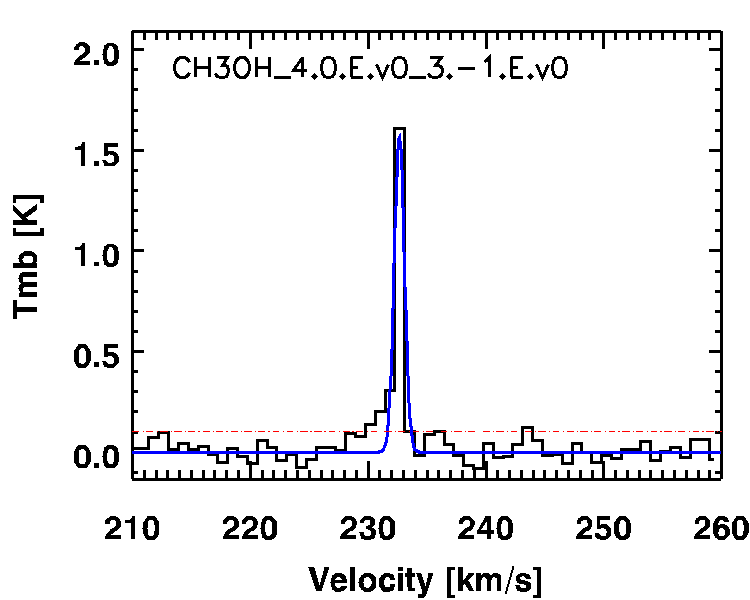}
\end{minipage}
%\hskip -1.2 cm
\begin{minipage}{0.25\textwidth}
\includegraphics[width=\textwidth]{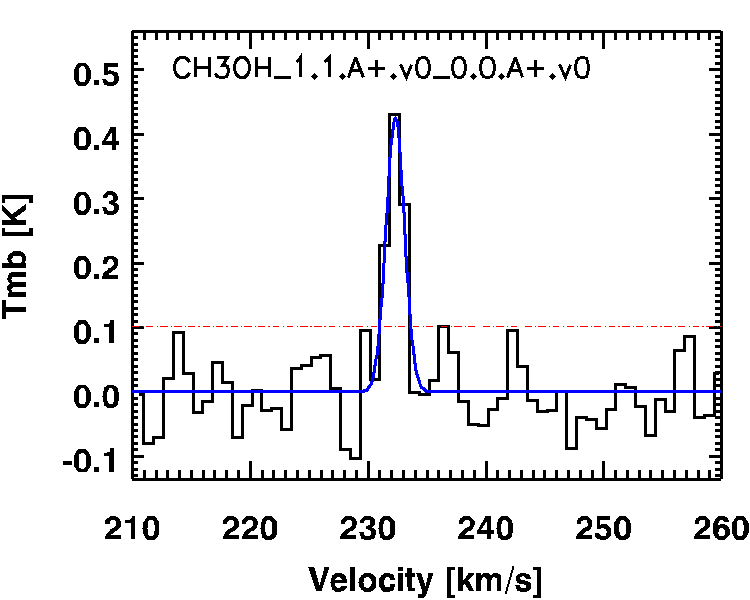}
\end{minipage}
\hskip -1.2 cm
\begin{minipage}{0.25\textwidth}
\includegraphics[width=\textwidth]{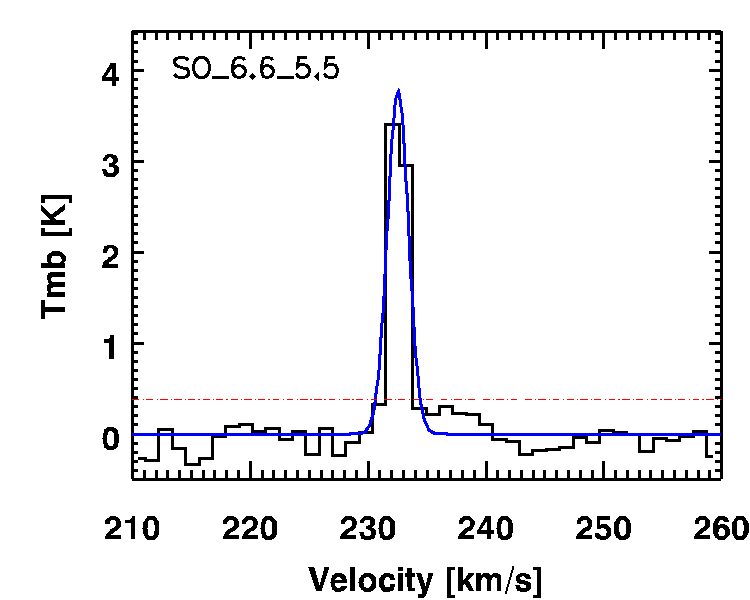}
\end{minipage}
\hskip -1.2 cm
\begin{minipage}{0.25\textwidth}
\includegraphics[width=\textwidth]{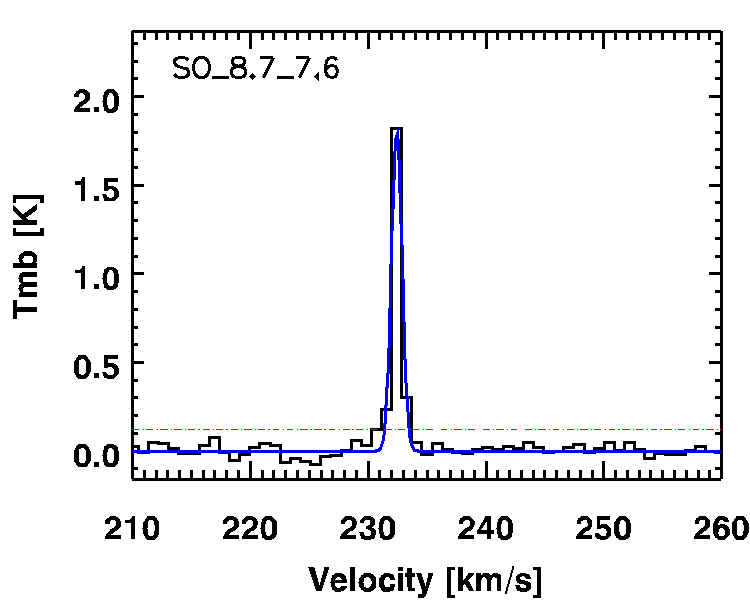}
\end{minipage}
\hskip -1.2 cm
\begin{minipage}{0.25\textwidth}
\includegraphics[width=\textwidth]{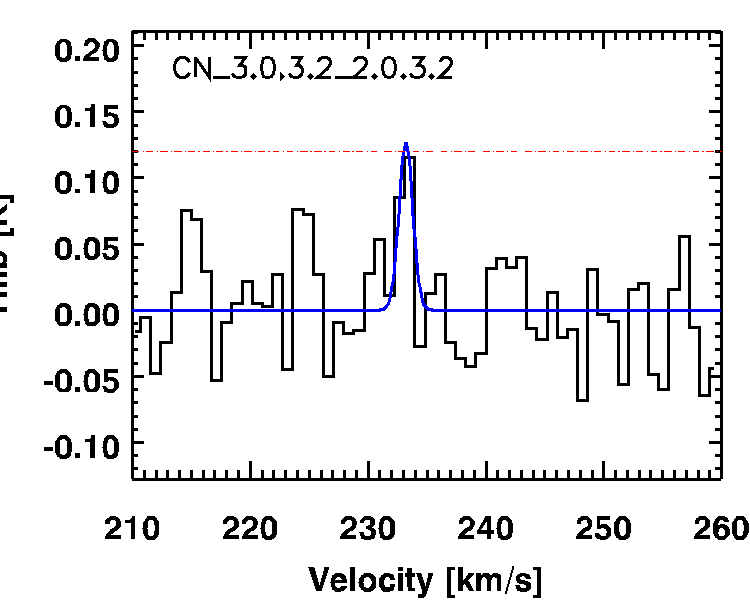}
\end{minipage}
%\hskip -1.2 cm
\begin{minipage}{0.25\textwidth}
\includegraphics[width=\textwidth]{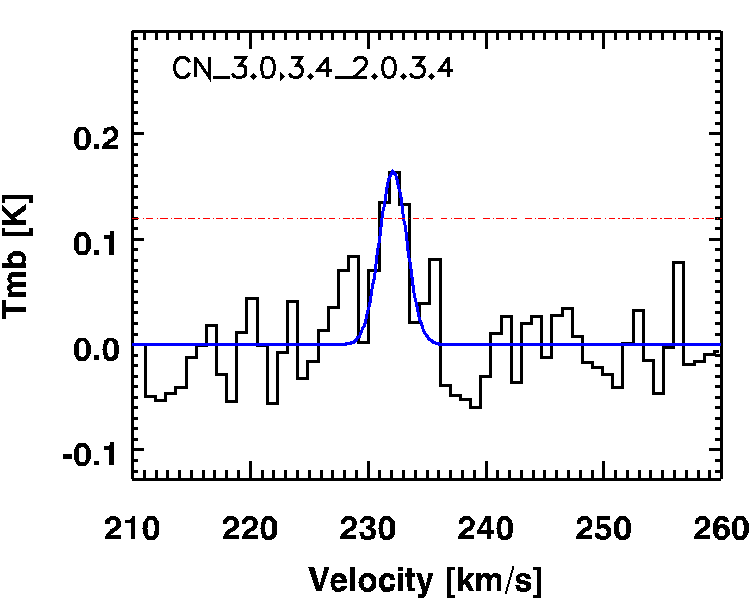}
\end{minipage}
\hskip -1.2 cm
\begin{minipage}{0.25\textwidth}
\includegraphics[width=\textwidth]{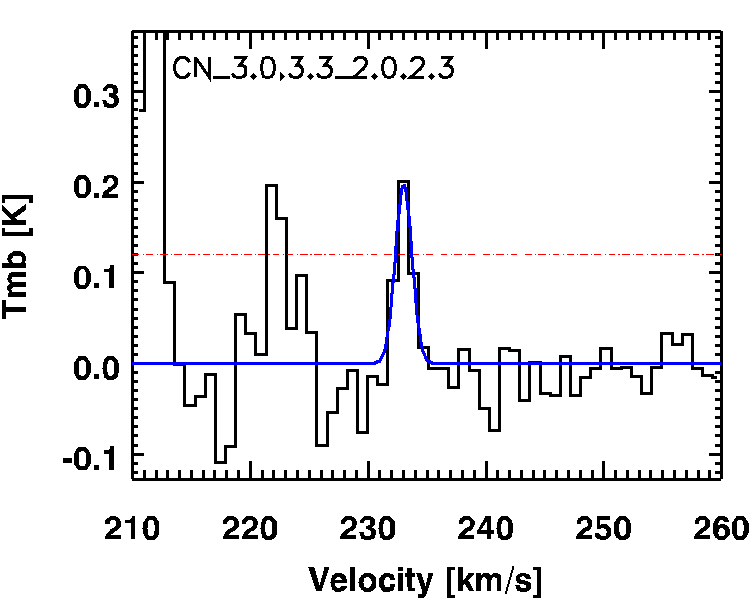}
\end{minipage}
\hskip -1.2 cm
\begin{minipage}{0.25\textwidth}
\includegraphics[width=\textwidth]{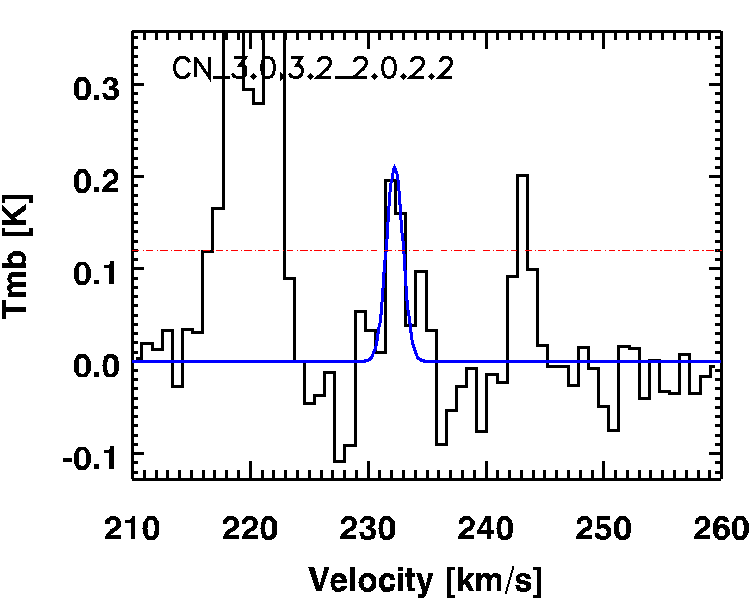}
\end{minipage}
\hskip -1.2 cm
\begin{minipage}{0.25\textwidth}
\includegraphics[width=\textwidth]{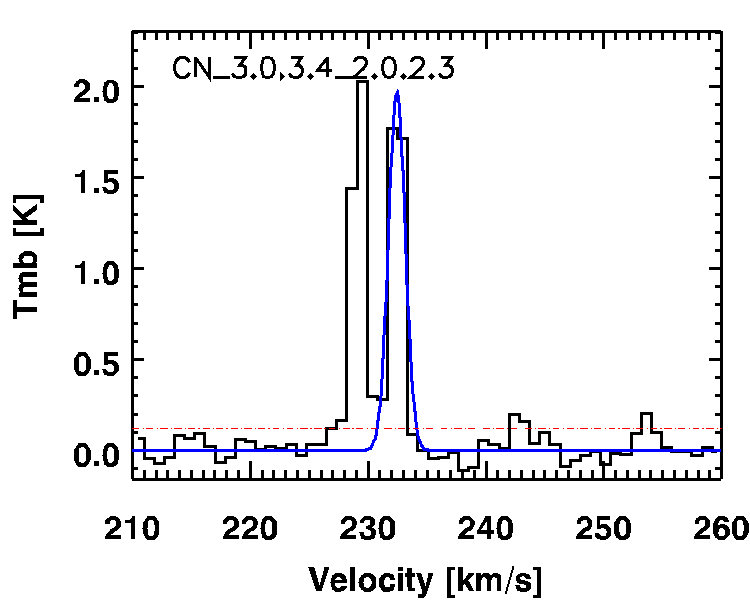}
\end{minipage}
%\hskip -1.2 cm
\begin{minipage}{0.25\textwidth}
\includegraphics[width=\textwidth]{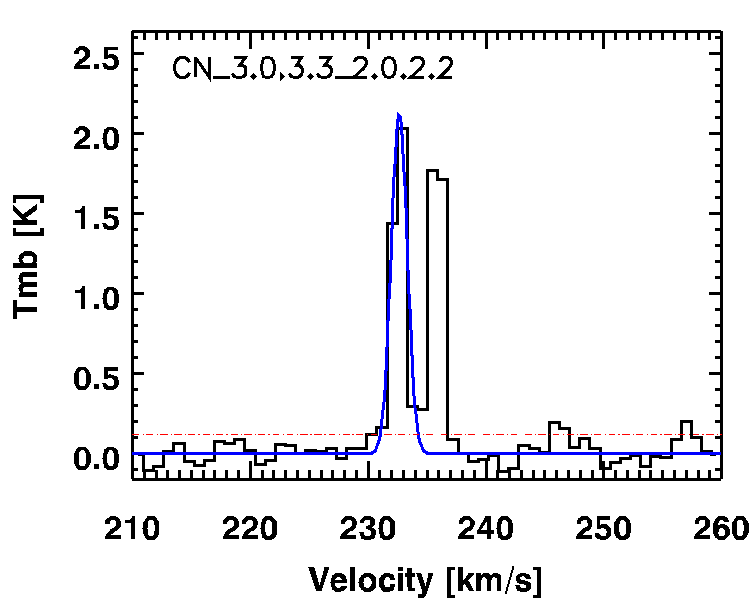}
\end{minipage}
\hskip -0.1 cm
\begin{minipage}{0.25\textwidth}
\includegraphics[width=\textwidth]{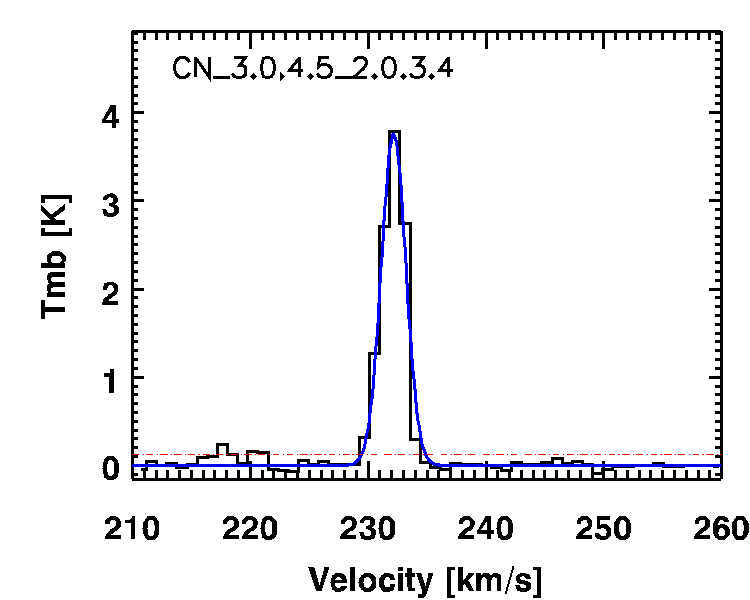}
\end{minipage}
\hskip -0.1 cm
\begin{minipage}{0.25\textwidth}
\includegraphics[width=\textwidth]{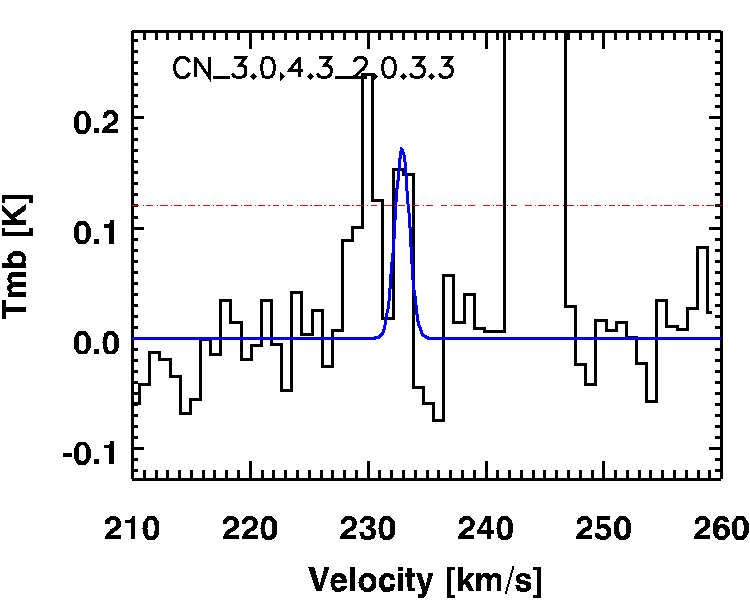}
\end{minipage}
\hskip -0.222 cm
\begin{minipage}{0.25\textwidth}
\includegraphics[width=\textwidth]{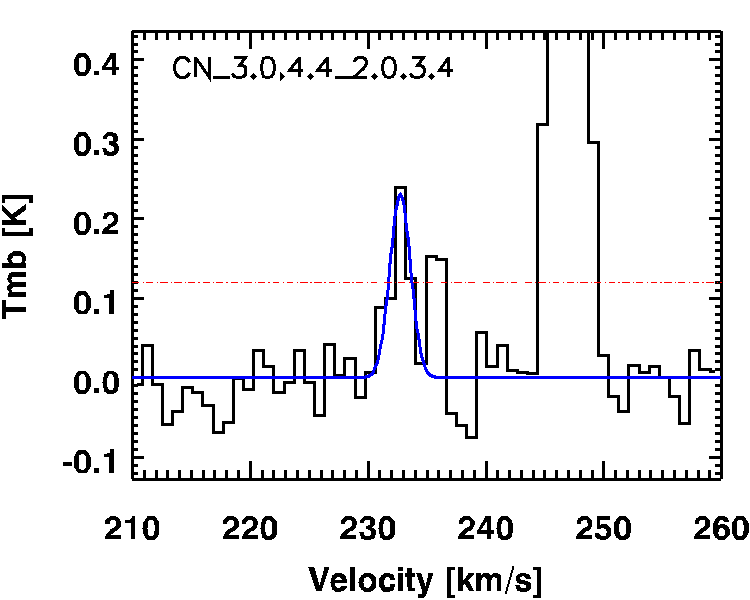}
\end{minipage}
\caption{Spectra of CH$_3$OH and CN emission lines extracted from 0$\farcs$42 (0.1 pc) diameter region centered at CH3OH peak emission in N79S-1. The blue lines represent Gaussian profiles fitted to the observed spectra (solid black lines). The dashed red line shows the 3 $\sigma$ rms noise.}
\label{Fig:1specS1}
\end{figure*}

\begin{figure*}
\begin{minipage}{0.25\textwidth}
\includegraphics[width=\textwidth]{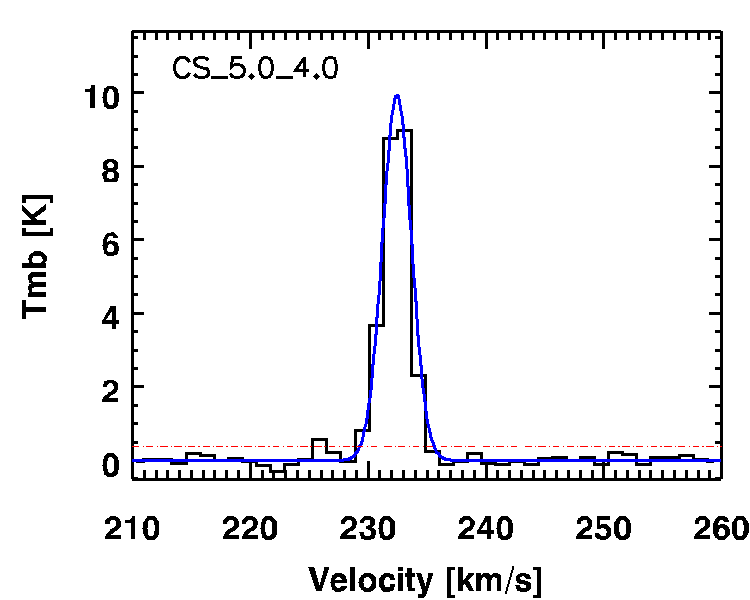}
\end{minipage}
\hskip -1.2cm
\begin{minipage}{0.25\textwidth}
\includegraphics[width=\textwidth]{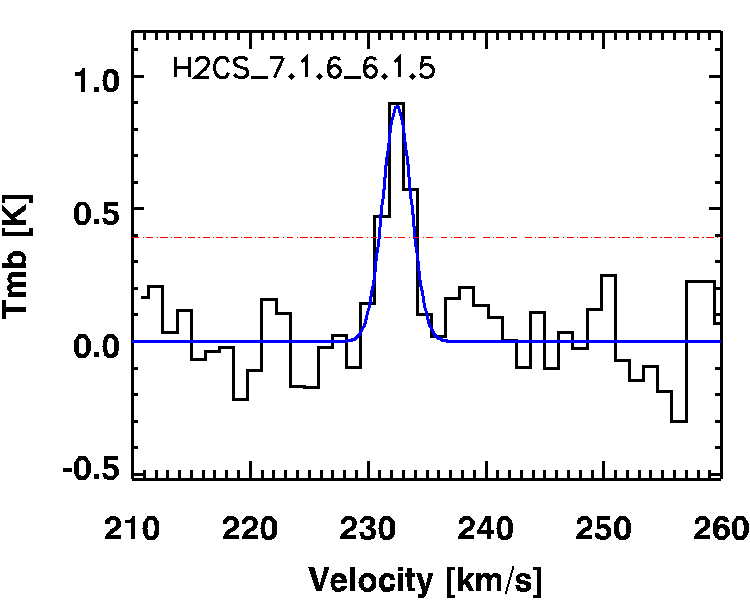}
\end{minipage}
\hskip -1.2cm
\begin{minipage}{0.25\textwidth}
\includegraphics[width=\textwidth]{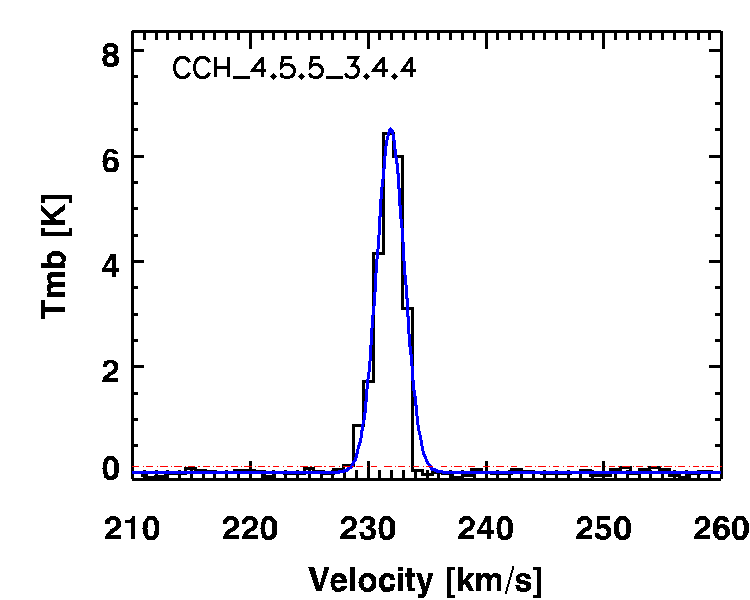}
\end{minipage}
\hskip -1.2cm
\begin{minipage}{0.25\textwidth}
\includegraphics[width=\textwidth]{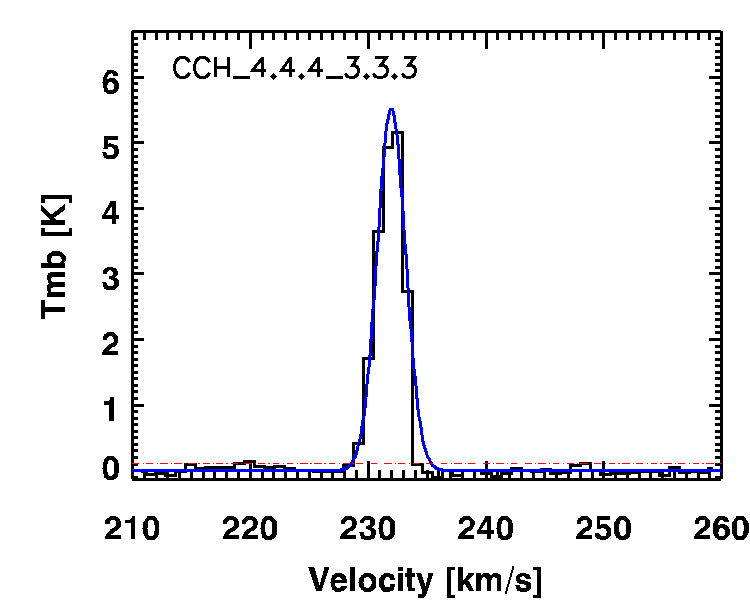}
\end{minipage}
%\hskip 1.2cm
\begin{minipage}{0.25\textwidth}
\includegraphics[width=\textwidth]{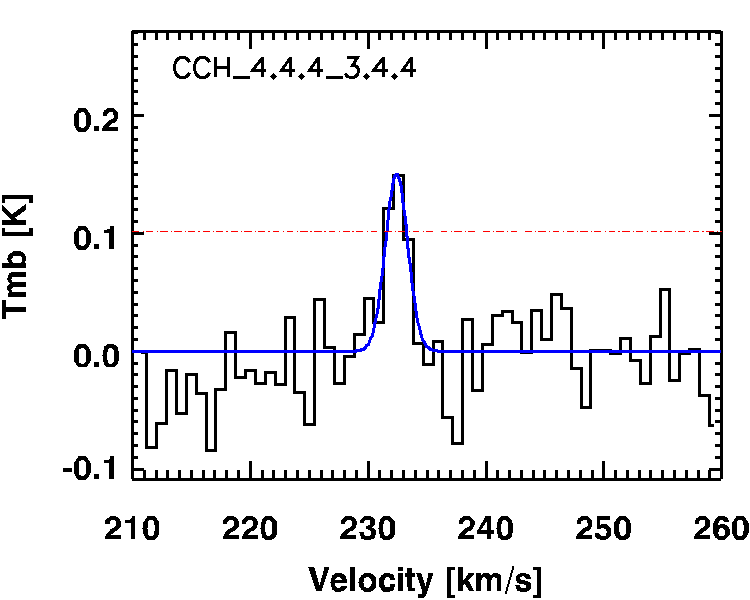}
\end{minipage}
\begin{minipage}{0.25\textwidth}
\includegraphics[width=\textwidth]{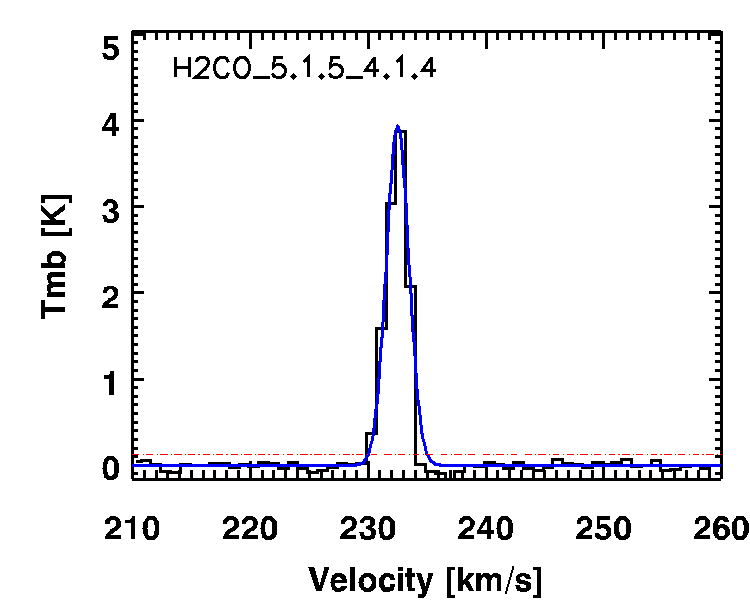}
\end{minipage}
\caption{Same as in Figure \ref{Fig:1specS1} but molecular emission of CS, H$_2$CS, CCH andH$_2$CO.}
\label{Fig:2specS1}
\end{figure*}

\begin{figure*}
\begin{minipage}{0.25\textwidth}
\includegraphics[width=\textwidth]{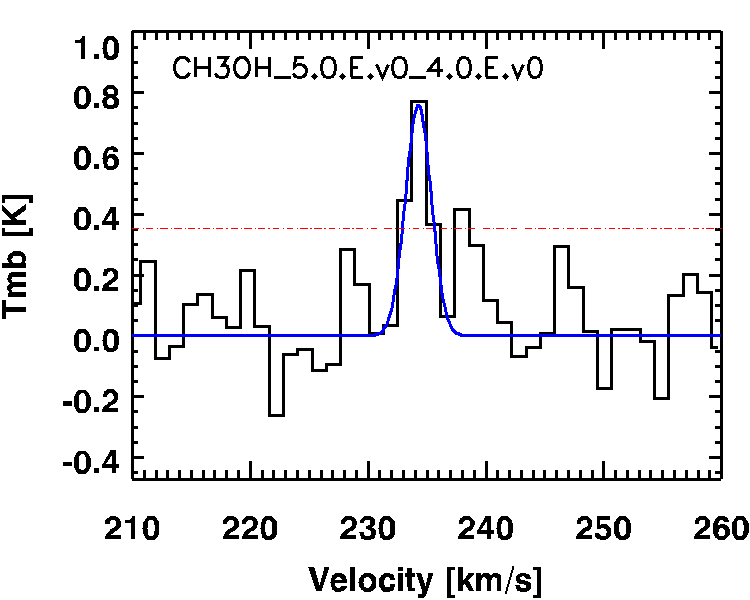}
\end{minipage}
\hskip -1.2 cm
\begin{minipage}{0.25\textwidth}
\includegraphics[width=\textwidth]{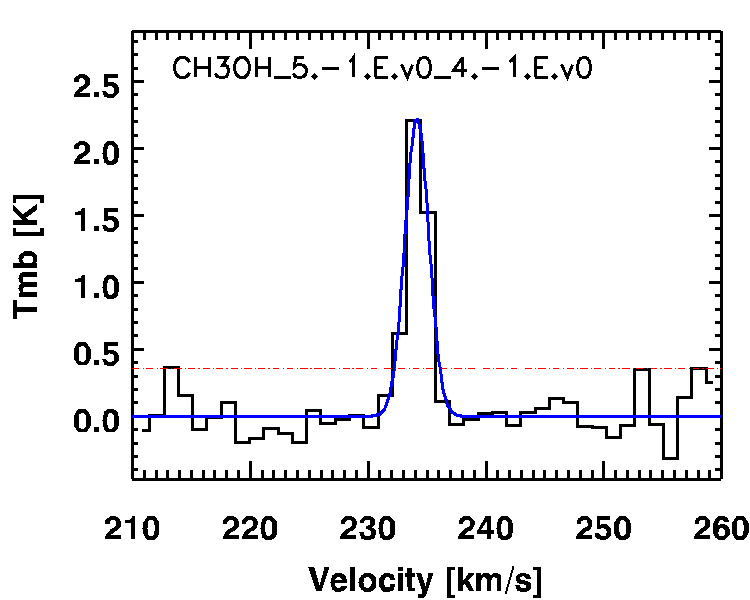}
\end{minipage}
\hskip -1.2 cm
\begin{minipage}{0.25\textwidth}
\includegraphics[width=\textwidth]{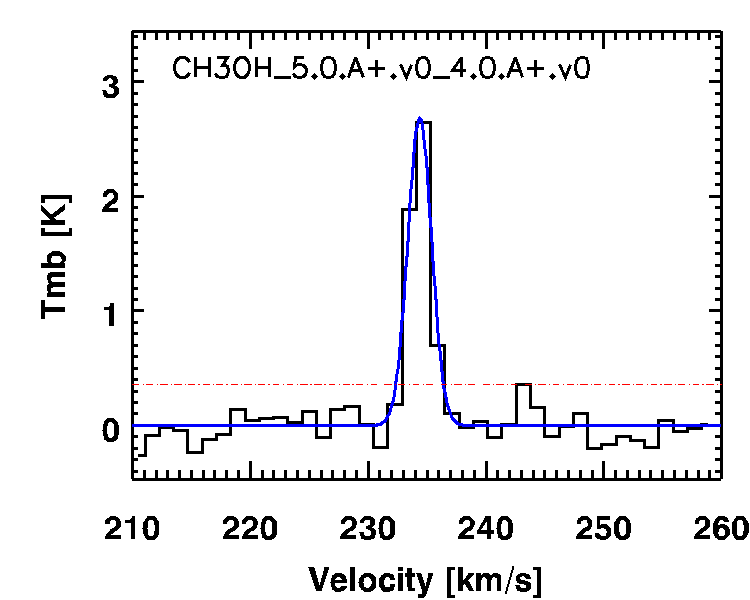}
\end{minipage}
\hskip -1.2 cm
\begin{minipage}{0.25\textwidth}
\includegraphics[width=\textwidth]{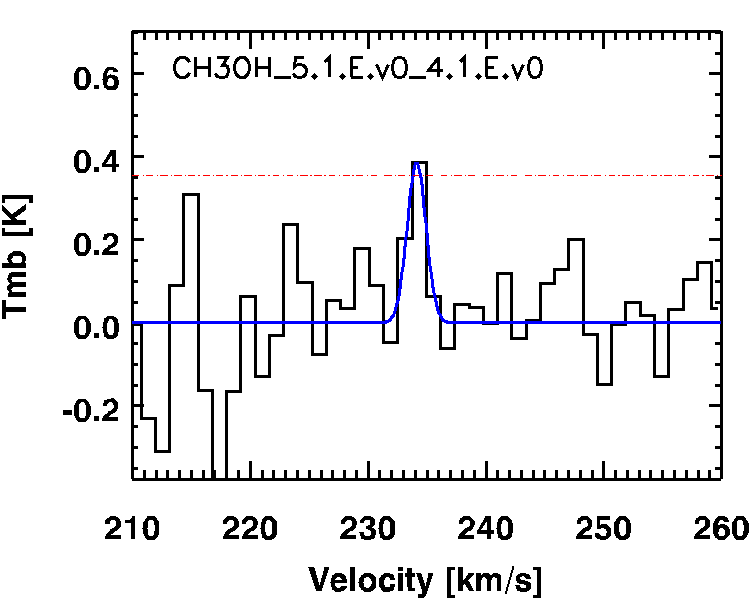}
\end{minipage}
\begin{minipage}{0.25\textwidth}
\includegraphics[width=\textwidth]{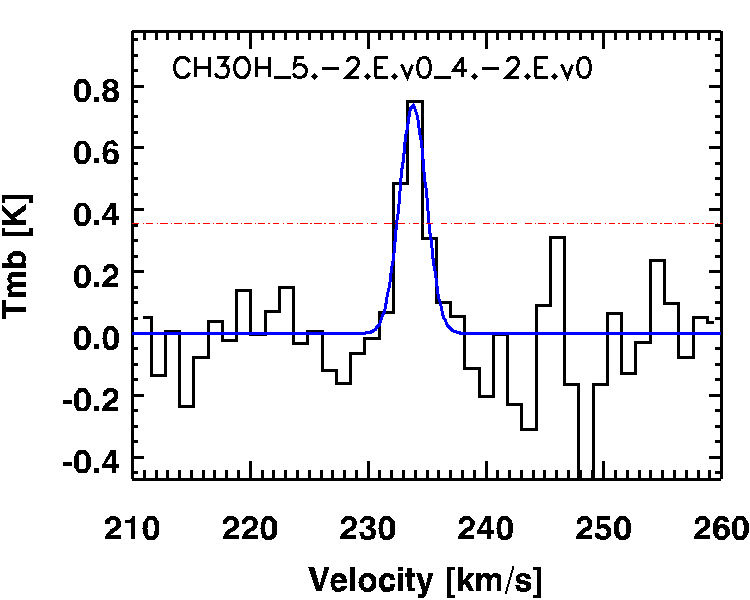}
\end{minipage}
\hskip -1.2 cm
\begin{minipage}{0.25\textwidth}
\includegraphics[width=\textwidth]{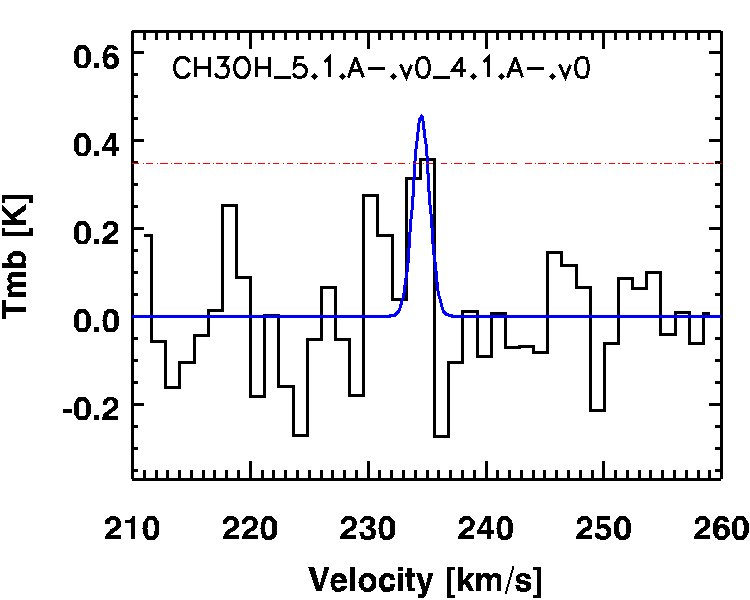}
\end{minipage}
\hskip -1.2 cm
\begin{minipage}{0.25\textwidth}
\includegraphics[width=\textwidth]{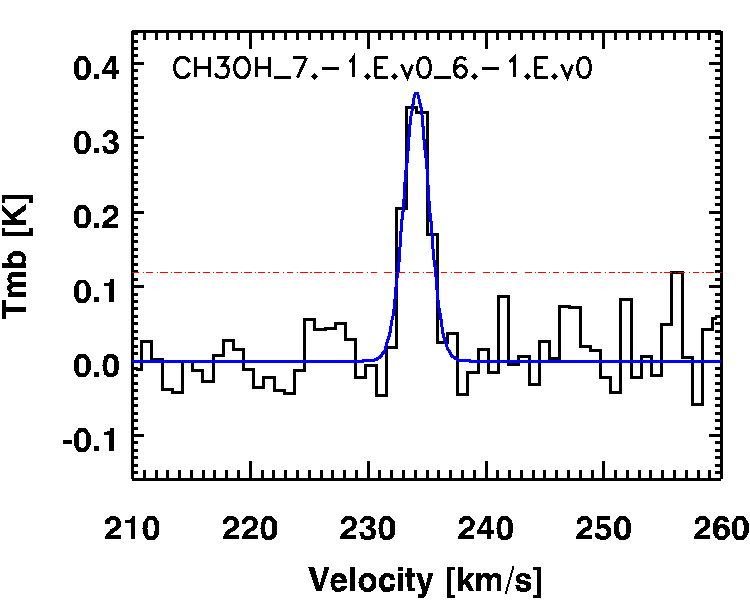}
\end{minipage}
\hskip -1.2 cm
\begin{minipage}{0.25\textwidth}
\includegraphics[width=\textwidth]{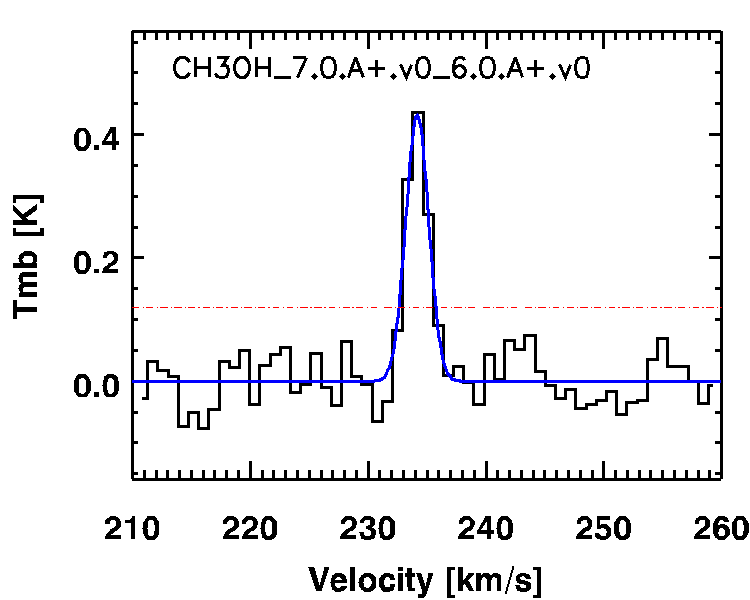}
\end{minipage}
\begin{minipage}{0.25\textwidth}
\includegraphics[width=\textwidth]{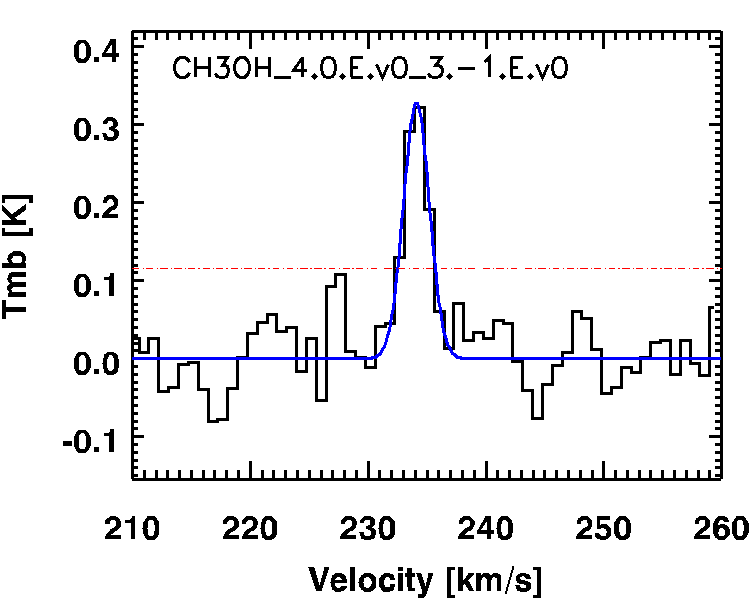}
\end{minipage}
\hskip -1.2 cm
\begin{minipage}{0.25\textwidth}
\includegraphics[width=\textwidth]{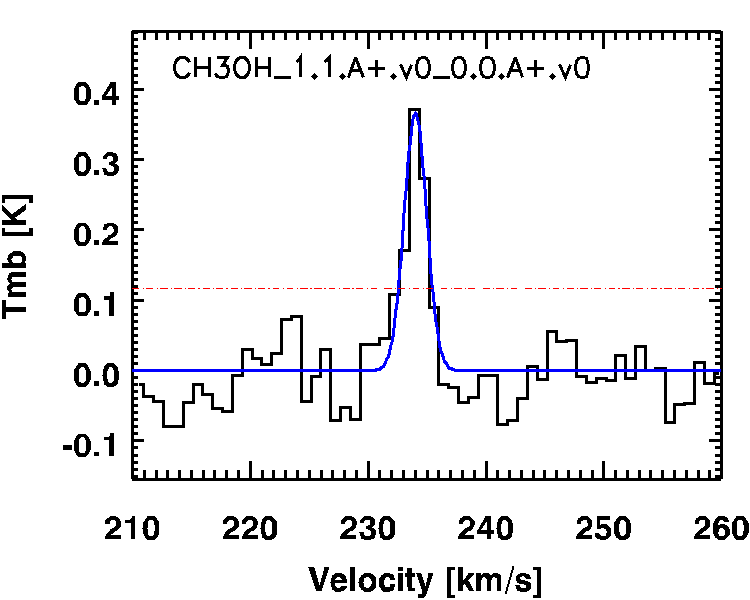}
\end{minipage}
\hskip -1.2 cm
\begin{minipage}{0.25\textwidth}
\includegraphics[width=\textwidth]{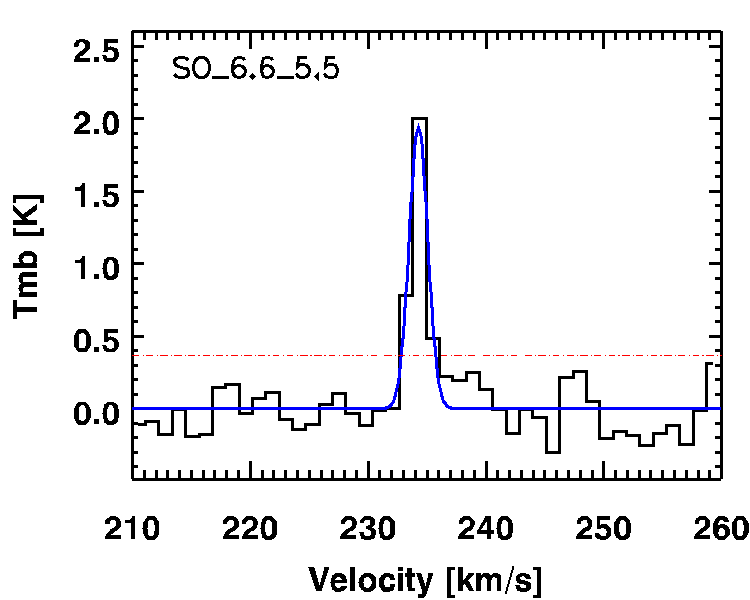}
\end{minipage}
\hskip -1.2 cm
\begin{minipage}{0.25\textwidth}
\includegraphics[width=\textwidth]{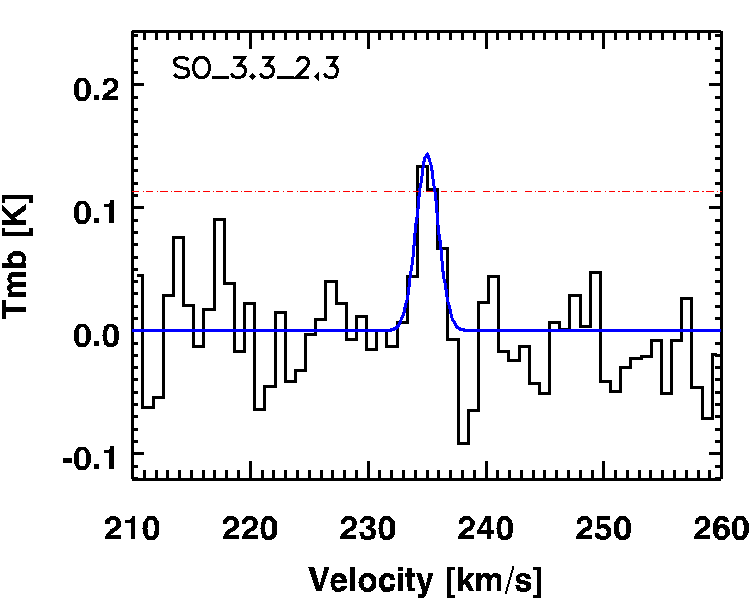}
\end{minipage}
%\hskip -1.2 cm
\begin{minipage}{0.25\textwidth}
\includegraphics[width=\textwidth]{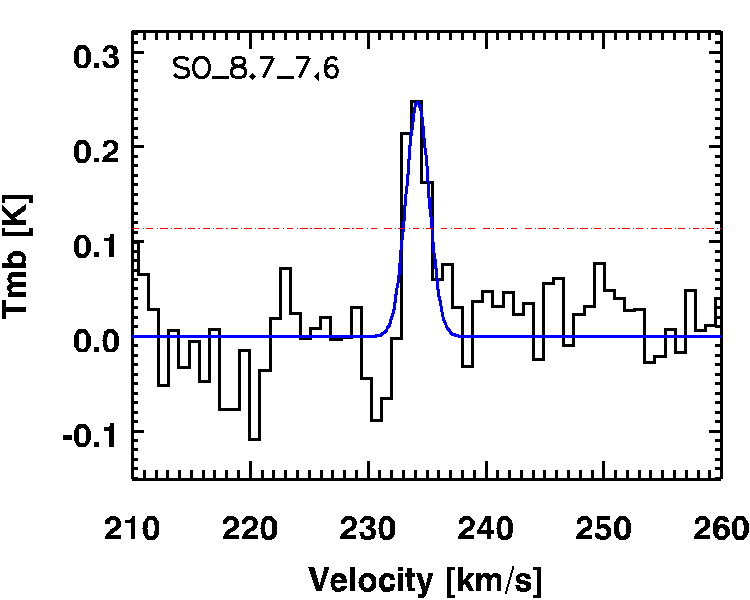}
\end{minipage}
\hskip -1.2 cm
\begin{minipage}{0.25\textwidth}
\includegraphics[width=\textwidth]{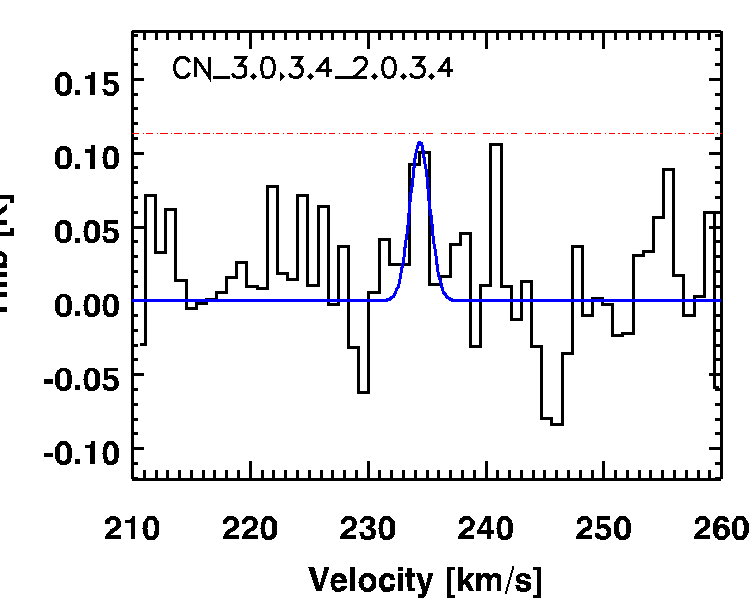}
\end{minipage}
\hskip -1.2 cm
\begin{minipage}{0.25\textwidth}
\includegraphics[width=\textwidth]{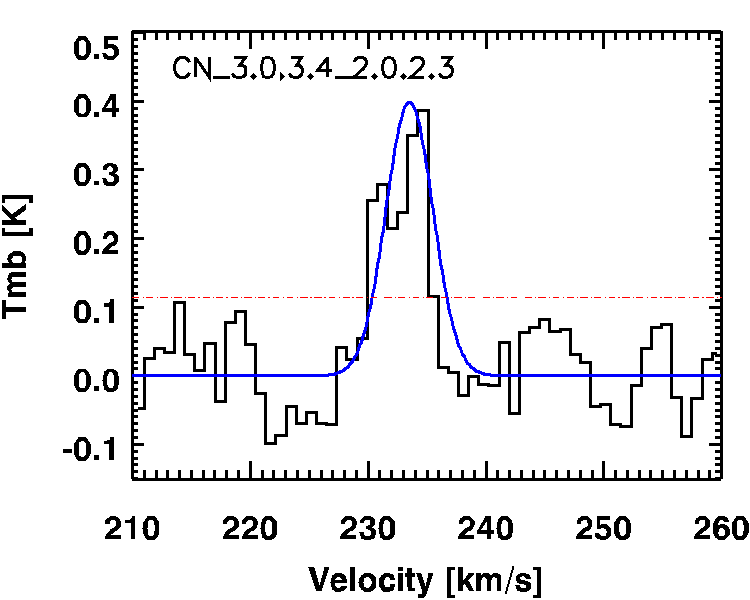}
\end{minipage}
\hskip -1.2 cm
\begin{minipage}{0.25\textwidth}
\includegraphics[width=\textwidth]{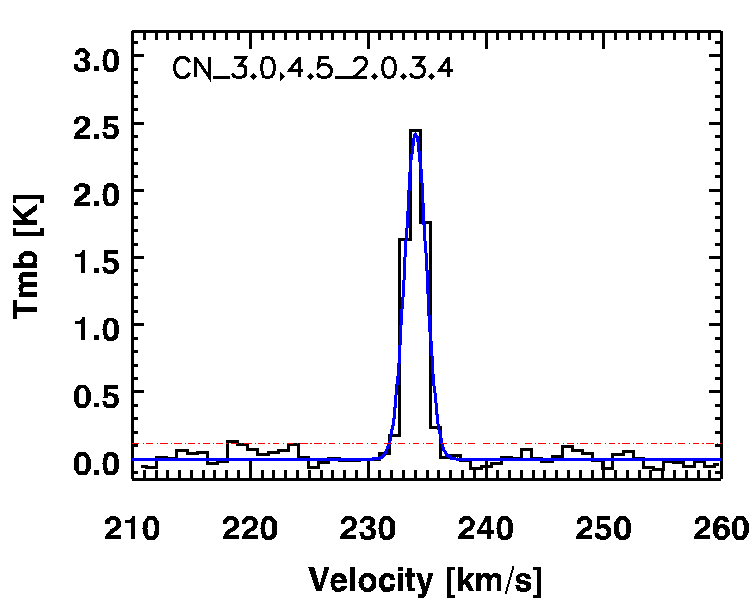}
\end{minipage}
%\hskip -1.2 cm
\begin{minipage}{0.25\textwidth}
\includegraphics[width=\textwidth]{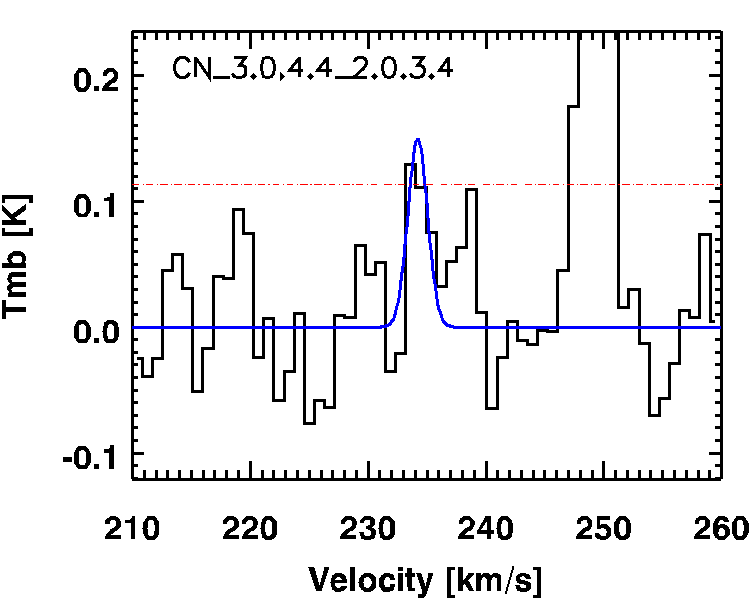}
\end{minipage}
\hskip -1.2 cm
\begin{minipage}{0.25\textwidth}
\includegraphics[width=\textwidth]{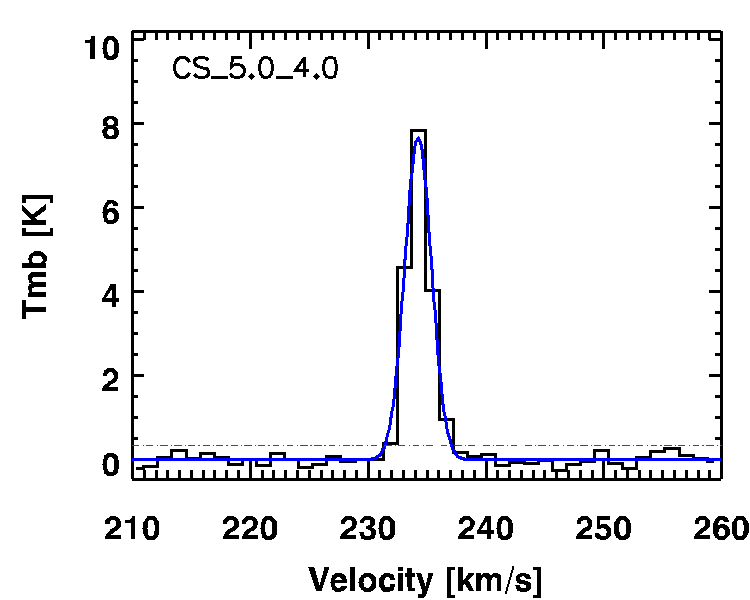}
\end{minipage}
\hskip -1.2 cm
\begin{minipage}{0.25\textwidth}
\includegraphics[width=\textwidth]{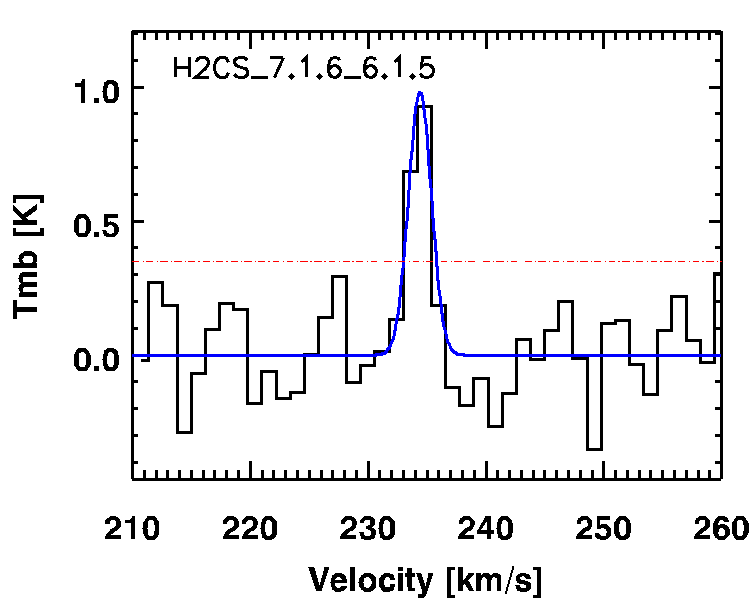}
\end{minipage}
\hskip -1.2 cm
\begin{minipage}{0.25\textwidth}
\includegraphics[width=\textwidth]{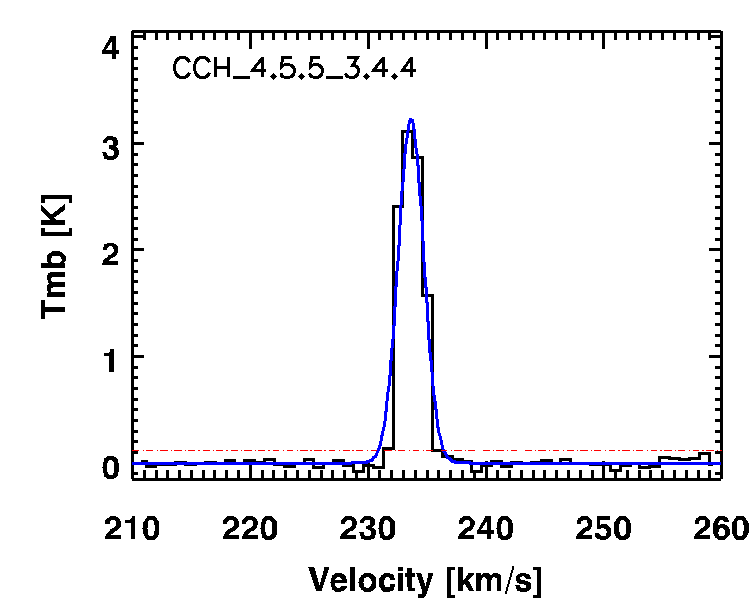}
\end{minipage}
%\hskip -1.2 cm
\begin{minipage}{0.25\textwidth}
\includegraphics[width=\textwidth]{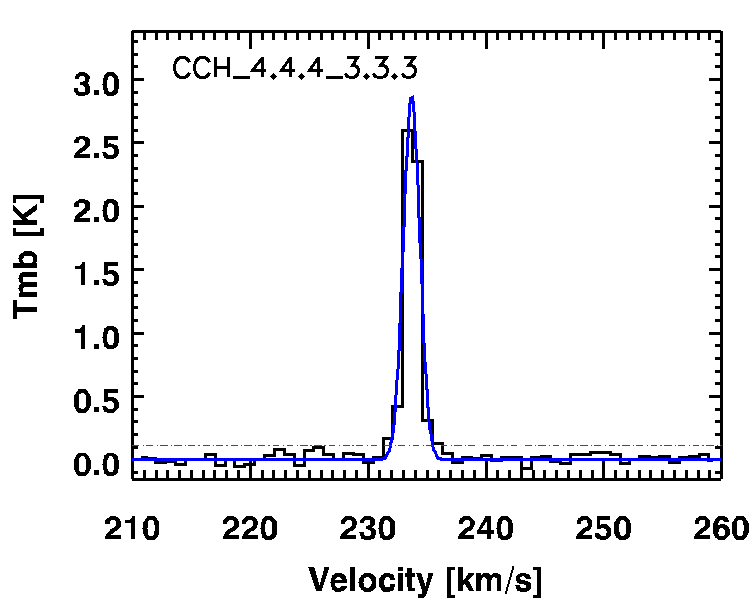}
\end{minipage}
\hskip -0.15cm
\begin{minipage}{0.25\textwidth}
\includegraphics[width=\textwidth]{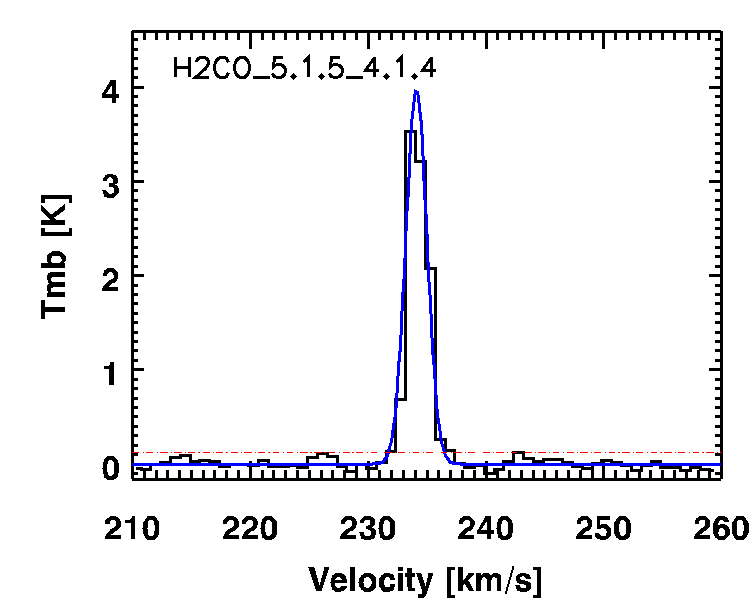}
\end{minipage}
\caption{Same as in Figure \ref{Fig:1specS1} but molecular emission lines detected from CH$_3$OH Peak emission at N79S-2.}
\label{Fig:3specS2}
\end{figure*}

Figures \ref{Fig:1specS1}, \ref{Fig:2specS1} and \ref{Fig:3specS2} show the observed and fitted to the observed transitions toward N79S-1 and  N79S-2 (see Section for details \ref{sec_spectra}).

%%%%%%%%%%%%%%%%%%%%%%%%%%%
%%%%%%%%%%%%%%%%%%%%%%%%%%%%%%%%%%%%%%%%%%%%%%%%%%%%%%%%%%%%%%%%%

%%%%%%%%%%%%%%%%%%%%%%%%%%%%%%%%%%%%%%%%%%%%%%%%%%%%%%%%%%%
\begin{figure*}
\begin{minipage}{0.25\textwidth}
\includegraphics[width=\textwidth]{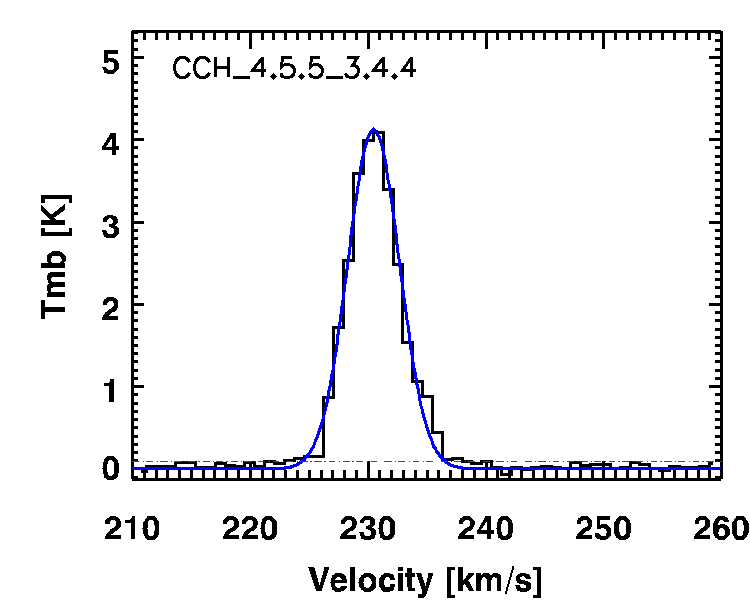}
\end{minipage}
\hskip -0.34 cm
\begin{minipage}{0.25\textwidth}
\includegraphics[width=\textwidth]{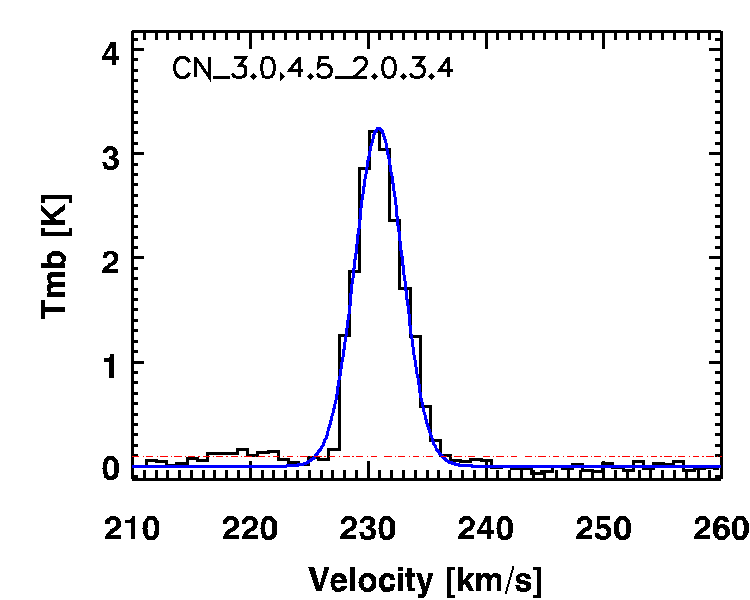}
\end{minipage}
\hskip -0.33 cm
\begin{minipage}{0.25\textwidth}
\includegraphics[width=\textwidth]{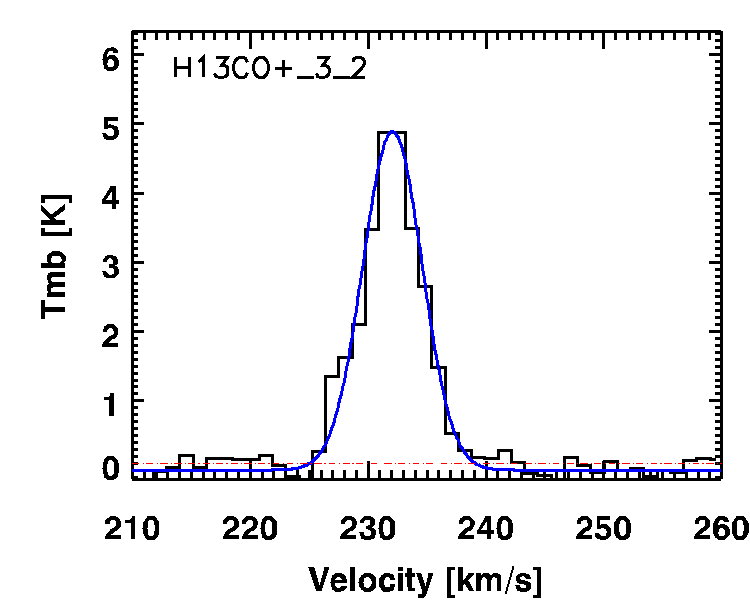}
\end{minipage}
%\hskip -1.0 cm
\begin{minipage}{0.25\textwidth}
\includegraphics[width=\textwidth]{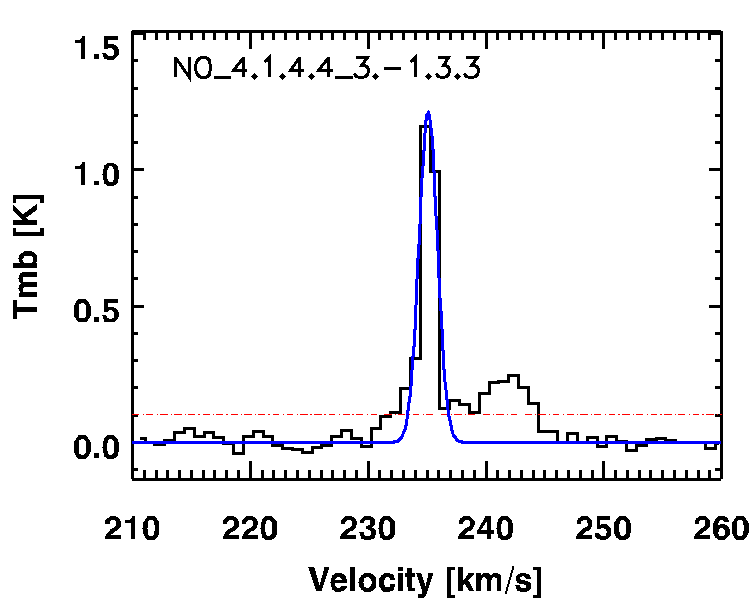}
\end{minipage}
\hskip -0.33 cm
\begin{minipage}{0.25\textwidth}
\includegraphics[width=\textwidth]{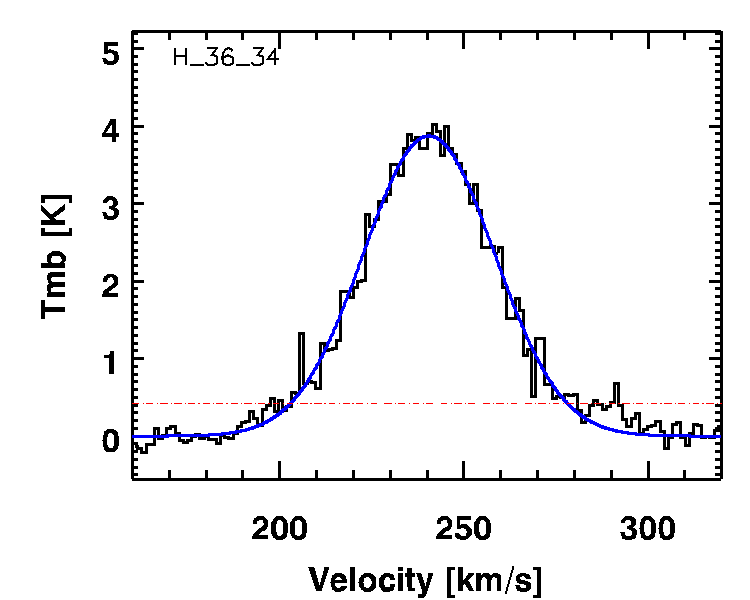}
\end{minipage}
%\hskip -1.0 cm
\begin{minipage}{0.25\textwidth}
\includegraphics[width=\textwidth]{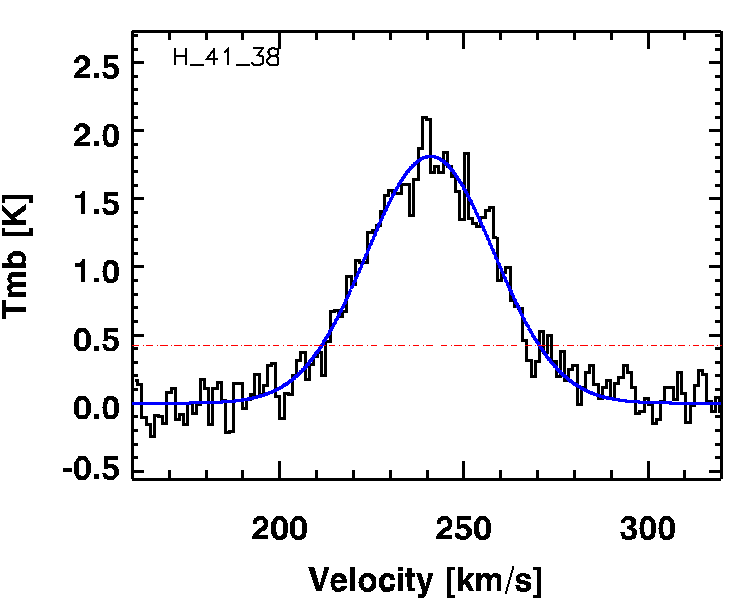}
\end{minipage}
\caption{Spectra of emission lines of CN, CCH, H$^{13}$CO$^+$, NO and hydrogen recombination lines are detected around the  SSC continuum peak/SSC candidate.}
\label{Fig:specac1}
\end{figure*}
%%%%%%%%%%%%%%%%%%%%%%%%%%%%%%%%%%%%%%%%%%%%%%%%%%%%%%%%%%%
\clearpage
\restartappendixnumbering
\section{line image \label{Fig:moment}}
\begin{figure*} 
\hspace*{0.3cm}
%\begin{minipage}{0.32\textwidth}
 %\begin{tikzpicture}
%\node[anchor=south west, inner sep=0] (image1) at (0,0)
%{\includegraphics[width=\textwidth]{CH3OH_stack6.pdf}};

%\begin{scope}[x={(image1.south east)},y={(image1.north west)}]
     % \node[black, font=\scriptsize\bfseries] at (0.55, 1.025) {CH$_3$OH (34 K$<$E$_u$$>$61 K)};
%\end{scope}
%\end{tikzpicture}
%\end{minipage}
%\hskip -1.0 cm
%\begin{minipage}{0.32\textwidth}
% \begin{tikzpicture}
%\node[anchor=south west, inner sep=0] (image1) at (0,0)
%{\includegraphics[width=\textwidth]{CH3OH_stack_16.pdf}};

%\begin{scope}[x={(image1.south east)},y={(image1.north west)}]
     % \node[black, font=\scriptsize\bfseries] at (0.55, 1.025) {CH$_3$OH (E$_u$ = 16 K)};
%\end{scope}
%\end{tikzpicture}
%\end{minipage}
%\begin{minipage}{0.32\textwidth}
% \begin{tikzpicture}
%\node[anchor=south west, inner sep=0] (image1) at (0,0)
%{\includegraphics[width=\textwidth]{CH3OH_stack_7.pdf}};
% Draw red lines to zoom-in areas
%\begin{scope}[x={(image1.south east)},y={(image1.north west)}]
      %\node[black, font=\scriptsize\bfseries] at (0.55, 1.025) {CH$_3$OH (64 K$<$E$_u$$>$91 K)};
%\end{scope}
%\end{tikzpicture}
%\end{minipage}
%\hspace*{0.3cm}
\begin{minipage}{0.33\textwidth}
 \begin{tikzpicture}
\node[anchor=south west, inner sep=0] (image1) at (0,0)
{\includegraphics[width=\textwidth]{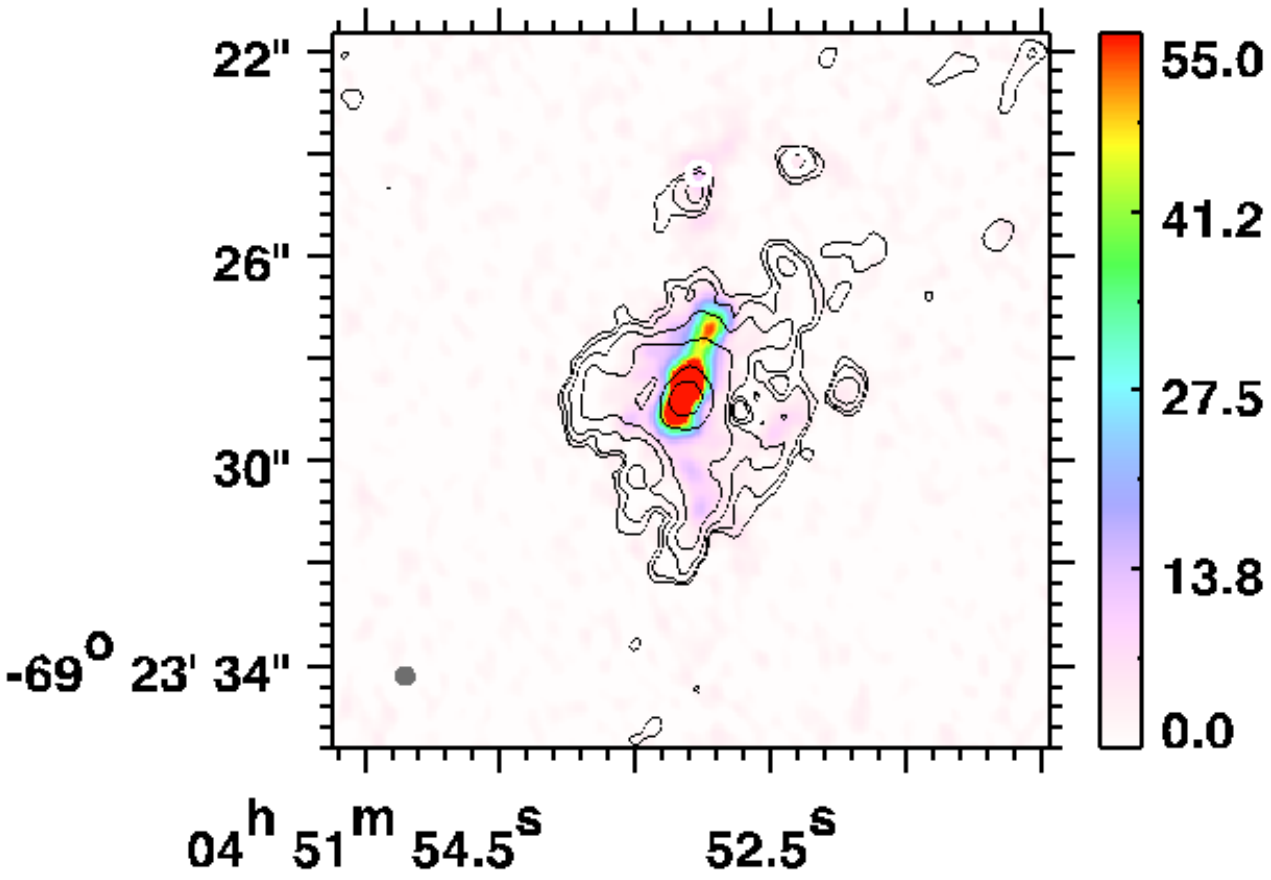}};
% Draw red lines to zoom-in areas
\begin{scope}[x={(image1.south east)},y={(image1.north west)}]
      \node[black, font=\scriptsize\bfseries] at (0.55, 1.025) {SO (E$_u$ = 56 K)};
\end{scope}
\end{tikzpicture}
\end{minipage}
\begin{minipage}{0.33\textwidth}
 \begin{tikzpicture}
\node[anchor=south west, inner sep=0] (image1) at (0,0)
{\includegraphics[width=\textwidth]{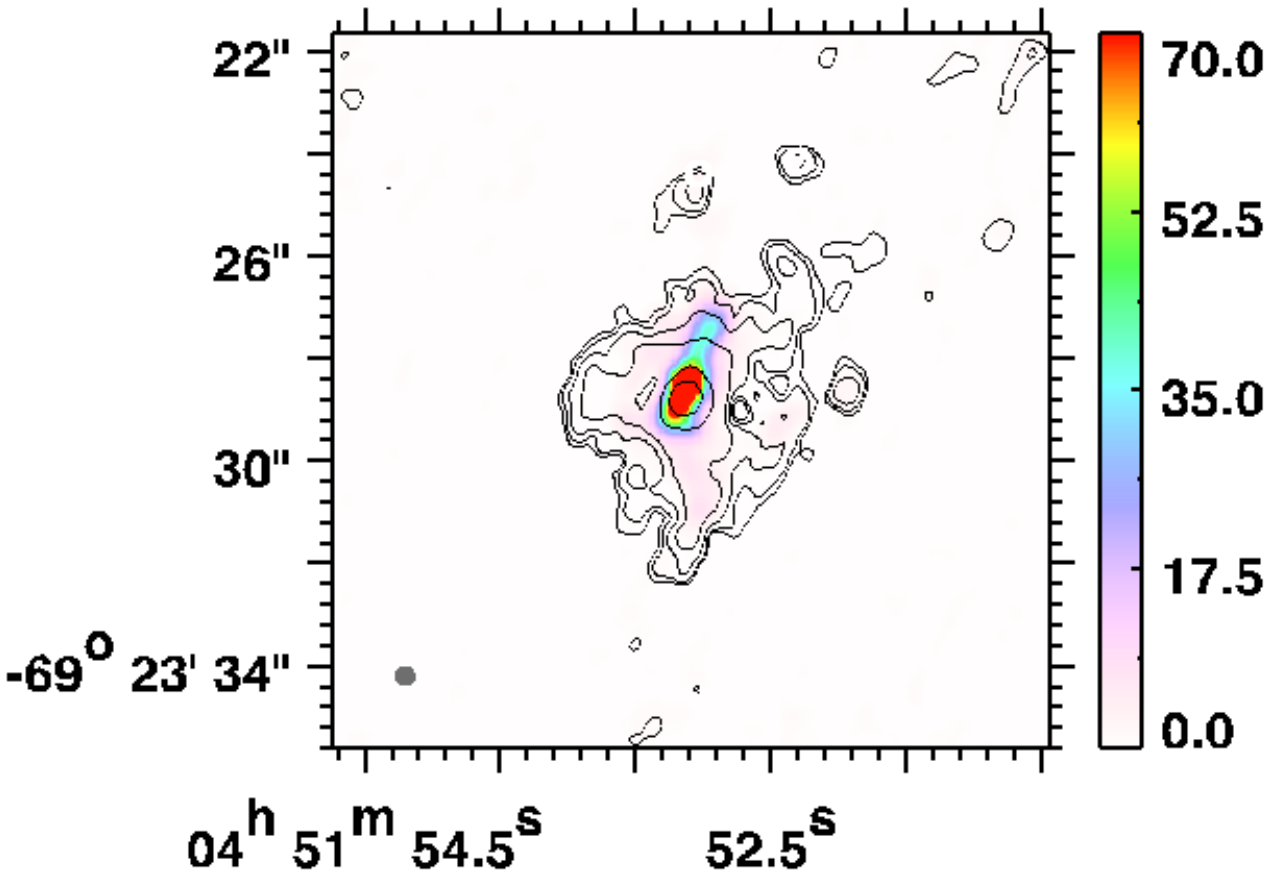}};
% Draw red lines to zoom-in areas
\begin{scope}[x={(image1.south east)},y={(image1.north west)}]
      \node[black, font=\scriptsize\bfseries] at (0.55, 1.025) {SO (E$_u$ = 81 K)};
\end{scope}
\end{tikzpicture}
\end{minipage}
\begin{minipage}{0.33\textwidth}
 \begin{tikzpicture}
\node[anchor=south west, inner sep=0] (image1) at (0,0)
{\includegraphics[width=\textwidth]{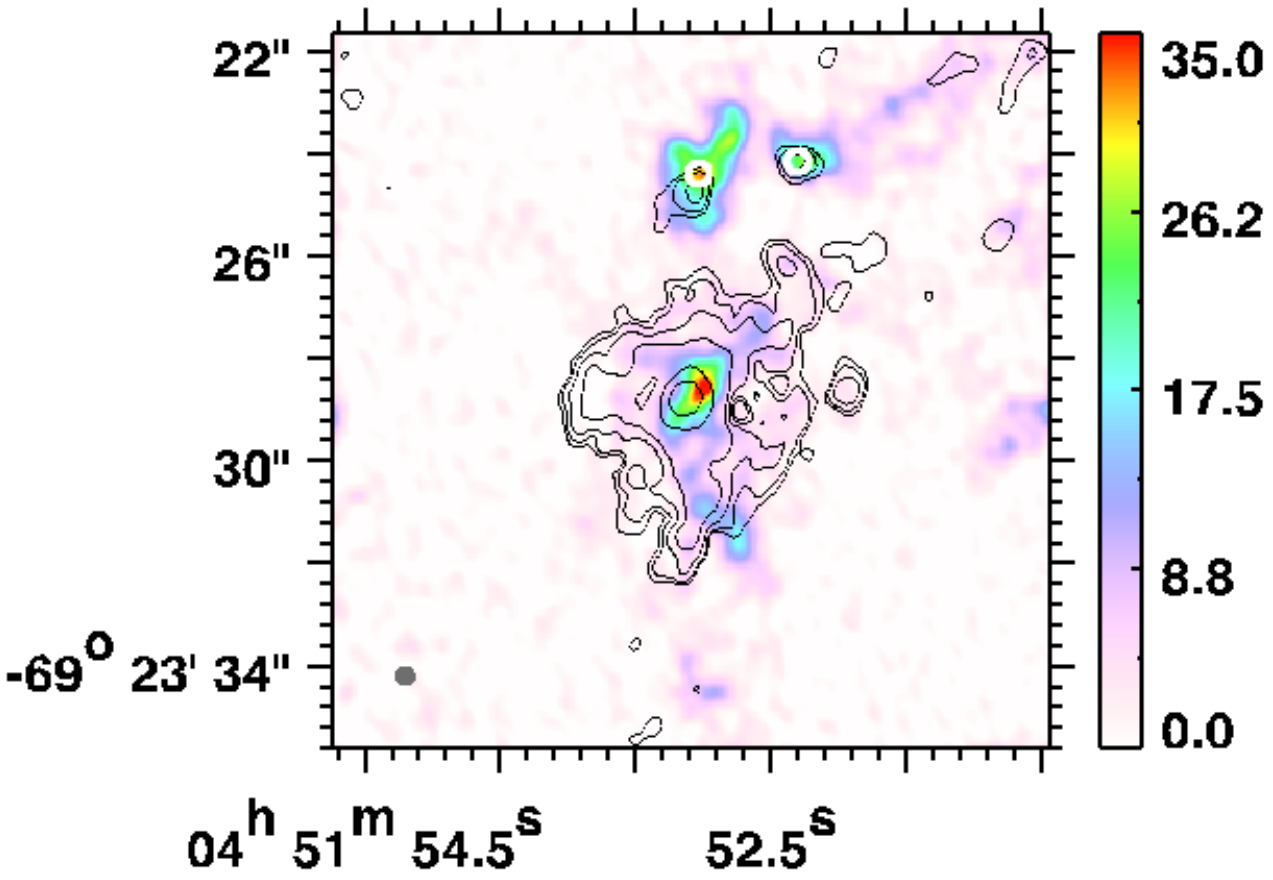}};
% Draw red lines to zoom-in areas
\begin{scope}[x={(image1.south east)},y={(image1.north west)}]
\node[black, font=\scriptsize\bfseries] at (0.55, 1.025) {CS (E$_u$ = 35) K};
\end{scope}
\end{tikzpicture}
\end{minipage}
\hspace*{0.3cm}
\begin{minipage}{0.33\textwidth}
 \begin{tikzpicture}
\node[anchor=south west, inner sep=0] (image1) at (0,0)
{\includegraphics[width=\textwidth]{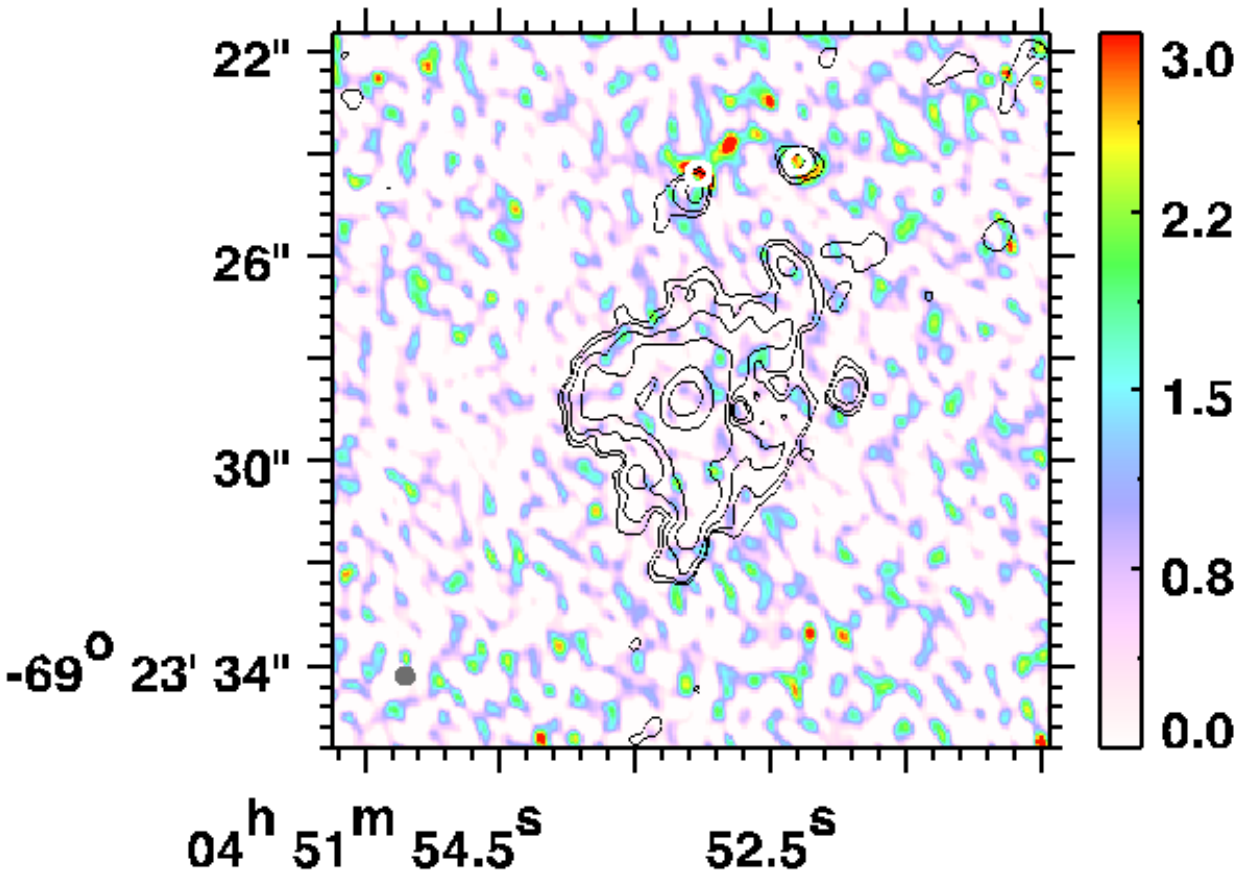}};
% Draw red lines to zoom-in areas
\begin{scope}[x={(image1.south east)},y={(image1.north west)}]
      \node[black, font=\scriptsize\bfseries] at (0.55, 1.025) {H$_2$CS (E$_u$ = 60 K)};
\end{scope}
\end{tikzpicture}
\end{minipage}
\begin{minipage}{0.33\textwidth}
 \begin{tikzpicture}
\node[anchor=south west, inner sep=0] (image1) at (0,0) 
{\includegraphics[width=\textwidth]{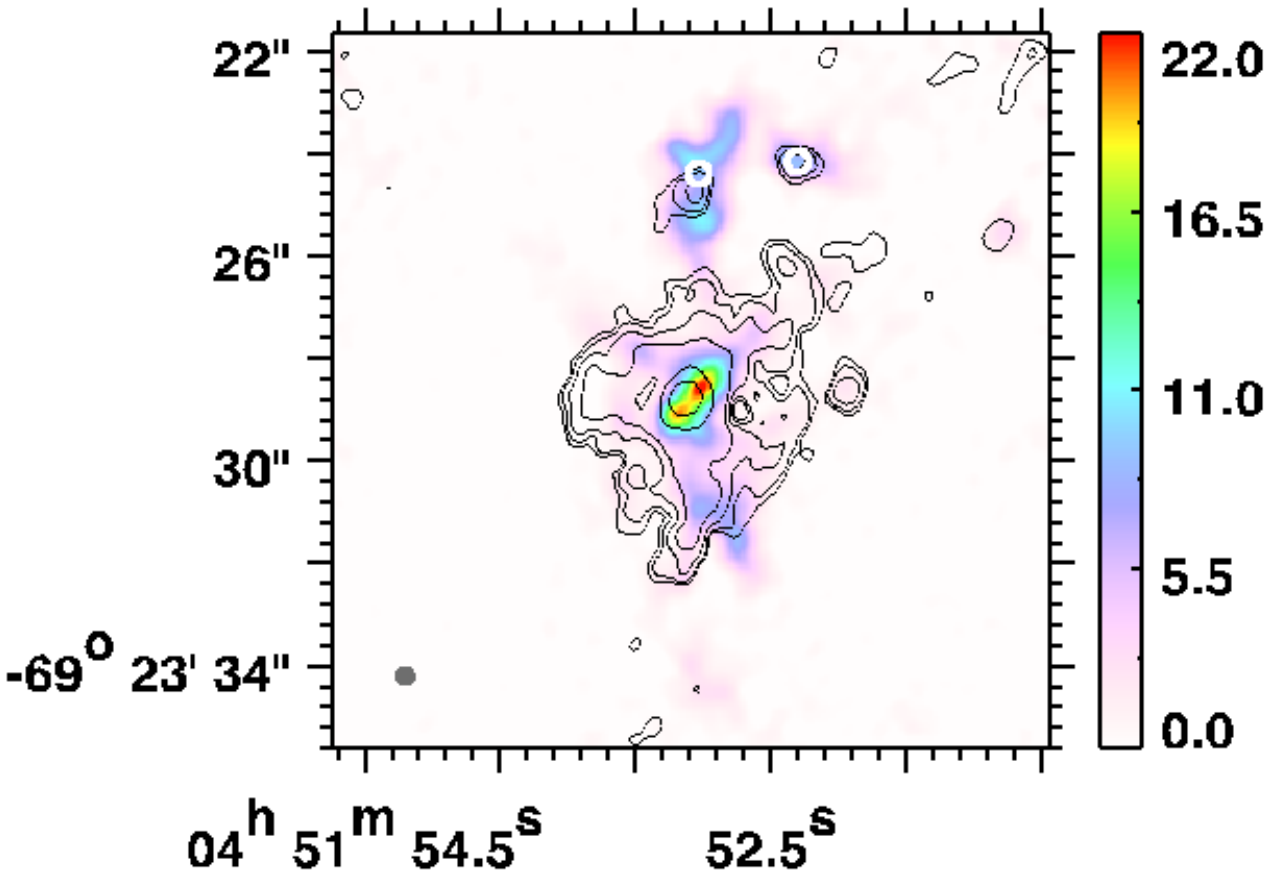}};
% Draw red lines to zoom-in areas
\begin{scope}[x={(image1.south east)},y={(image1.north west)}]
      \node[black, font=\scriptsize\bfseries] at (0.55, 1.025) {H$_2$CO (E$_u$ = 62 K)};
\end{scope}
\end{tikzpicture}
\end{minipage}
\begin{minipage}{0.33\textwidth}
 \begin{tikzpicture}
\node[anchor=south west, inner sep=0] (image1) at (0,0) 
{\includegraphics[width=\textwidth]{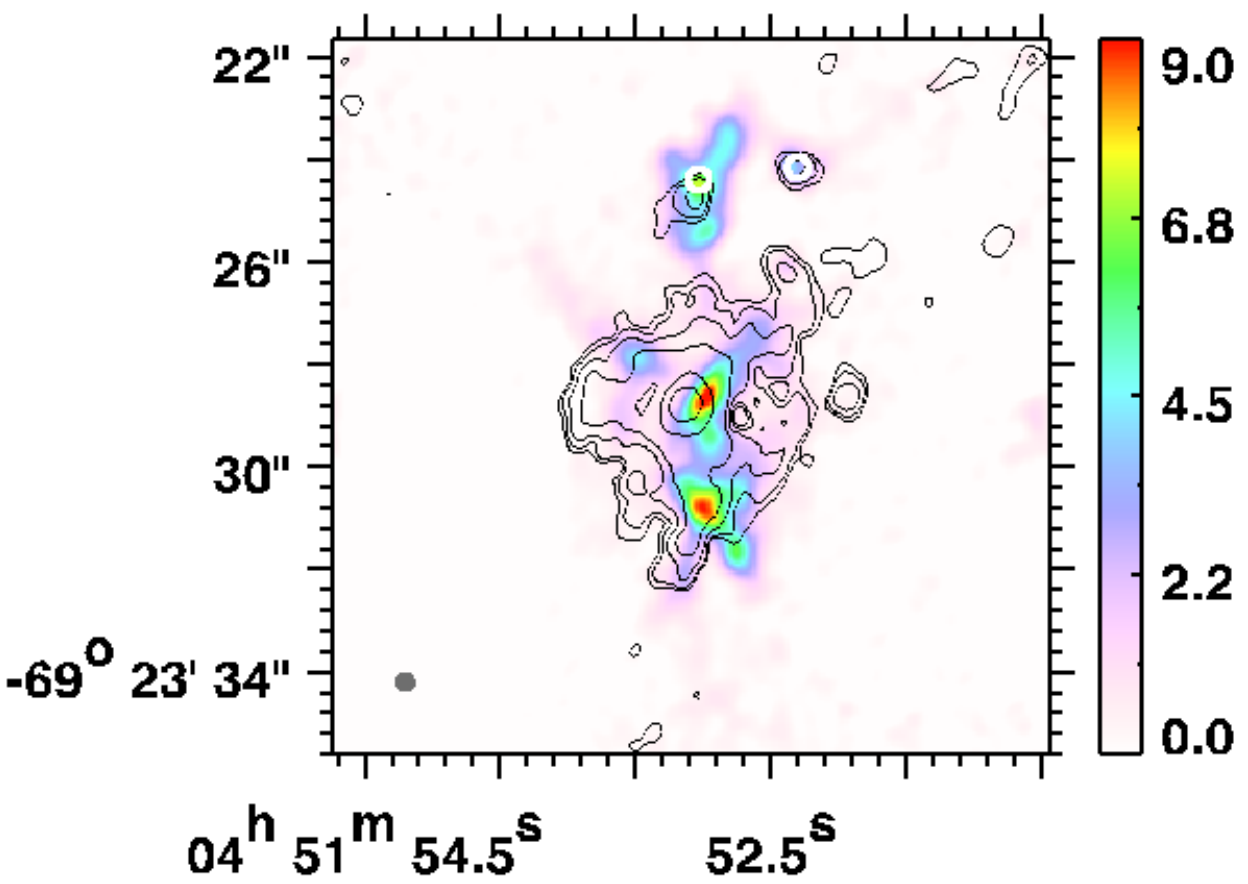}};
% Draw red lines to zoom-in areas
\begin{scope}[x={(image1.south east)},y={(image1.north west)}]
      \node[black, font=\scriptsize\bfseries] at (0.55, 1.025) {CN (E$_u$ = 32 K)};
\end{scope}
\end{tikzpicture}
\end{minipage}
\hspace*{0.3cm}
%\begin{minipage}{0.30cm}
%\vspace*{-0.8cm}
%\small
%\rotatebox{90}{\textcolor{black}{Declination (J2000)}}
%\end{minipage}
\begin{minipage}{0.33\textwidth}
 \begin{tikzpicture}
\node[anchor=south west, inner sep=0] (image1) at (0,0) 
 %\hspace{-5.0cm}
{\includegraphics[width=\textwidth]{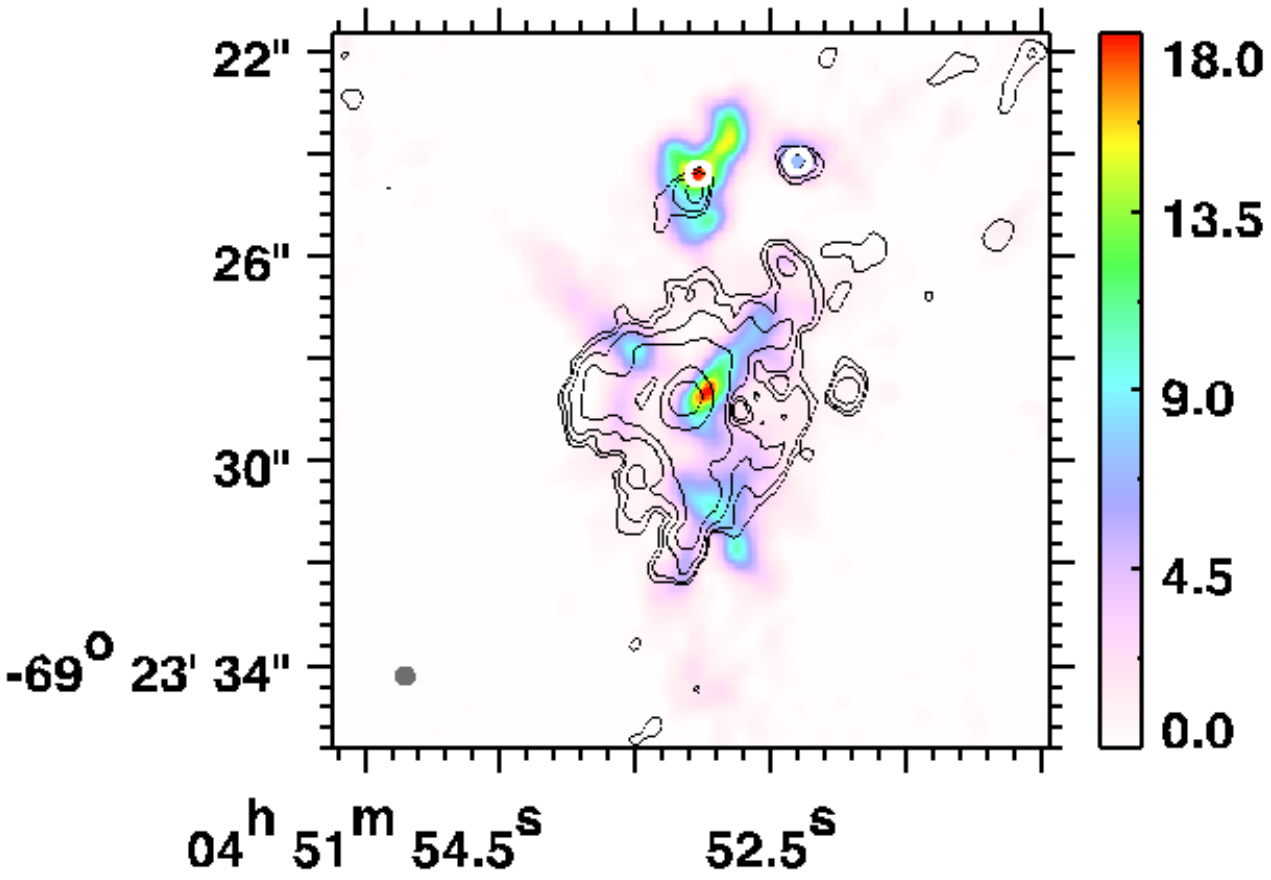}};
% Draw red lines to zoom-in areas
\begin{scope}[x={(image1.south east)},y={(image1.north west)}]
      \node[black, font=\scriptsize\bfseries] at (0.55, 1.025) {CCH (E$_u$ = 42) K};
\end{scope}
\end{tikzpicture}
%\vspace*{0.1cm}\hspace*{1.8cm}\small \textcolor{black}{Right Ascension (J2000)}
\end{minipage}
%\hskip -0.2 cm
\begin{minipage}{0.33\textwidth}
 \begin{tikzpicture}
\node[anchor=south west, inner sep=0] (image1) at (0,0) 
{\includegraphics[width=\textwidth]{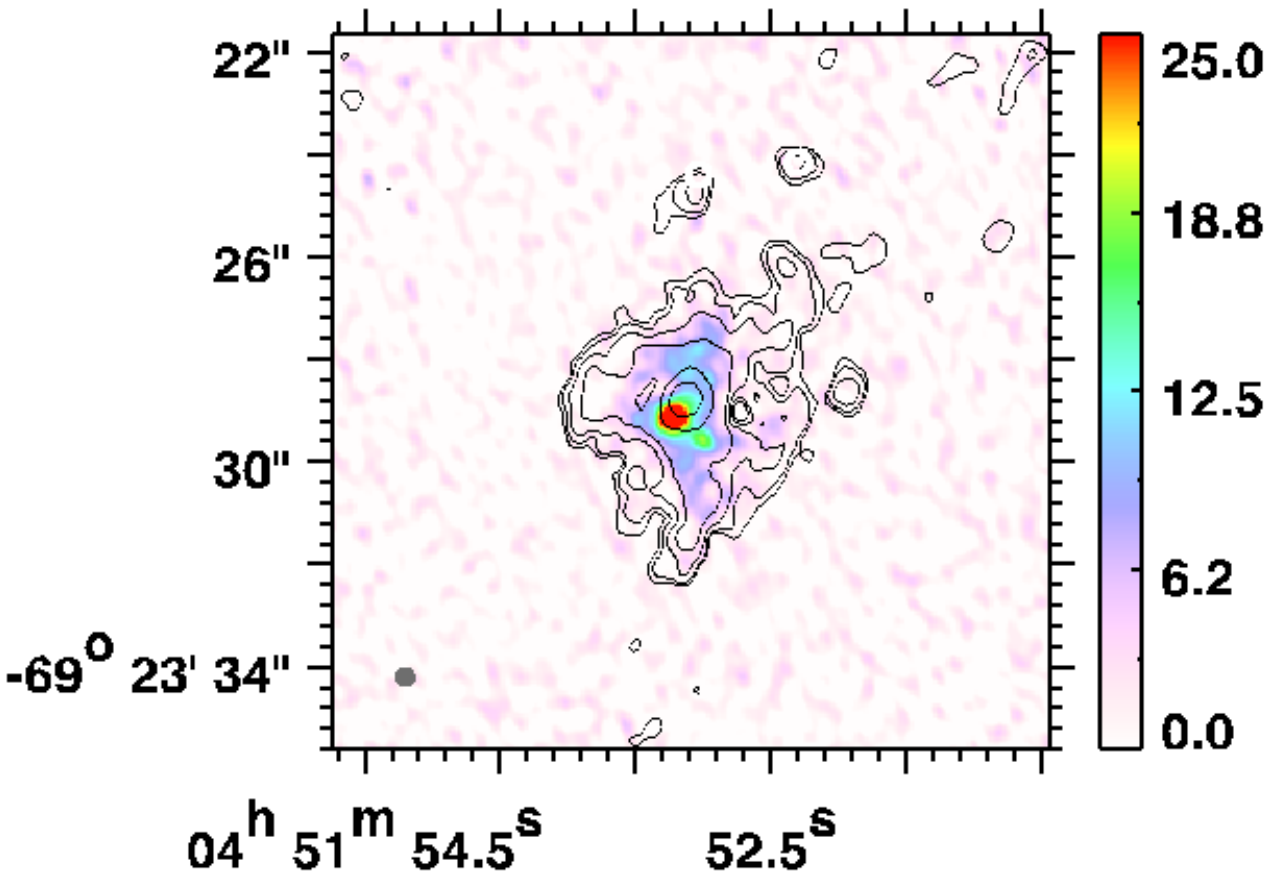}};
% Draw red lines to zoom-in areas
\begin{scope}[x={(image1.south east)},y={(image1.north west)}]
      \node[black, font=\scriptsize\bfseries] at (0.55, 1.025) {H$^{13}$CO$^+$ (E$_u$ = 25 K)};
\end{scope}
\end{tikzpicture}
\end{minipage}
%\hskip -0.2 cm
\begin{minipage}{0.33\textwidth}
 \begin{tikzpicture}
\node[anchor=south west, inner sep=0] (image1) at (0,0) 
{\includegraphics[width=\textwidth]{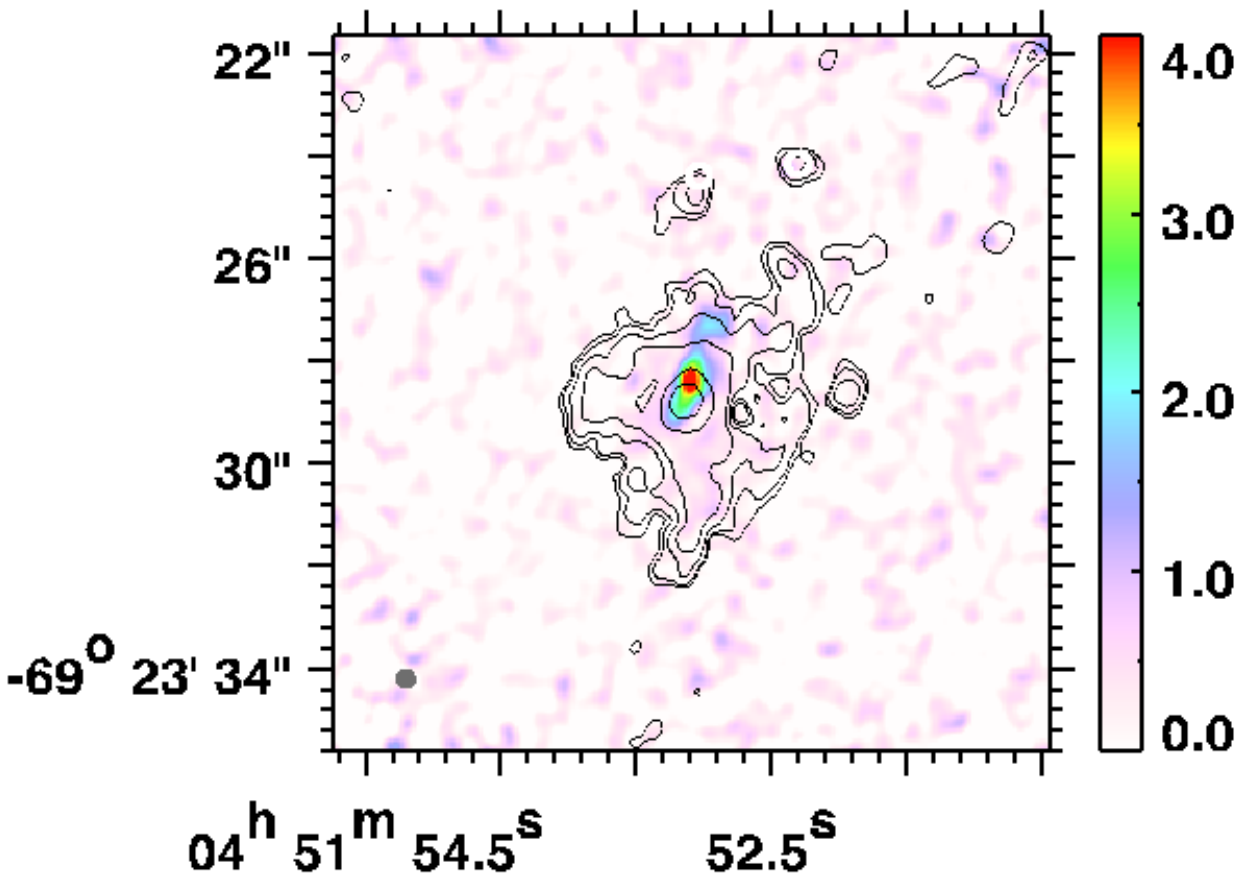}};
% Draw red lines to zoom-in areas
\begin{scope}[x={(image1.south east)},y={(image1.north west)}]
      \node[black, font=\scriptsize\bfseries] at (0.55, 1.025) {NO (E$_u$ = 36 K)};
\end{scope}
\end{tikzpicture}
\end{minipage}
%\hskip -0.2 cm
%\begin{minipage}{0.30cm}
%\hspace*{-0.3cm}
%\vspace*{-0.8cm}
%\small
%\rotatebox{270}{\textcolor{black}{K km/s}}
%\end{minipage}
\begin{minipage}{0.30cm}
\vspace*{-0.8cm}
\small
\rotatebox{90}{\textcolor{black}{Declination (J2000)}}
\end{minipage}
\begin{minipage}{0.33\textwidth}
 \begin{tikzpicture}
\node[anchor=south west, inner sep=0] (image1) at (0,0) 
 %\hspace{-5.0cm}
{\includegraphics[width=\textwidth]{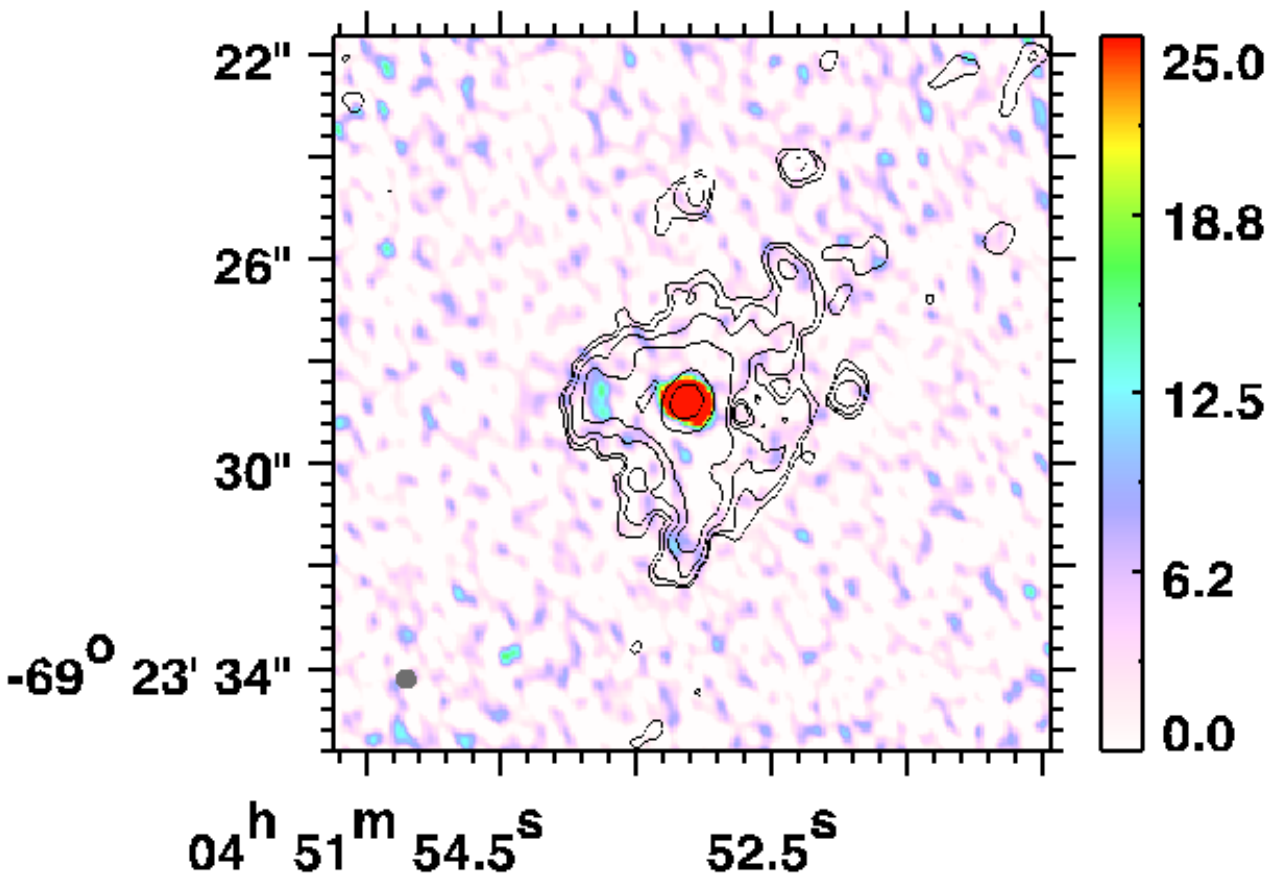}};
% Draw red lines to zoom-in areas
\begin{scope}[x={(image1.south east)},y={(image1.north west)}]
      \node[black, font=\scriptsize\bfseries] at (0.55, 1.025) {H36$\beta$ (260.032 GHz)};
\end{scope}
\end{tikzpicture}
\vspace*{0.1cm}\hspace*{1.8cm}\small \textcolor{black}{Right Ascension (J2000)}
\end{minipage}
\begin{minipage}{0.33\textwidth}
 \begin{tikzpicture}
\node[anchor=south west, inner sep=0] (image1) at (0,0) 
{\includegraphics[width=\textwidth]{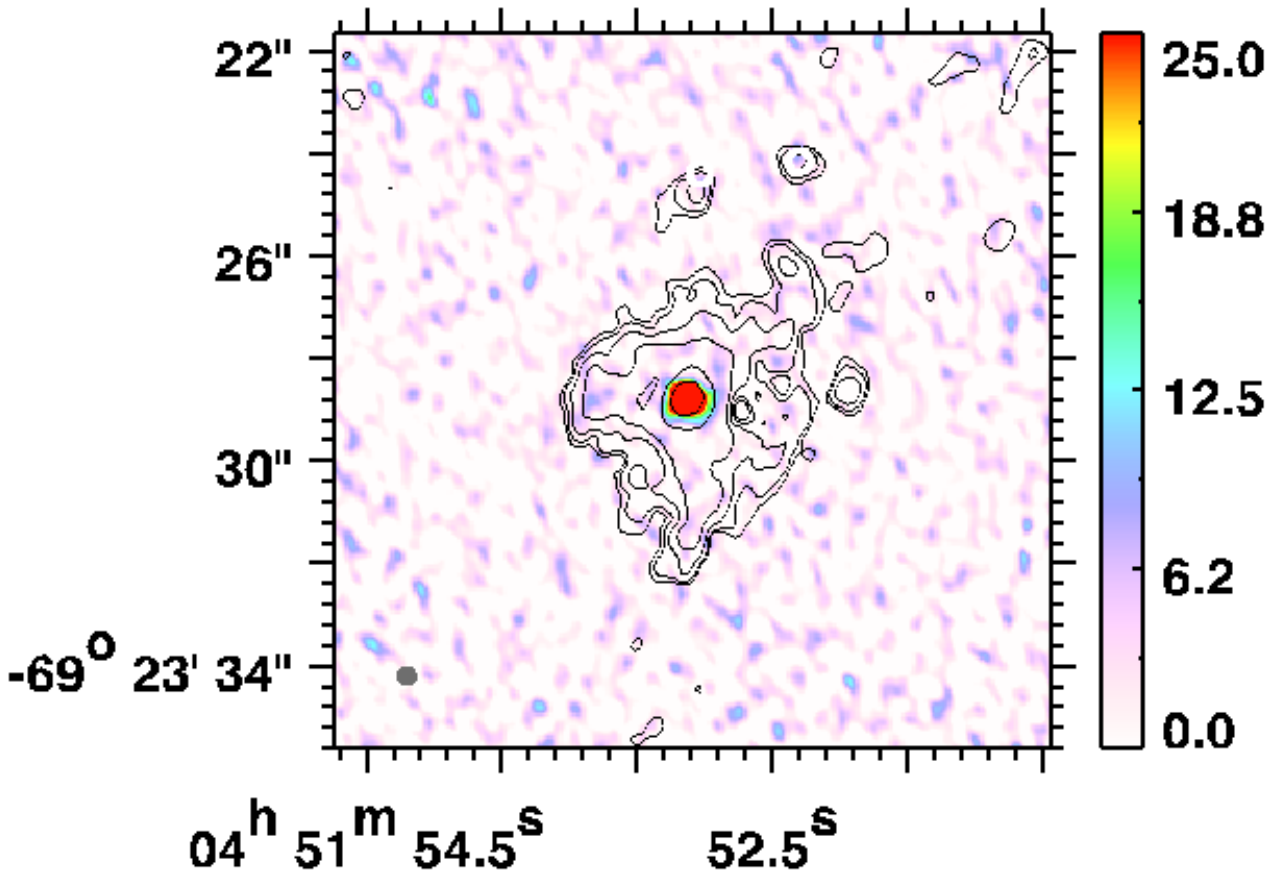}};
% Draw red lines to zoom-in areas
\begin{scope}[x={(image1.south east)},y={(image1.north west)}]
      \node[black, font=\scriptsize\bfseries] at (0.55, 1.025) {H41$\gamma$ (257.635 GHz)};
\end{scope}
\end{tikzpicture}
\end{minipage}
\hskip -0.2 cm
\begin{minipage}{0.30cm}
\hspace*{0.05cm}
\vspace*{1.1cm}
\small
\rotatebox{270}{\textcolor{black}{K km/s}}
\end{minipage}
\caption{Integrated intensity distributions of the  H$_2$CS, H$_2$CO, H$^{13}$CO$^+$, SO, NO, CN, CCH, and hydrogen recombination lines. Contours represent the distribution of the 0.87 mm continuum (see the details in Figure \ref{Fig:cont-imag}). Two white open circles represent positions from where spectra are extracted in this work. The synthesized beam size is shown by the gray-filled circle in each panel.}
\label{fig:mom1}
\end{figure*}

\begin{figure*}
\hspace*{0.3cm}
%\begin{minipage}{0.33\textwidth}
% \begin{tikzpicture}
%\node[anchor=south west, inner sep=0] (image1) at (0,0)
%{\includegraphics[width=\textwidth]{N79S-CH3OH_stack_6.pdf}};
% Draw red lines to zoom-in areas
%\begin{scope}[x={(image1.south east)},y={(image1.north west)}]
     % \node[black, font=\scriptsize\bfseries] at (0.55, 1.025) {CH$_3$OH (34 K$<$E$_u$$>$61 K)};
%\end{scope}
%\end{tikzpicture}
%\end{minipage}
%\hskip -1.0 cm
%\begin{minipage}{0.33\textwidth}
 %\begin{tikzpicture}
%\node[anchor=south west, inner sep=0] (image1) at (0,0)
%{\includegraphics[width=\textwidth]{N79S-CH3OH_stack_16.pdf}};
% Draw red lines to zoom-in areas
%\begin{scope}[x={(image1.south east)},y={(image1.north west)}]
     % \node[black, font=\scriptsize\bfseries] at (0.55, 1.025) {CH$_3$OH (E$_u$ = 16 K)};
%\end{scope}
%\end{tikzpicture}
%\end{minipage}
%\begin{minipage}{0.33\textwidth}
% \begin{tikzpicture}
%\node[anchor=south west, inner sep=0] (image1) at (0,0)
%{\includegraphics[width=\textwidth]{N79S-CH3OH_stack_7.pdf}};
% Draw red lines to zoom-in areas
%\begin{scope}[x={(image1.south east)},y={(image1.north west)}]
    %  \node[black, font=\scriptsize\bfseries] at (0.55, 1.025) {CH$_3$OH (64 K$<$E$_u$$>$91 K)};
%\end{scope}
%\end{tikzpicture}
%\end{minipage}
%\hspace*{0.3cm}
\begin{minipage}{0.33\textwidth}
 \begin{tikzpicture}
\node[anchor=south west, inner sep=0] (image1) at (0,0)
{\includegraphics[width=\textwidth]{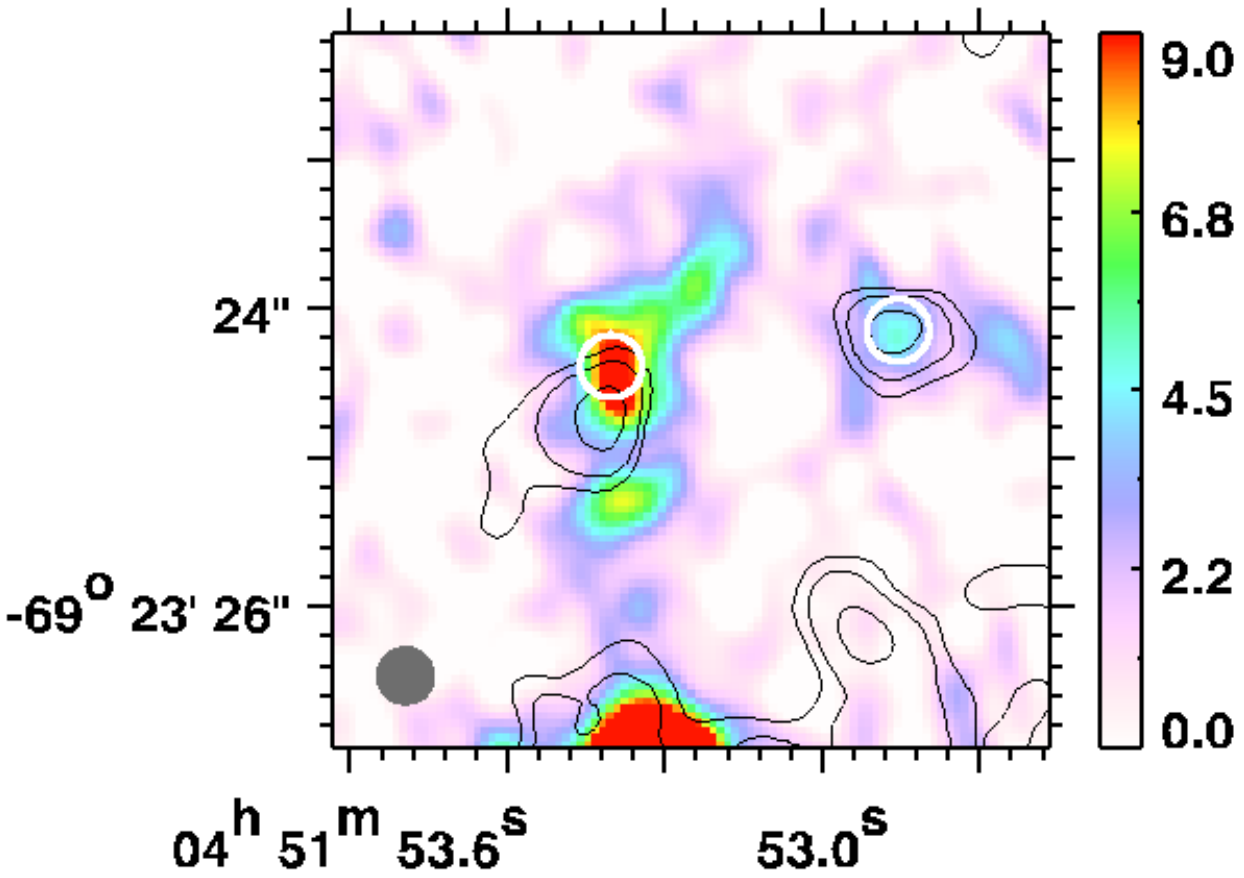}};
% Draw red lines to zoom-in areas
\begin{scope}[x={(image1.south east)},y={(image1.north west)}]
      \node[black, font=\scriptsize\bfseries] at (0.55, 1.025) {SO (E$_u$ = 56 K)};
\end{scope}
\end{tikzpicture}
\end{minipage}
\begin{minipage}{0.33\textwidth}
 \begin{tikzpicture}
\node[anchor=south west, inner sep=0] (image1) at (0,0)
{\includegraphics[width=\textwidth]{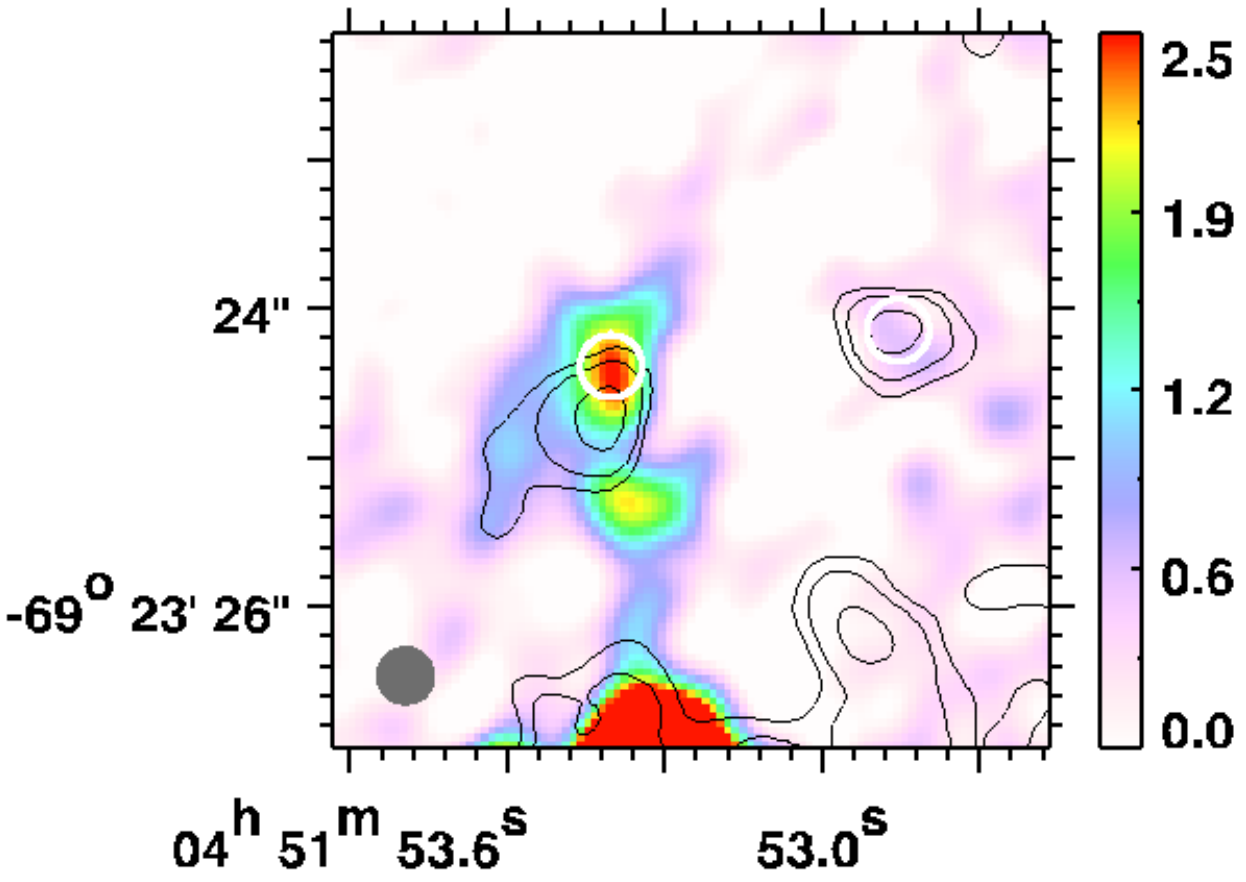}};
% Draw red lines to zoom-in areas
\begin{scope}[x={(image1.south east)},y={(image1.north west)}]
      \node[black, font=\scriptsize\bfseries] at (0.55, 1.025) {SO (E$_u$ = 81 K)};
\end{scope}
\end{tikzpicture}
\end{minipage}
\begin{minipage}{0.33\textwidth}
 \begin{tikzpicture}
\node[anchor=south west, inner sep=0] (image1) at (0,0)
{\includegraphics[width=\textwidth]{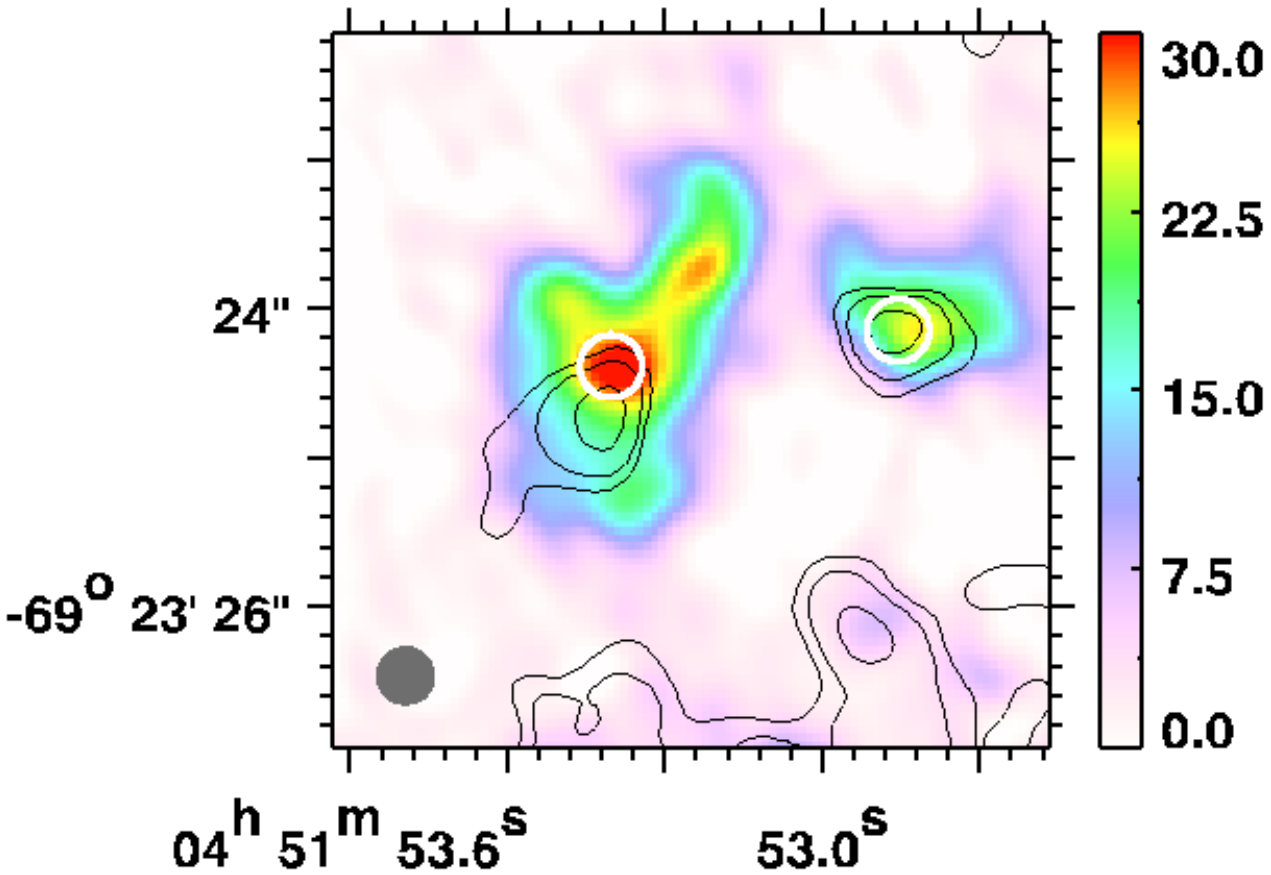}};
% Draw red lines to zoom-in areas
\begin{scope}[x={(image1.south east)},y={(image1.north west)}]
\node[black, font=\scriptsize\bfseries] at (0.55, 1.025) {CS (E$_u$ = 35) K};
\end{scope}
\end{tikzpicture}
\end{minipage}
\hspace*{0.3cm}
\begin{minipage}{0.33\textwidth}
 \begin{tikzpicture}
\node[anchor=south west, inner sep=0] (image1) at (0,0)
{\includegraphics[width=\textwidth]{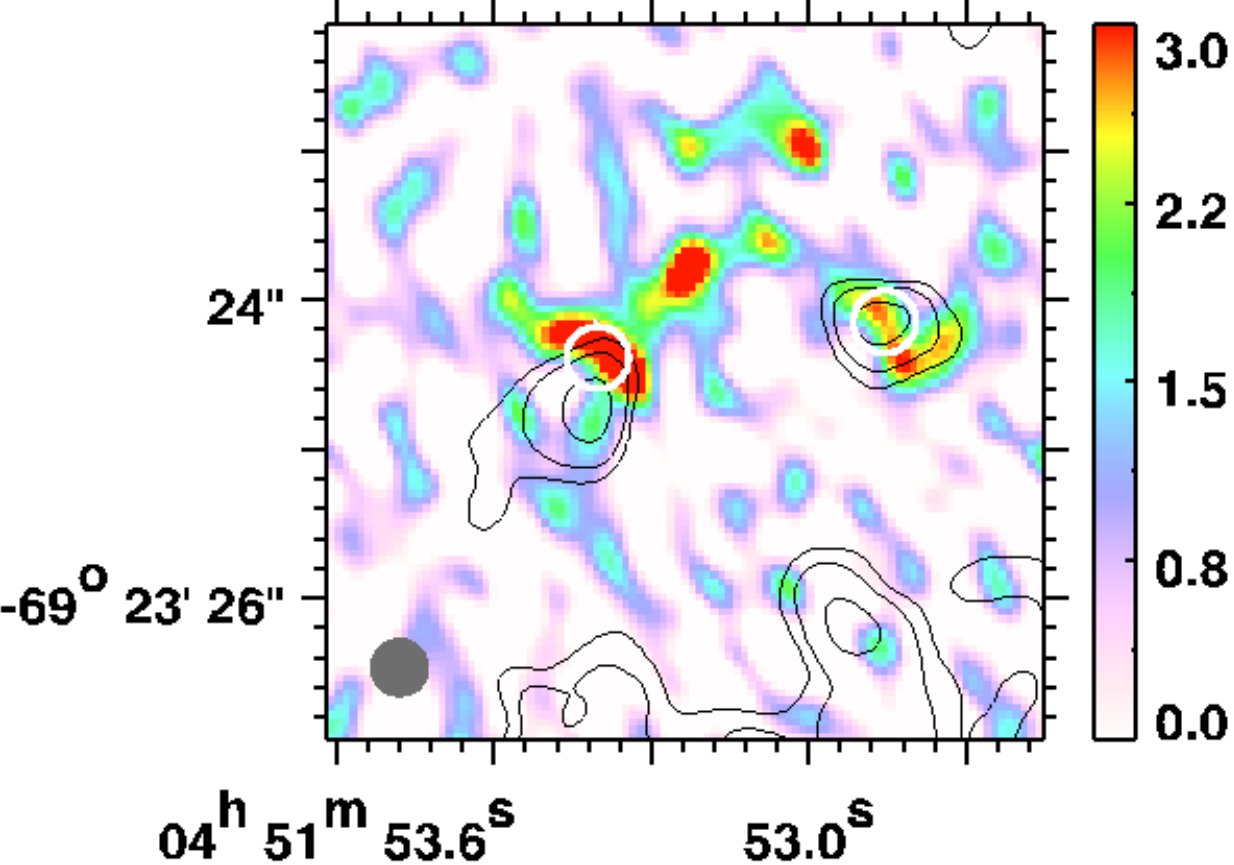}};
% Draw red lines to zoom-in areas
\begin{scope}[x={(image1.south east)},y={(image1.north west)}]
      \node[black, font=\scriptsize\bfseries] at (0.55, 1.025) {H$_2$CS (E$_u$ = 60 K)};
\end{scope}
\end{tikzpicture}
\end{minipage}
\begin{minipage}{0.33\textwidth}
 \begin{tikzpicture}
\node[anchor=south west, inner sep=0] (image1) at (0,0) 
{\includegraphics[width=\textwidth]{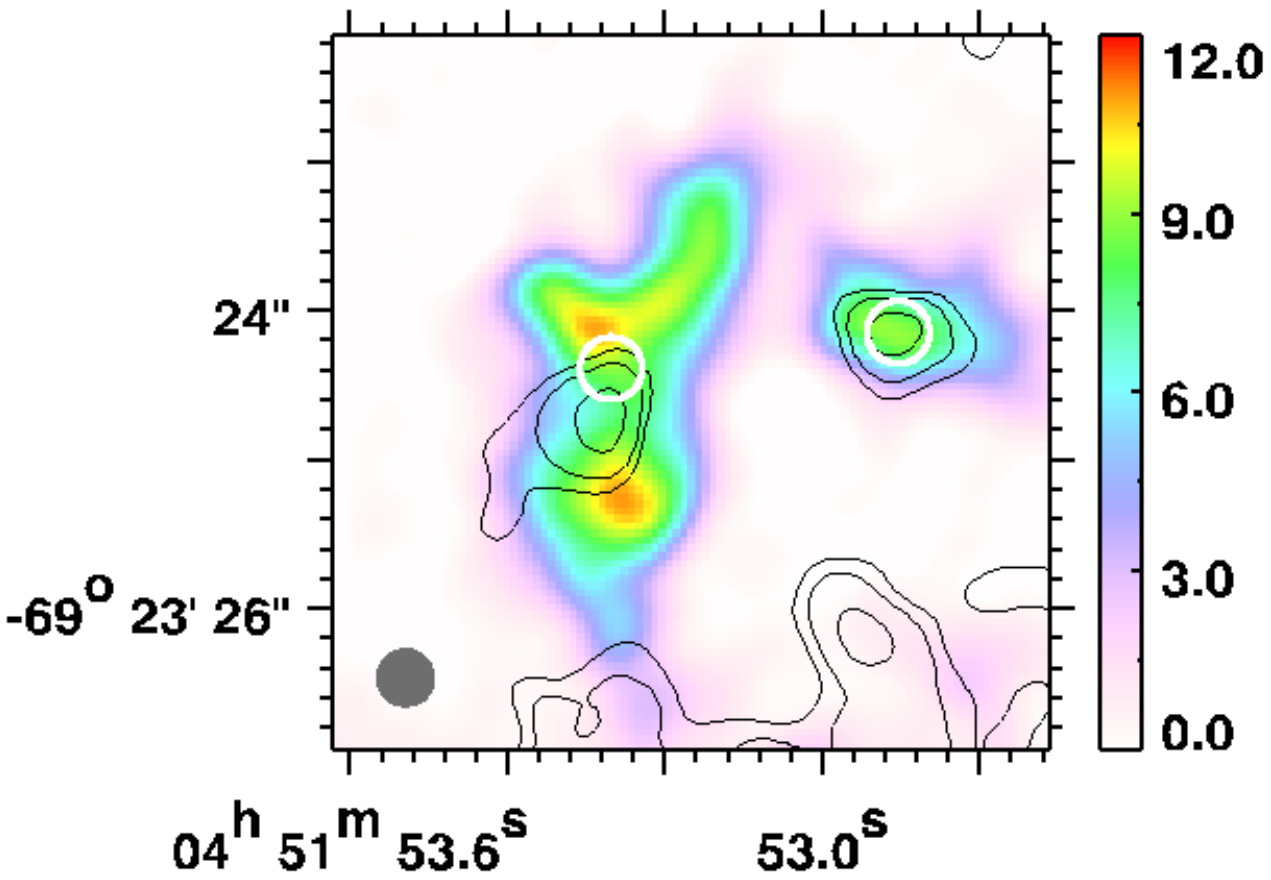}};
% Draw red lines to zoom-in areas
\begin{scope}[x={(image1.south east)},y={(image1.north west)}]
      \node[black, font=\scriptsize\bfseries] at (0.55, 1.025) {H$_2$CO (E$_u$ = 62 K)};
\end{scope}
\end{tikzpicture}
\end{minipage}
\begin{minipage}{0.33\textwidth}
 \begin{tikzpicture}
\node[anchor=south west, inner sep=0] (image1) at (0,0) 
{\includegraphics[width=\textwidth]{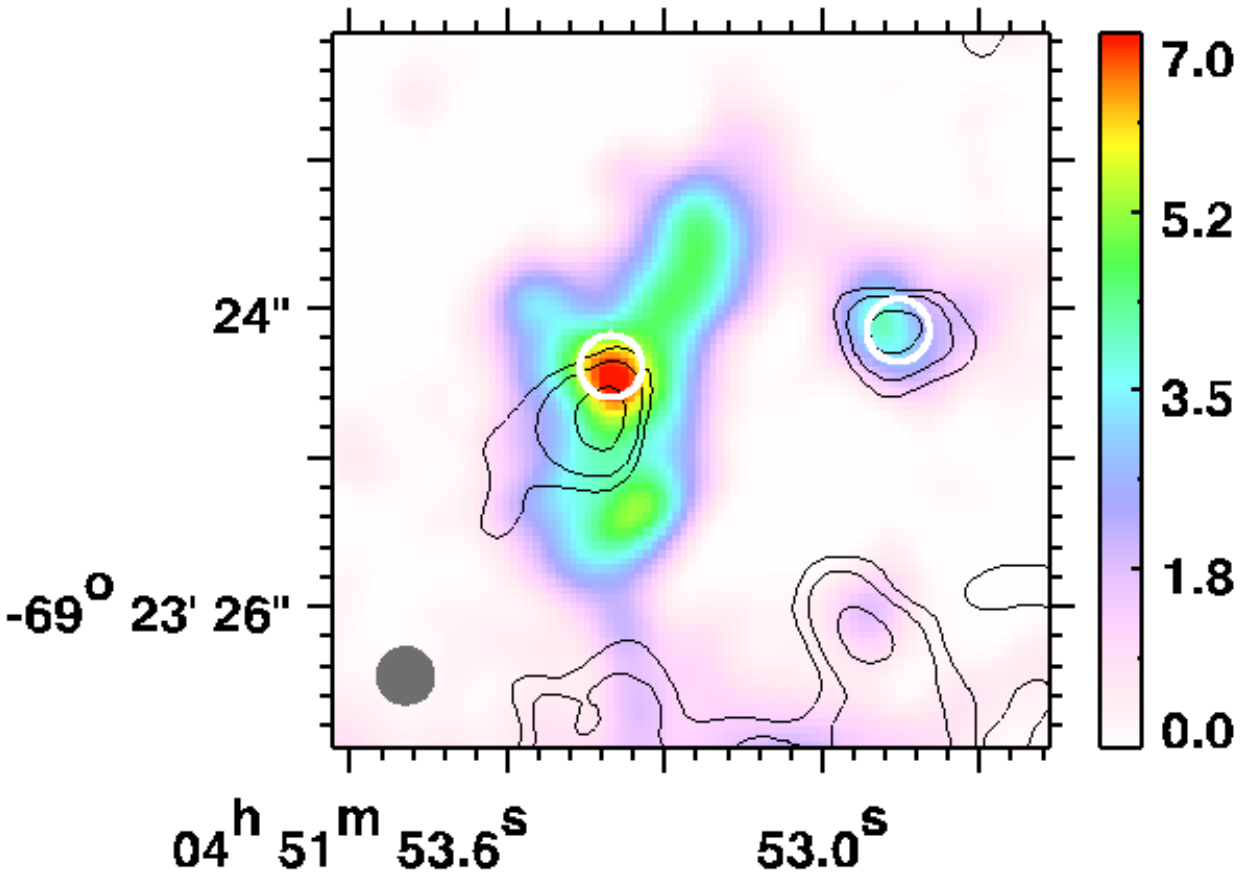}};
% Draw red lines to zoom-in areas
\begin{scope}[x={(image1.south east)},y={(image1.north west)}]
      \node[black, font=\scriptsize\bfseries] at (0.55, 1.025) {CN (E$_u$ = 32 K)};
\end{scope}
\end{tikzpicture}
\end{minipage}
\begin{minipage}{0.30cm}
\vspace*{-0.8cm}
\small
\rotatebox{90}{\textcolor{black}{Declination (J2000)}}
\end{minipage}
\begin{minipage}{0.33\textwidth}
 \begin{tikzpicture}
\node[anchor=south west, inner sep=0] (image1) at (0,0) 
 %\hspace{-5.0cm}
{\includegraphics[width=\textwidth]{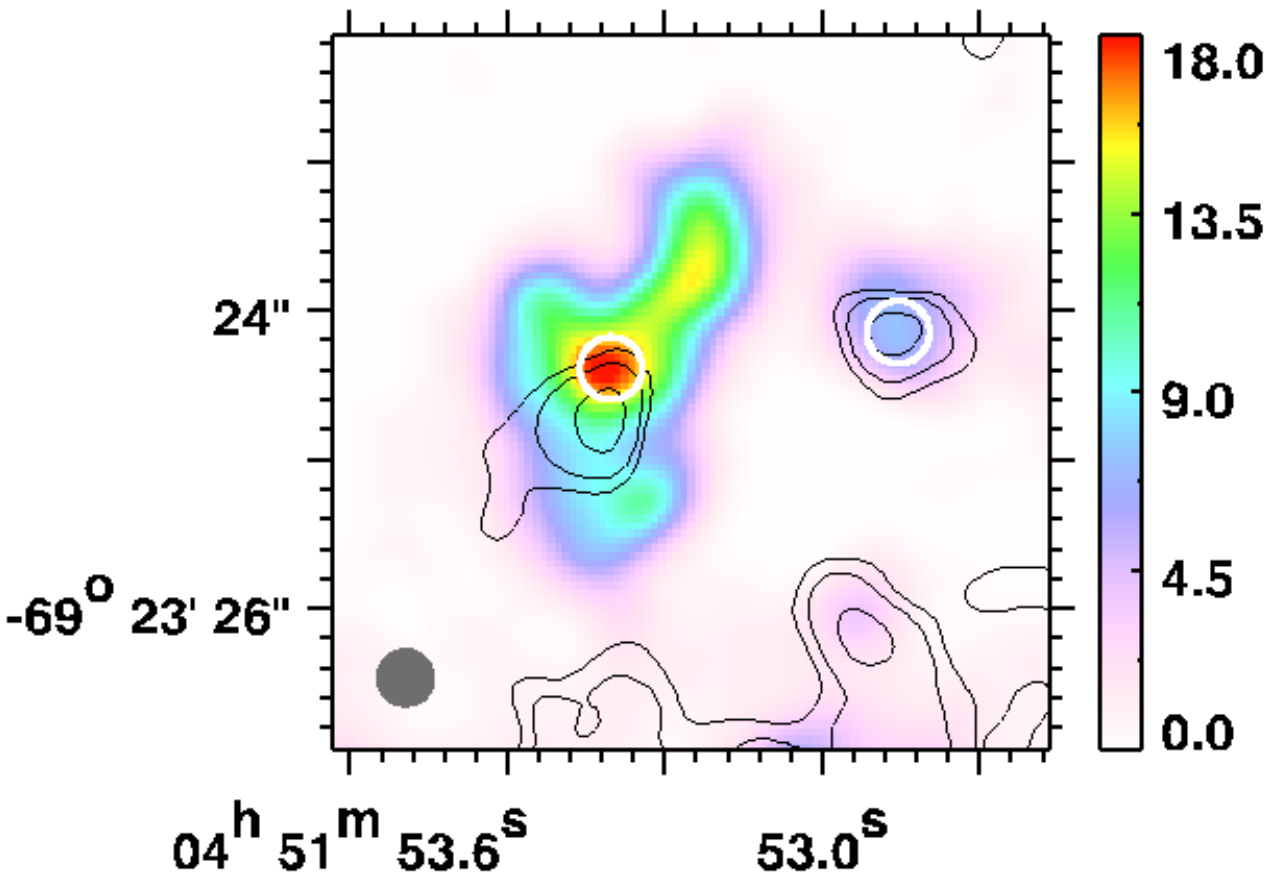}};
% Draw red lines to zoom-in areas
\begin{scope}[x={(image1.south east)},y={(image1.north west)}]
      \node[black, font=\scriptsize\bfseries] at (0.55, 1.025) {CCH (E$_u$ = 42) K};
\end{scope}
\end{tikzpicture}
\vspace*{0.1cm}\hspace*{1.8cm}\small \textcolor{black}{Right Ascension (J2000)}
\end{minipage}
\begin{minipage}{0.30cm}
\vspace*{-0.8cm}
\small
\rotatebox{270}{\textcolor{black}{K km/s}}
\end{minipage}
\caption{ Same as in Figure \ref{fig:mom1}. A zoom-in view of molecular distribution in N79S-1 and N79S-2.}%A zoom-in version of each panel in Figure \ref{fig:mom1} }
\label{fig:mom2}
\end{figure*}

\begin{figure*}
\begin{minipage}{0.30cm}
\vspace*{-0.2cm}
\small
\hspace*{1.7cm}\rotatebox{90}{\textcolor{black}{Declination (J2000)}}
\end{minipage}
\begin{minipage}{0.4\textwidth}
\includegraphics[width=\textwidth]{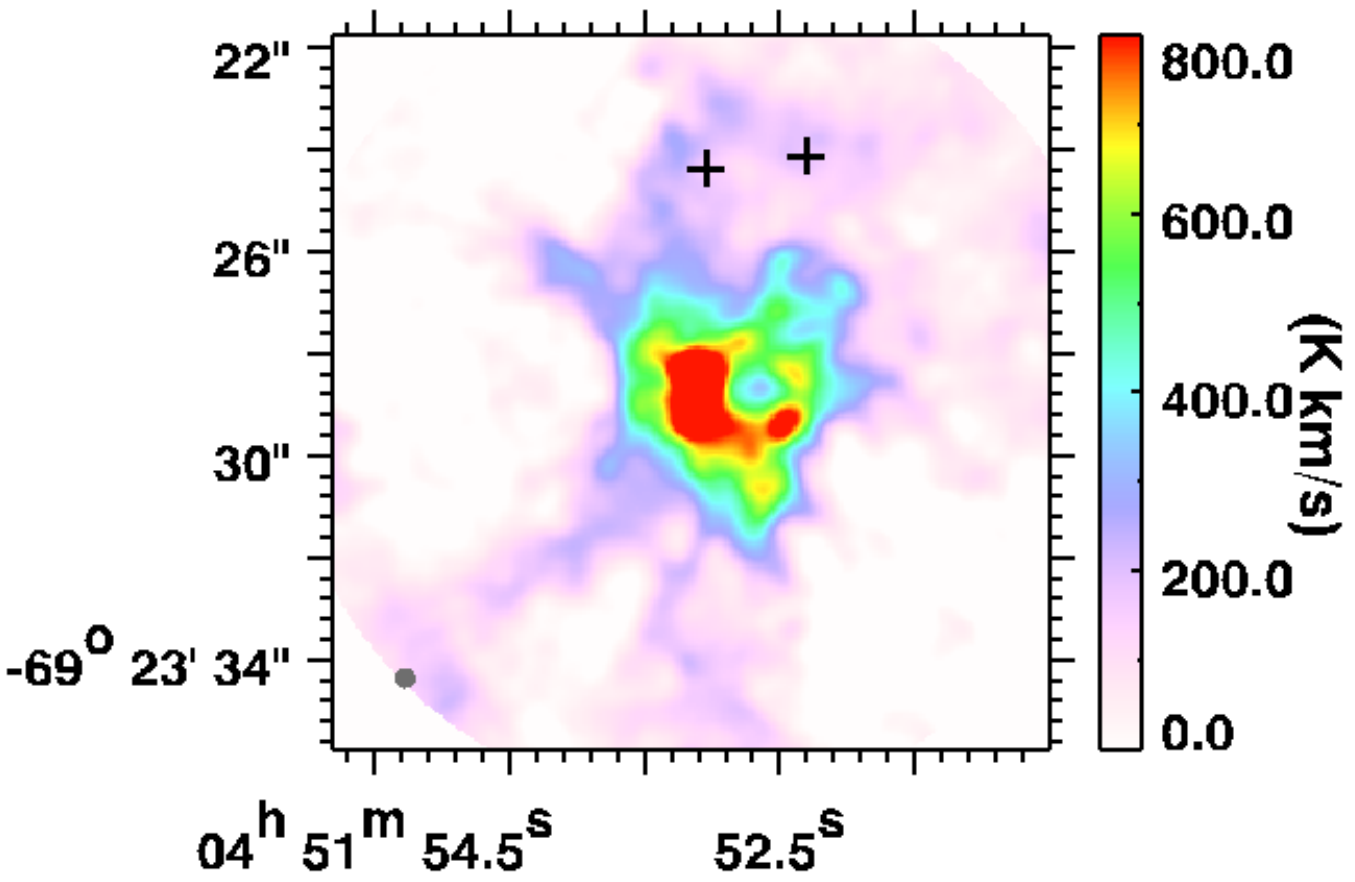}
\vspace*{0.1cm}\hspace*{2.3cm}\small \textcolor{black}{Right Ascension (J2000)}
\end{minipage}
%\hskip -1.0 cm
\begin{minipage}{0.4\textwidth}
\includegraphics[width=\textwidth]{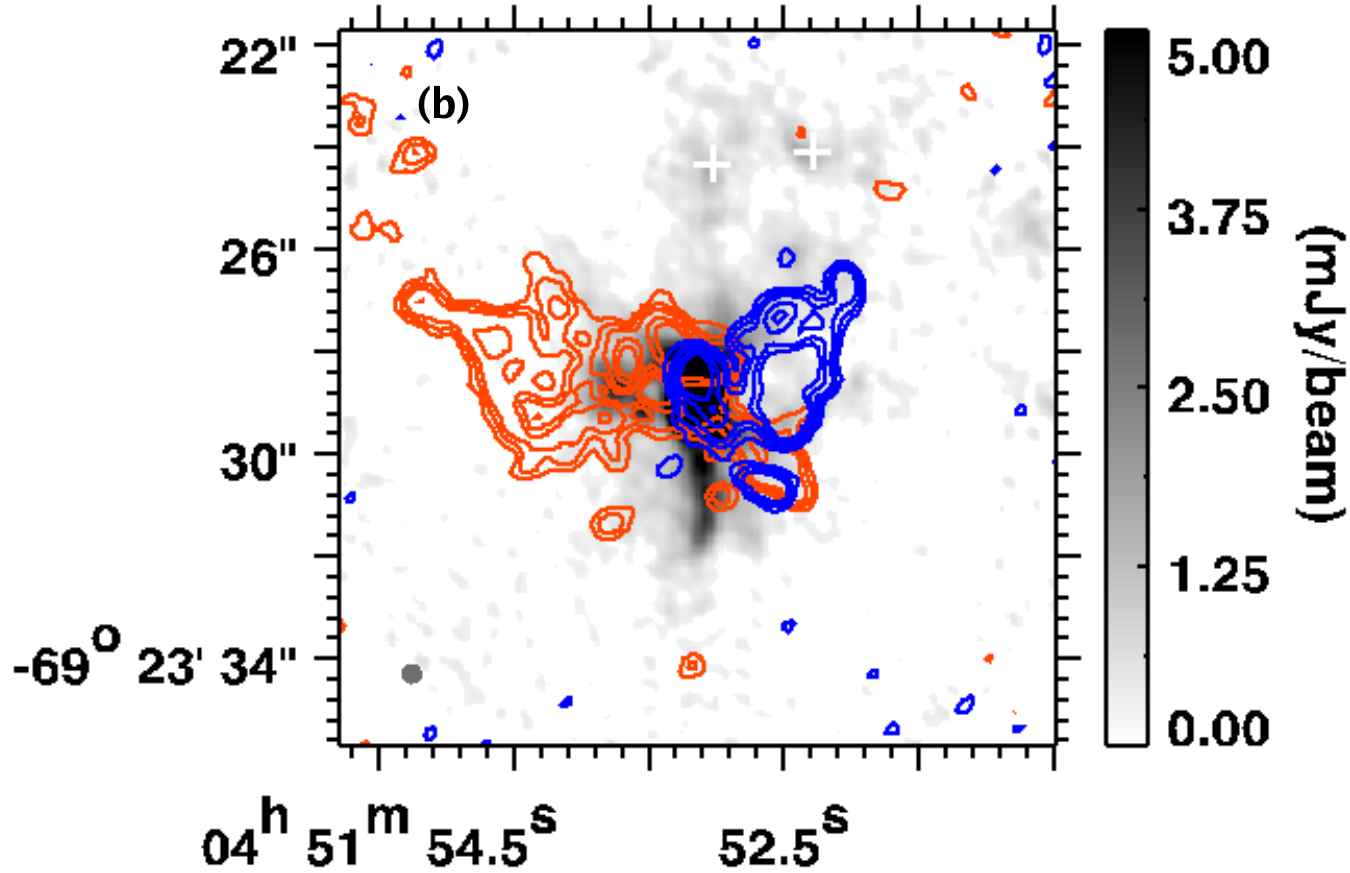}
\end{minipage}
\begin{minipage}{0.33\textwidth}
\includegraphics[width=\textwidth]{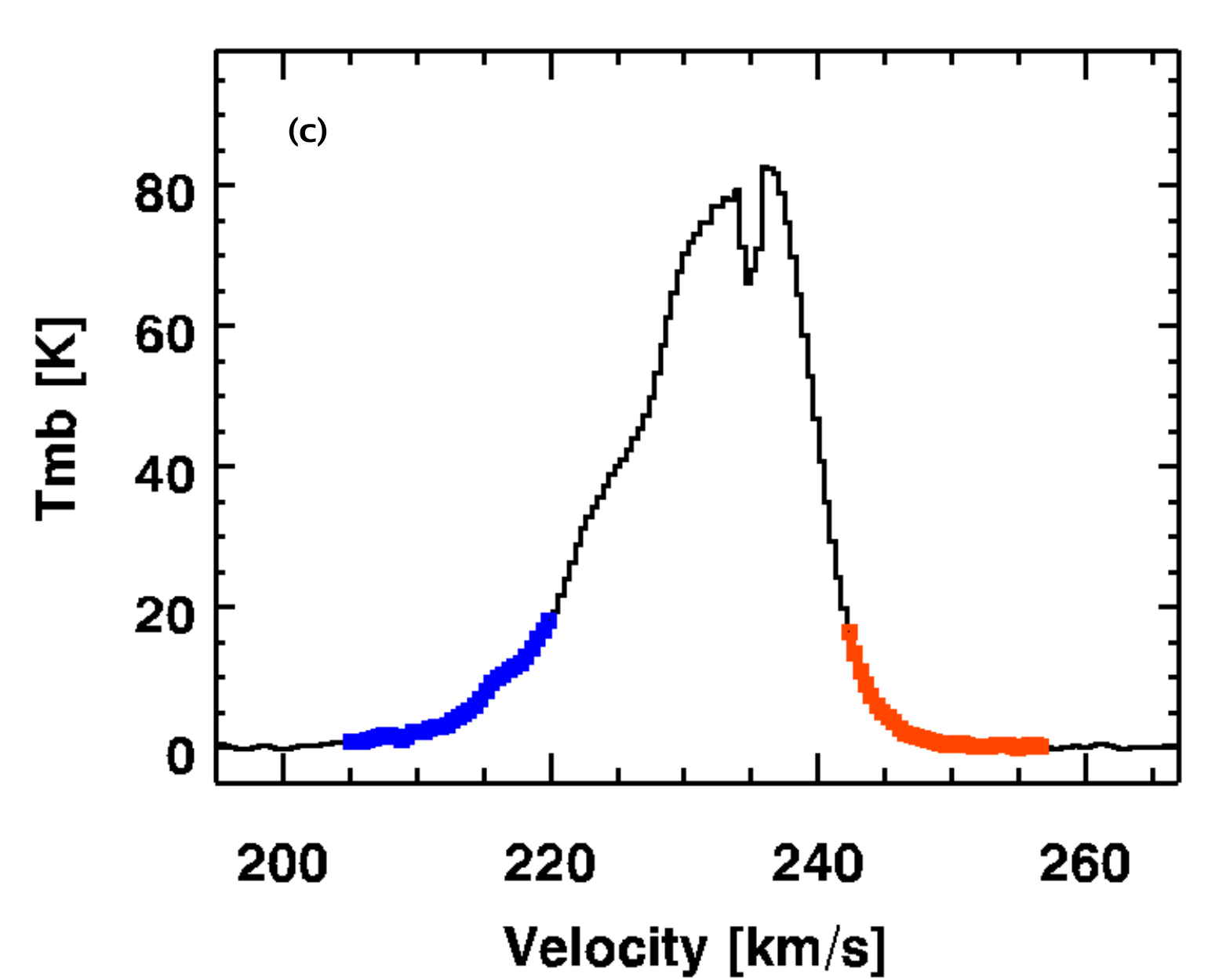}
\end{minipage}
\hskip -0.1 cm
\begin{minipage}{0.33\textwidth}
\includegraphics[width=\textwidth]{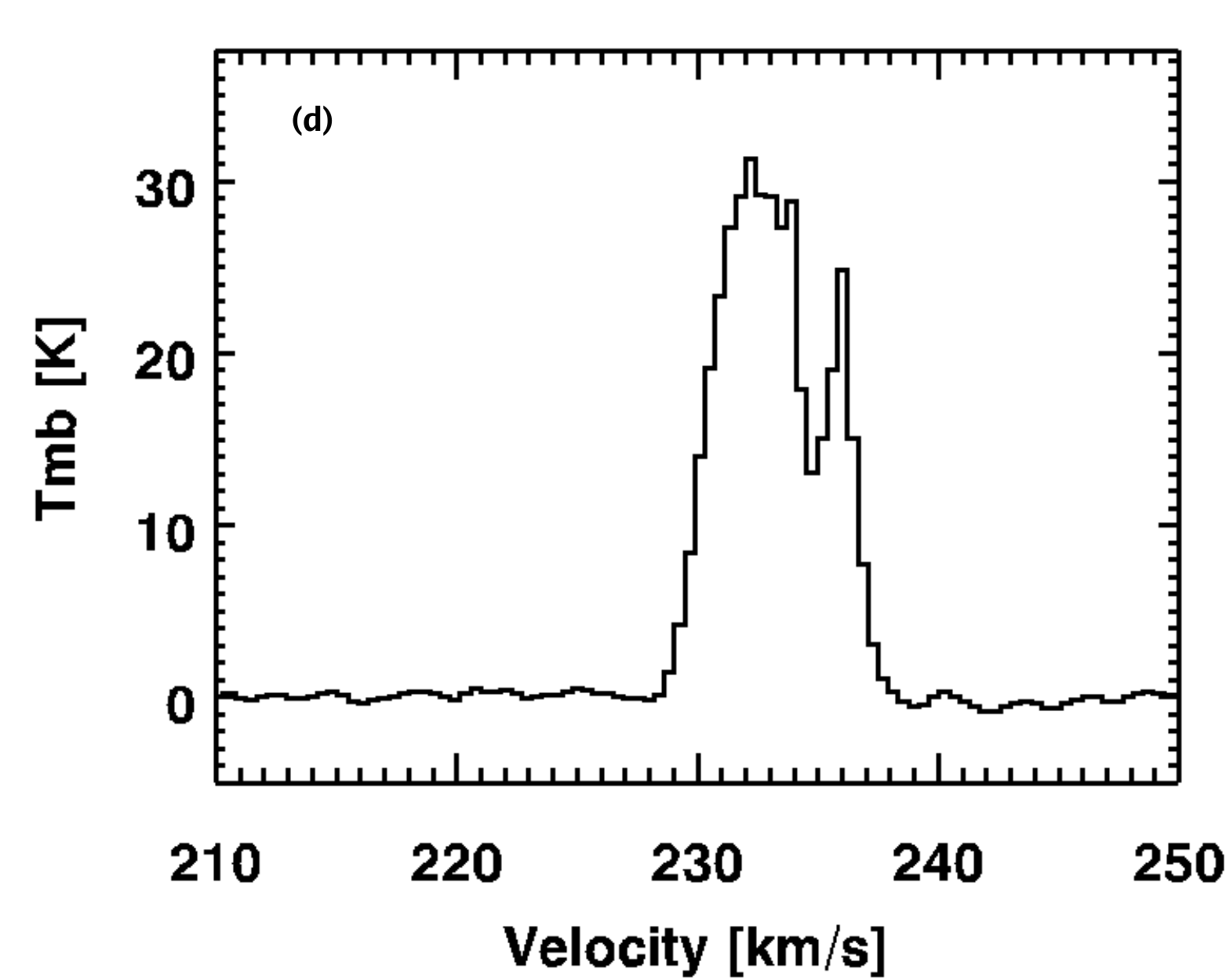}
\end{minipage}
\hskip -0.1 cm
\begin{minipage}{0.33\textwidth}
\includegraphics[width=\textwidth]{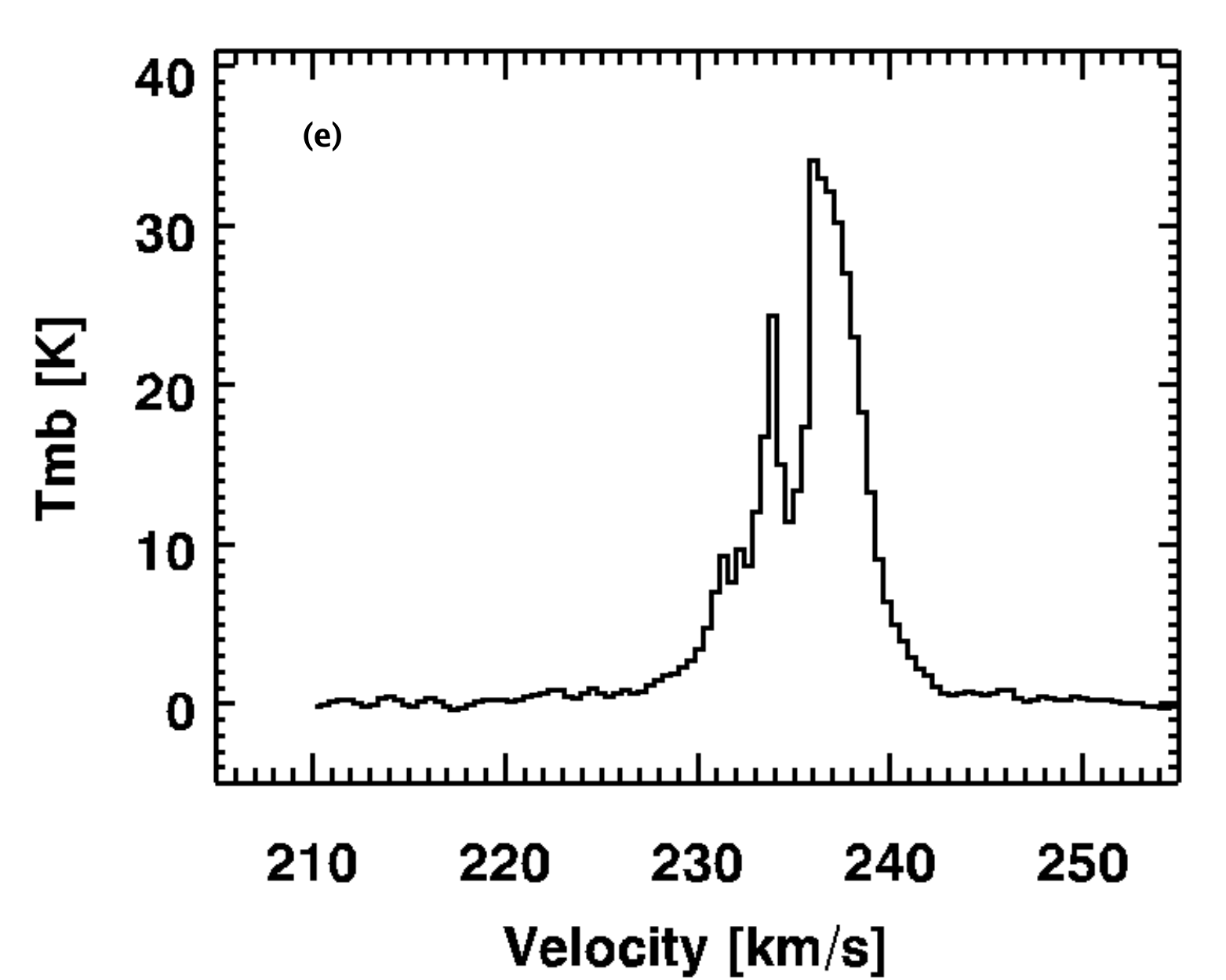}
\end{minipage}
\caption{(a) Integrated intensity distributions of CO(3-2) lines. White crosses represent the CH$_3$OH emission peak at N79S-1 and N79S-2. (b) Spatial distribution of blue-shifted and red-shifted emission of CO(3-2) lines. Blue contours represent the blue-shifted component (integrated over 205-220 km s$^{-1}$), while red contours represent the red-shifted component (integrated over 242-257 km s$^{-1}$) and the systemic velocity is 234.5 km s$^{-1}$. The background is the 0.87 mm continuum flux. (c) Spectral line profile of CO(3-2) lines extracted from SSC continuum peak. The velocity ranges of the blue-shifted and red-shifted components are indicated by blue and red, respectively. (d) Spectral line profile of CO(3-2) lines extracted from CH$_3$OH peak emission at N79S-1 and the systemic velocity is 232.5 km s$^{-1}$ (e) Spectral line profile of CO(3-2) lines extracted from CH$_3$OH peak emission at N79S-2 and the systemic velocity is 234.0 km s$^{-1}$ }  
\label{fig:CO_mom}
\end{figure*}

The integrated intensity distribution of all observed molecules is shown in Figures \ref{fig:mom1}, \ref{fig:mom2} and \ref{fig:CO_mom} (see the section for details \ref{sec:line_image}).

%%%%%%%%%%%%%%%%%%%%%%%%%%%%%%%%%%%%%%%%%%%%%%%%%%%%%

%%%%%%%%%%%%%%%%%%%%%%%%%%%%%%%%%%%%%%%%%%%%%%%%%%%%%%%%%%%%%%%%%%
%%%%%%%%%%%%%%%%%%%

\begin{figure*}
%\hskip -0.5 cm
\begin{minipage}{0.22\textwidth}
\includegraphics[width=\textwidth,trim= 0.0cm 0.0cm 0.0cm 0.0cm]{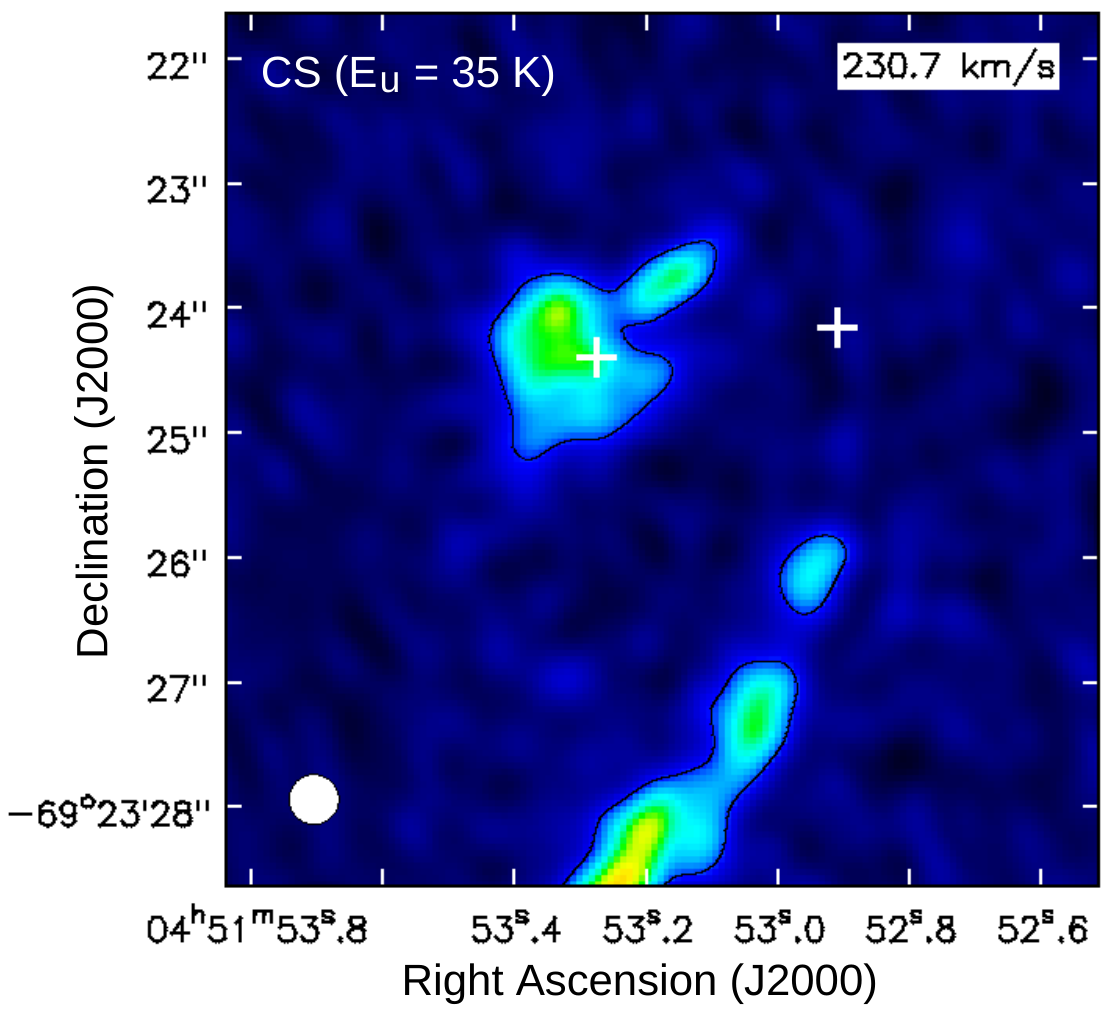}
\end{minipage}
%trim=left bottom right top
\hskip -0.2 cm
\begin{minipage}{0.1767\textwidth}
\includegraphics[width=\textwidth,trim= 0.00cm 0.0cm 0.0cm 1.76cm]{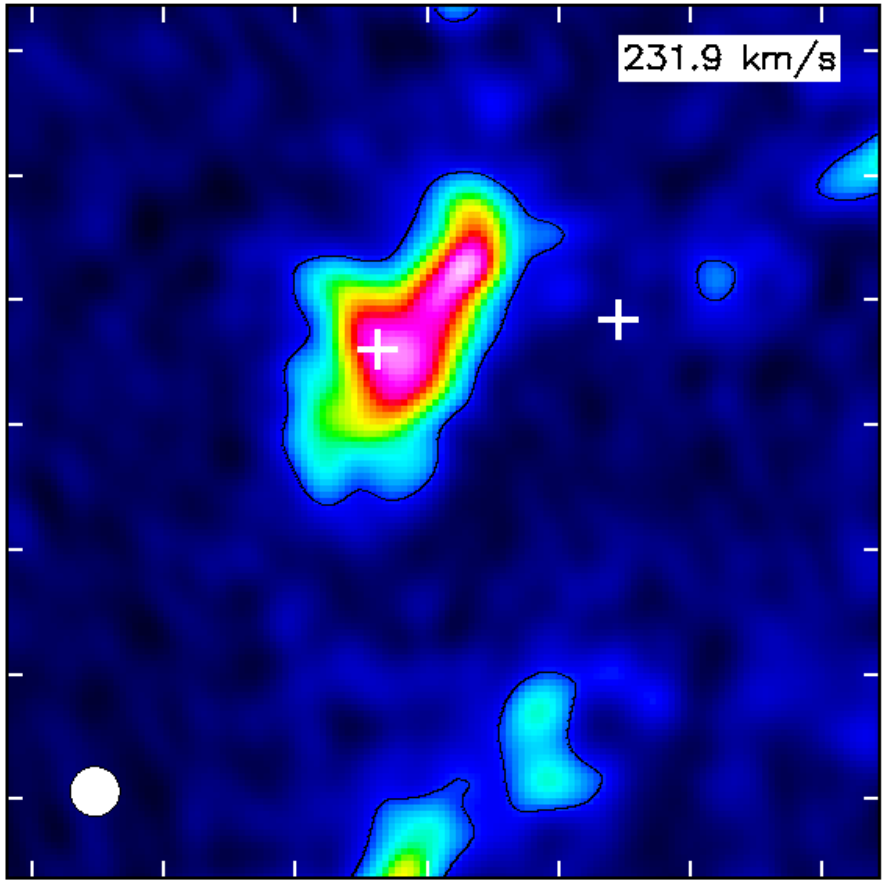}
\end{minipage}
\hskip -0.2cm
\begin{minipage}{0.1767\textwidth}
\includegraphics[width=\textwidth,trim= 0.00cm 0.0cm 0.0cm 1.76CMcm]{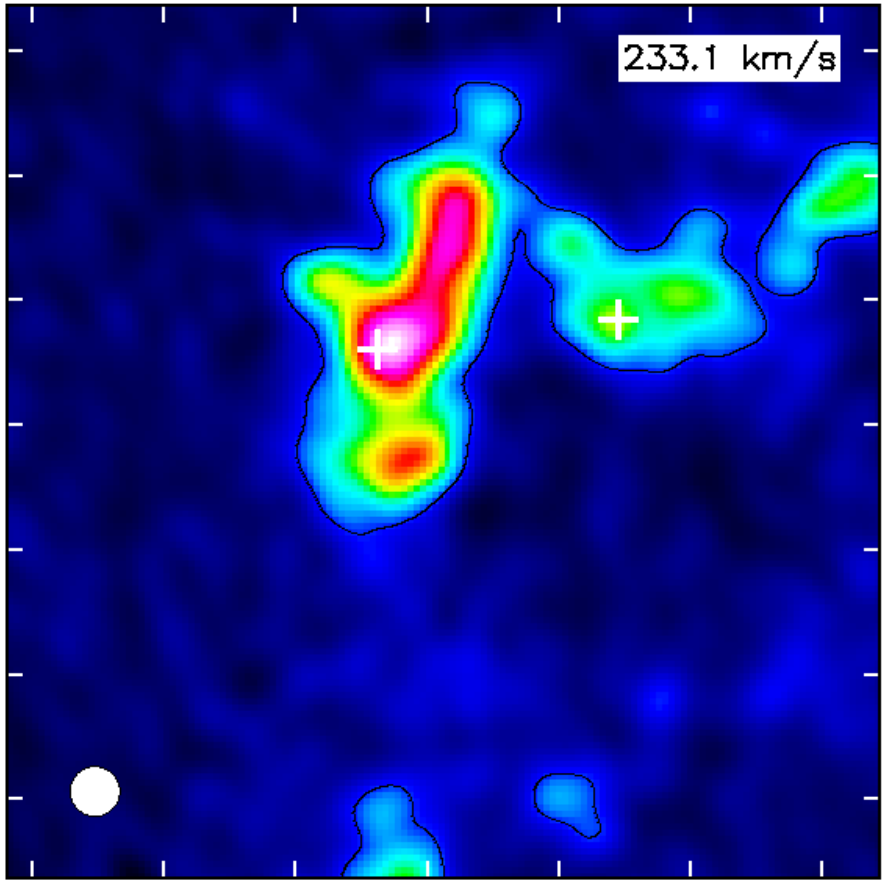}
\end{minipage}
\hskip -0.18 cm
\begin{minipage}{0.1767\textwidth}
\includegraphics[width=\textwidth,trim= 0.00cm 0.0cm 0.0cm 1.76cm]{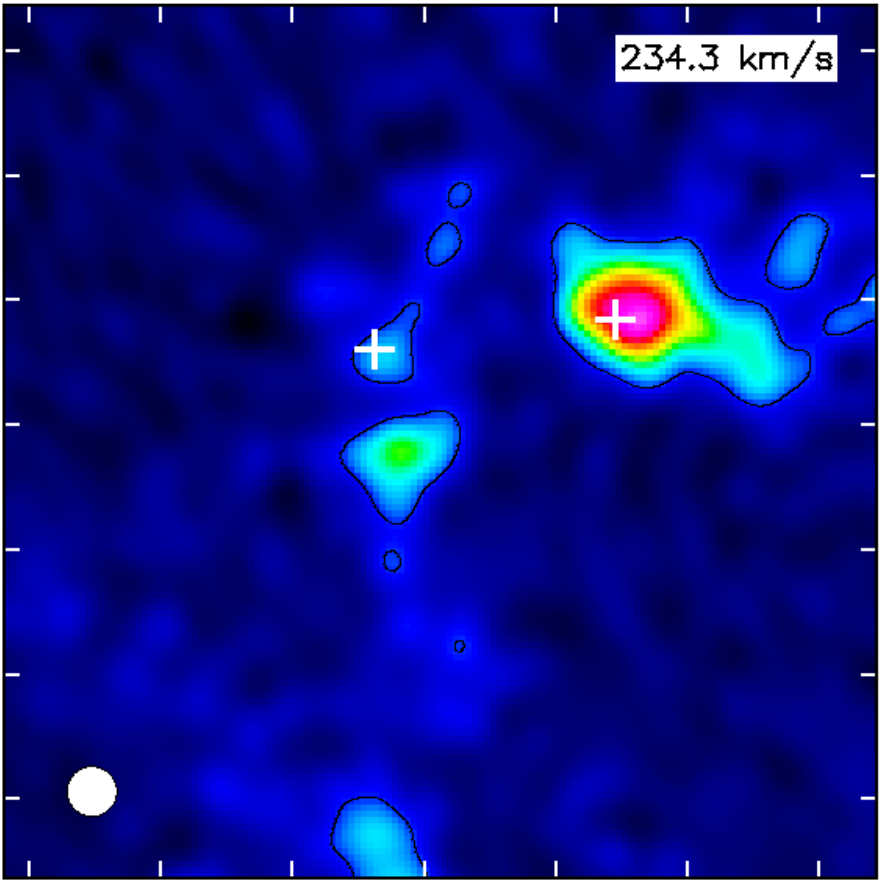}
\end{minipage}
\hskip -0.18 cm
\begin{minipage}{0.23\textwidth}
\includegraphics[width=\textwidth,trim= 0.00cm 0.0cm 0.0cm 1.76cm]{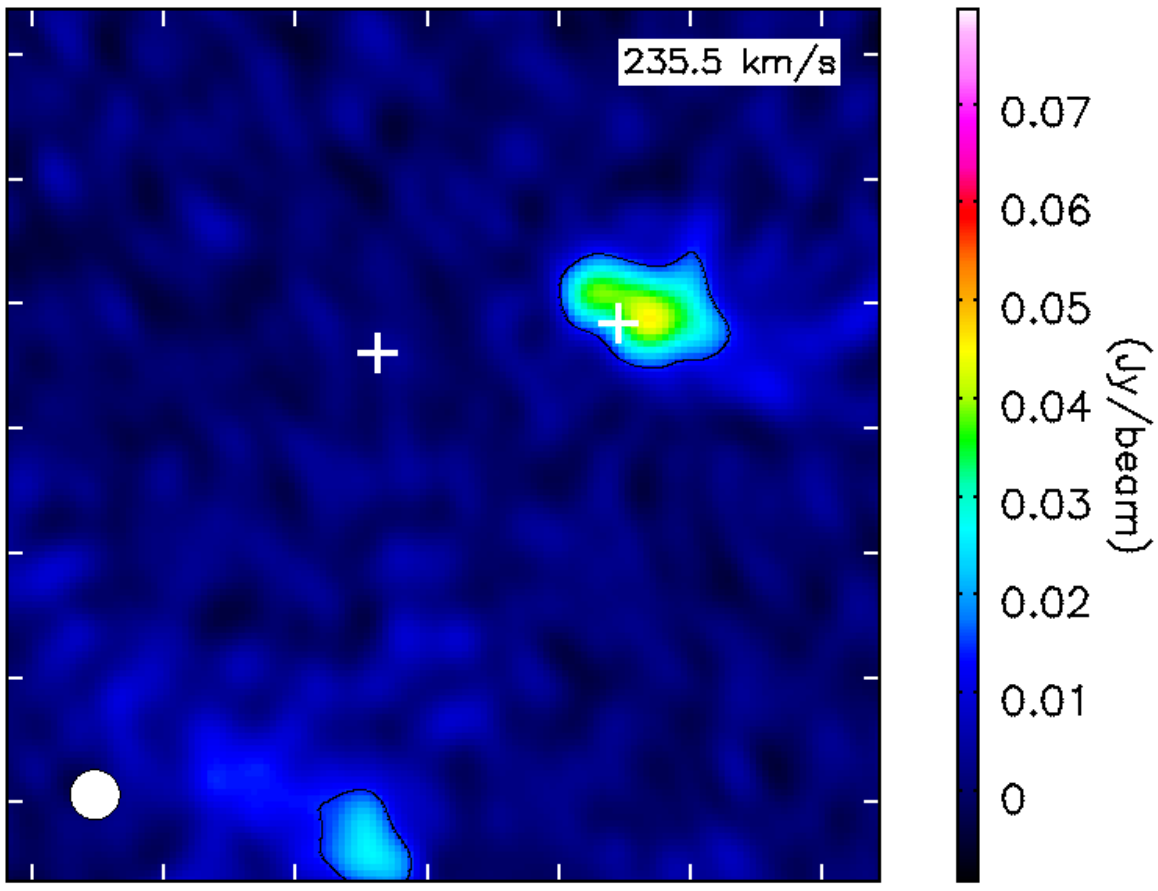}
\end{minipage}
%\hskip -2.0 cm
\begin{minipage}{0.22\textwidth}
\includegraphics[width=\textwidth,trim= 0.0cm 0.0cm 0.0cm 0.0cm]{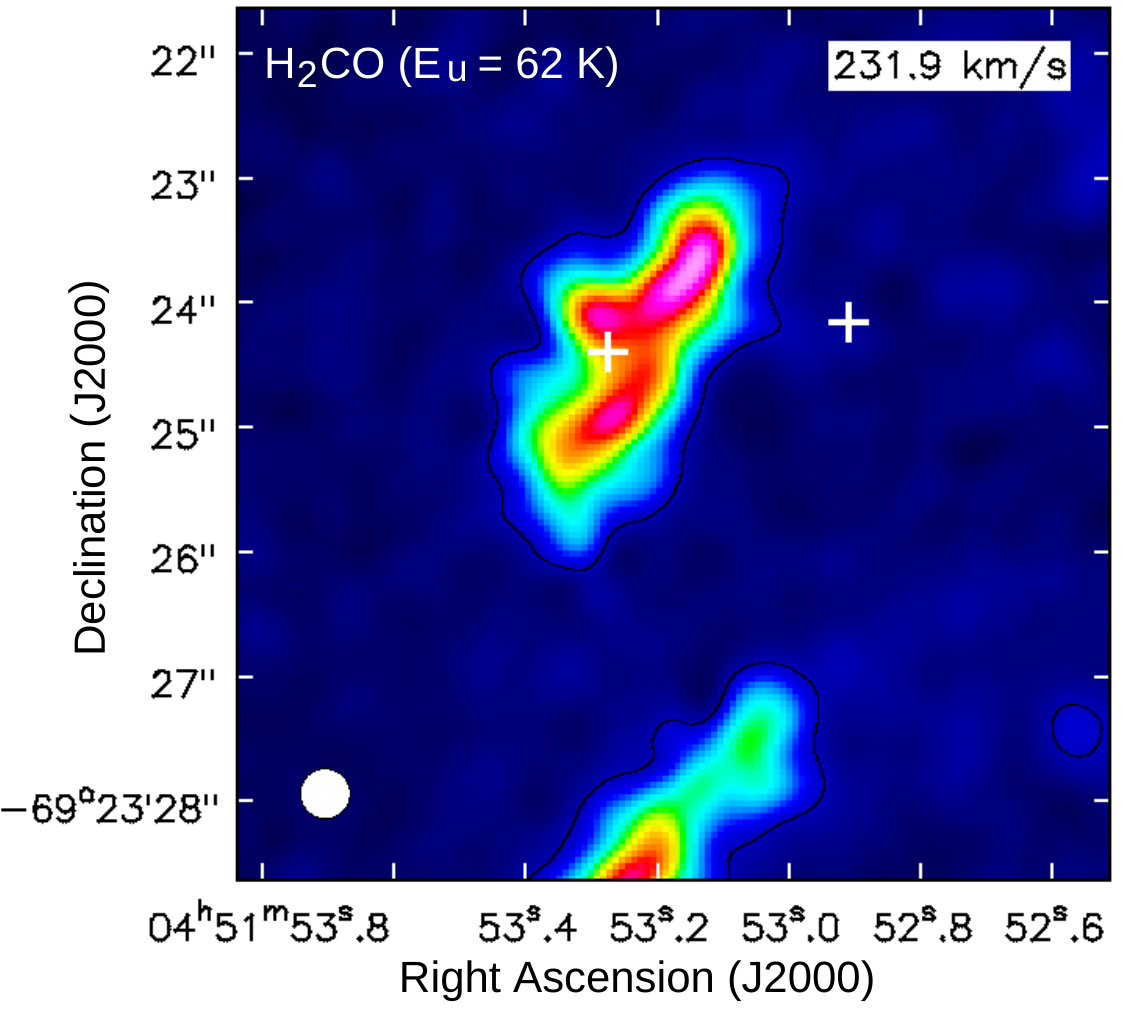}
\end{minipage}
%trim=left bottom right top
\hskip -0.1 cm
\begin{minipage}{0.1767\textwidth}
\includegraphics[width=\textwidth,trim= 0.00cm 0.0cm 0.0cm 1.85cm]{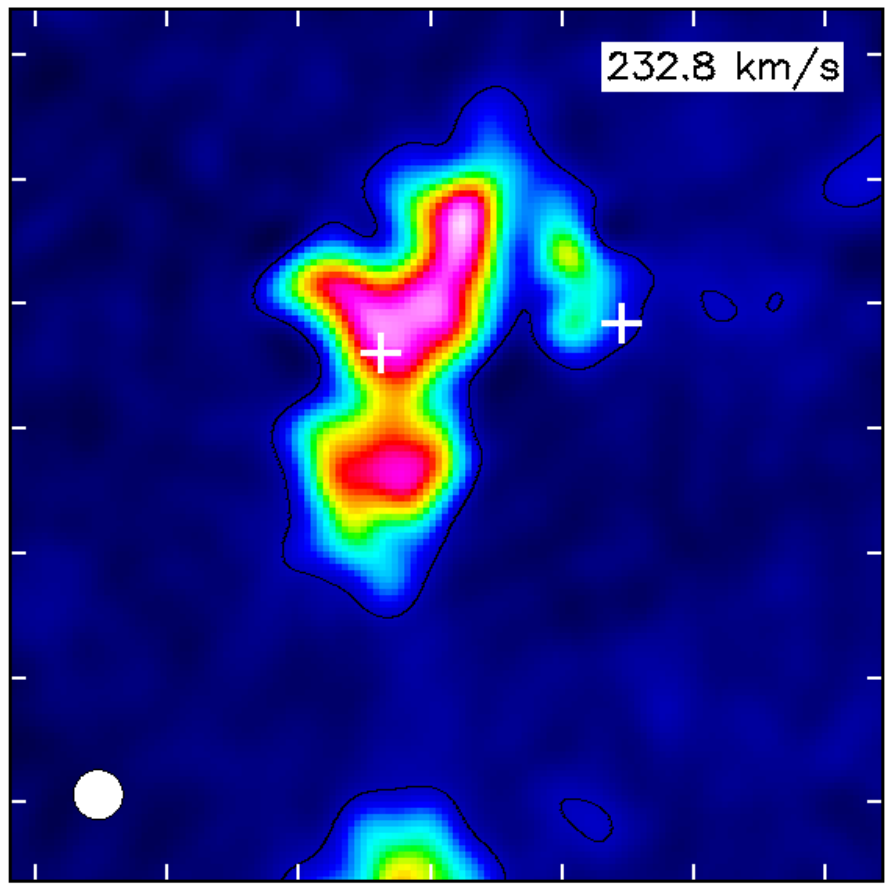}
\end{minipage}
\hskip -0.1cm
\begin{minipage}{0.1767\textwidth}
\includegraphics[width=\textwidth,trim= 0.00cm 0.0cm 0.0cm 1.85CMcm]{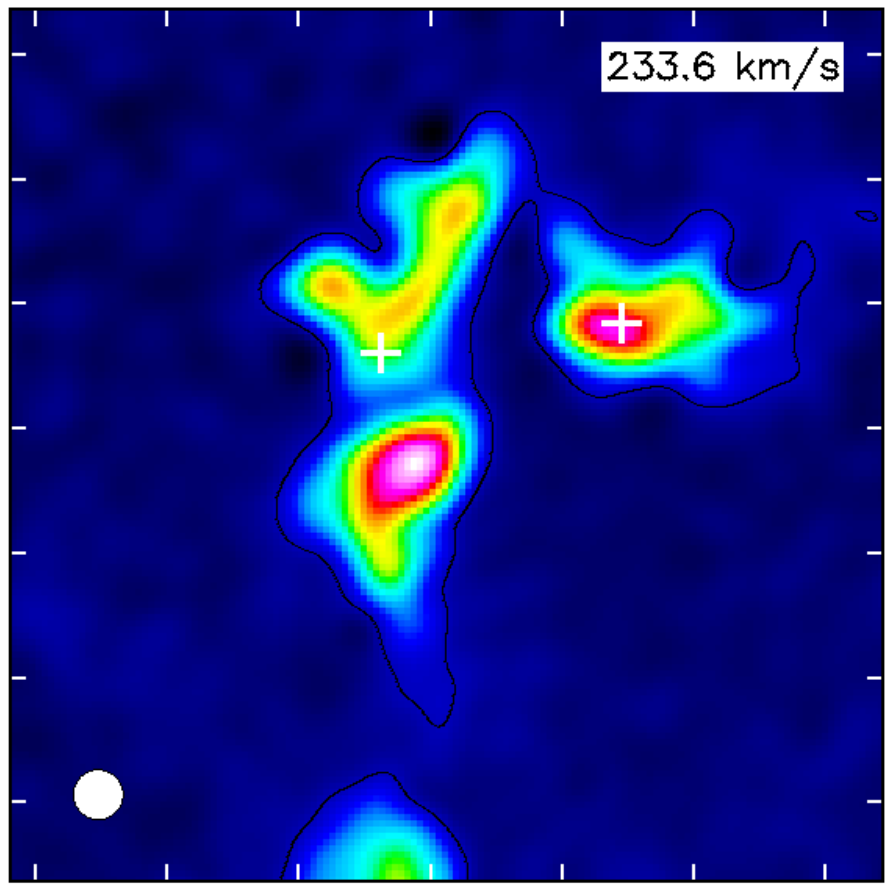}
\end{minipage}
\hskip -0.08 cm
\begin{minipage}{0.1767\textwidth}
\includegraphics[width=\textwidth,trim= 0.00cm 0.0cm 0.0cm 1.85cm]{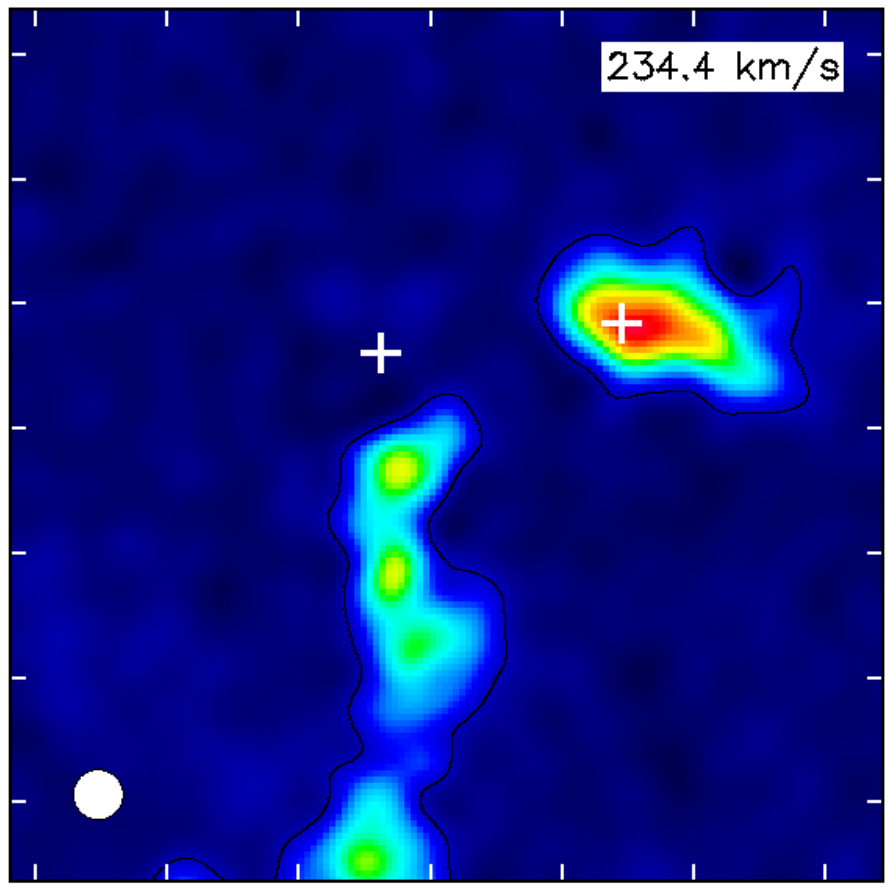}
\end{minipage}
\hskip -0.08 cm
\begin{minipage}{0.23\textwidth}
\includegraphics[width=\textwidth,trim= 0.00cm 0.0cm 0.0cm 1.85cm]{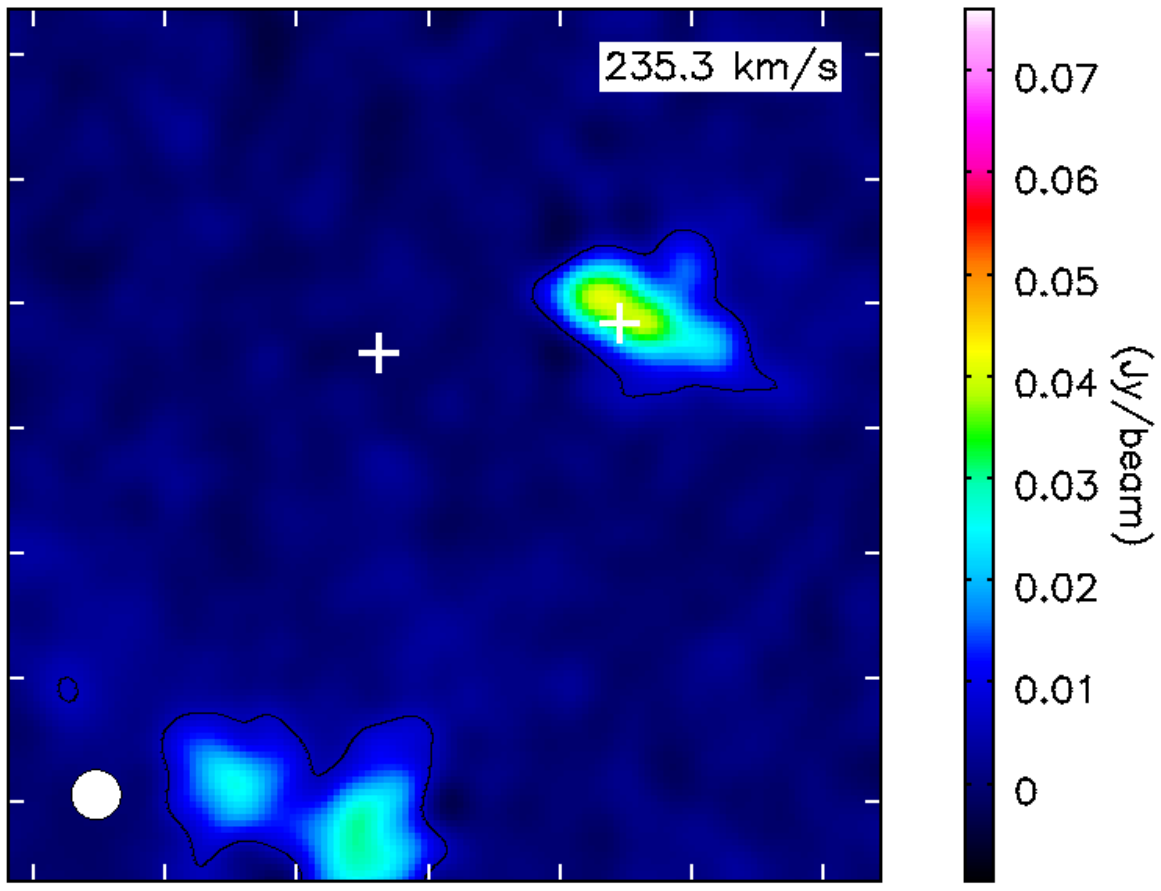}
\end{minipage}
\caption{Channel maps for CS (5-4) and H$_2$CO ({5$_{1,5}$-4$_{1,4}$}) (velocity ranges for CS (top panels): 230.7-235.5 km/s and velocity ranges for H$_2$CO (bottom panels) :231.9 - 235.3 km/s). White crosses represent the CH$_3$OH peak emission at N79S-1 (left cross) and N79S-2 (right cross). The contours represent the 5$\sigma$ of the rms noise. The synthesized beam size is shown by the white-filled circle in each panel.}
\label{fig:chan_map}
\end{figure*}
\clearpage
\bibliography{main_suman}{}
\bibliographystyle{aasjournal}
\end{document}